\documentclass[a4paper,fleqn]{cas-dc}

\usepackage[numbers,sort&compress]{natbib}
\usepackage{makecell}
\usepackage{array}
\usepackage{booktabs}
\usepackage{lscape}
\usepackage{tikz}
\usepackage[flushleft]{threeparttable}
\def\checkmark{\tikz\fill[scale=0.4](0,.35) -- (.25,0) -- (1,.7) -- (.25,.15) -- cycle;}

\def\tsc#1{\csdef{#1}{\textsc{\lowercase{#1}}\xspace}}
\tsc{WGM}
\tsc{QE}
\tsc{EP}
\tsc{PMS}
\tsc{BEC}
\tsc{DE}
\usepackage[nonumberlist,toc=true]{glossaries} 
\makeglossaries
\newacronym{ui}{UI}{User Interface}
\newacronym{sgg}{SGG}{Serious Games and Gamification}
\newacronym{sdt}{SDT}{Self Determination Theory}
\newacronym{copd}{COPD}{Chronic Obstructive Pulmonary Disease}
\newacronym{ict}{ICT}{Information and Communication Technologies}
\newacronym{hcp}{HCP}{Healthcare Professional}
\newacronym{hcps}{HCPs}{Healthcare Professionals}
\newacronym{dhts}{DHTs}{Digital Health Technologies}
\newacronym{asd}{ASD}{Autism Spectrum Disorder}
\newacronym{cbt}{CBT}{Cognitive Behavioural Therapy}
\newacronym{scas}{SCAS}{Spence Children Anxiety Scale}
\newacronym{scasp}{SCAS-P}{Spence Children Anxiety Scale-Parents}
\newacronym{dsm}{DSM-V}{Diagnostic and Statistical Manual of Mental Disorders, Fifth Edition}
\newacronym{ocd}{OCD}{Obsessive-Compulsive Disorder}
\newacronym{gad}{GAD}{Generalised Anxiety Disorder}
\newacronym{pmr}{PMR}{Progressive Muscle Relaxation}
\newacronym{gi}{GI}{Guided Imagery}
\newacronym{nice}{NICE}{National Institute for Health and Care Excellence}
\newacronym{qol}{QOL}{quality of life}
\newacronym{adhd}{ADHD}{Attention Deficit Hyperactivity Disorder}
\newacronym{mse}{MSE}{Multi-Sensory Environments}
\newacronym{plb}{PLB}{Pursed Lip Breathing}
\newacronym{rcmas}{RCMAS}{Revised Children's Manifest Anxiety Scale }
\newacronym{cats}{CATS}{Children's Automatic Thoughts Scale}

\newacronym{mdao}{MDAO}{Mechanics, Dynamics, Aesthetics, and Outcomes}
\newacronym{sgh}{SGH}{Serious Games for Health}

\newacronym{bfi}{BFI-44}{Big Five Inventory}
\newacronym{iqr}{IQR}{interquartile range}
\newacronym{spv}{SPV}{Scale Point Variation}

\newacronym{hapa}{HAPA}{Health Action Process}
\newacronym{ncd}{NCD}{Non-communicable disease}
\newacronym{hbc}{hBC}{Mobile health Behavioural Change}
\newacronym{hbcss}{hBCSS}{Mobile health Behaviour Change Support Systems}
\newacronym{ipaq-lf}{IPAQ-LF}{International Physical Activity Questionnaire- Long Form}

\newacronym{usn}{USN}{Unilateral Spatial Neglect}
\newacronym{vr}{VR}{Virtual Reality}
\newacronym{ar}{AR}{Augmented Reality}
\newacronym{hmd}{HMD}{Head Mounted Display}
\newacronym{cbs}{CBS}{Catherine Bergego Scale}
\newacronym{tap}{TAP}{Test for Attention Performance}
\newacronym{vst}{VST}{Visual Scanning Training}
\newacronym{oks}{OKS}{Optokinetic Stimulation}
\newacronym{nvm}{NVM}{Neck-Muscle Vibration}
\newacronym{cvs}{CVS}{Caloric Vestibular Stimulation}
\newacronym{gvs}{GVS}{Galvanic Vestibular Stimulation}
\newacronym{bit}{BIT}{Behavioural Inattention Test}
\newacronym{pa}{PA}{Prism Adaptation}
\newacronym{la}{LA}{Limb Activation}
\newacronym{tms}{TMS}{Transcranial Magnetic Stimulation}
\newacronym{tdcs}{tDCS}{transcranial Direct Current Stimulation}
\newacronym{sat}{SAT}{Sustained Attention Training}
\newacronym{nmt}{NMT}{Neurologic Music Therapy}
\newacronym{mnt}{MNT}{Music Neglect Therapy}
\newacronym{who}{WHO}{World Health Organisation}
\newacronym{gdpr}{GDPR}{General Data Protection Regulation}
\newacronym{gpu}{GPU}{Graphics Processing Unit}

\newacronym{nl}{NL}{Nutrition Literacy}
\newacronym{fl}{FL}{Food Literacy}
\newacronym{mlp}{MLP}{Multilayer Perceptron}
\newacronym{ann}{ANN}{artificial neural network}
\newacronym{rl}{RL}{reinforcement learning}
\newacronym{sgs}{SGs}{Serious Games}
\newacronym{sg}{SG}{Serious Game}
\newacronym{npcs}{NPCs}{Non-Player Characters}
\newacronym{ai}{AI}{Artificial Intelligence}
\newacronym{genai}{genAI}{Generative Artificial Intelligence}
\newacronym{sggdutch}{SGG}{Serious Games en Gamificatie}
\newacronym{ass}{ASS}{Autisme Spectrum Stoornis}
\newacronym{cgt}{CGT}{Cognitieve Gedragstherapie}
\newacronym{ml}{ML}{Machine Learning}
\newacronym{mcts}{MCTS}{Monte Carlo Tree Search}
\newacronym{eeg}{EEG}{Electroencephalogram}
\newacronym{mab}{MAB}{Multi-Armed Bandit}
\newacronym{cnn}{CNN}{Convolutional Neural Network}
\begin{document}
\let\WriteBookmarks\relax
\def\floatpagepagefraction{1}
\def\textpagefraction{.001}

\shorttitle{Personalization in Serious Games and Gamification for Healthcare}

\shortauthors{S. Carlier et~al.}

\title [mode = title]{Personalization in Serious Games and Gamification for Healthcare: A Three-Tiered Review of Models, Methods and Opportunities}                      



%
\author[1]{St\'{e}phanie Carlier}[
                        style=chinese,
                       orcid=0000-0001-6150-717X
                      ]

\cormark[1]


\ead{stephanie.carlier@ugent.be}



\affiliation[1]{organization={IDLab, iGent Tower—Department of Information Technology, Ghent University—imec},
    addressline={Technologiepark-Zwijnaarde 126}, 
    city={Ghent},
   postcode={B-9052}, 
    country={Belgium}}

\author[1]{Femke De Backere}[style=chinese]
\fnmark[1]

\author[1]{Filip De Turck}[style=chinese]
\fnmark[1]





\begin{abstract}
Serious games and gamification (SGG) have shown to have positive effects on health outcomes of eHealth applications. However, research has shown that a shift towards a personalized approach is needed, considering the diversity of users. This introduces new challenges to the domain of SGG as research is needed on how such personalization is achieved. A literature search was conducted to provide an overview of personalization strategies. In total, 50 articles were identified, 35 reported on a serious game and 15 focused on gamification. We introduce a three‑tiered classification model, including a model level, a personalization paradigm level, and algorithmic framework level to synthesize how personalization is implemented. Data‑driven approaches are most common overall (22/50), with knowledge‑driven and hybrid methods more prevalent in rehabilitation, reflecting safety and explainability requirements. Popular modeling choices include Hexad‑based player modeling and ontologies for expert knowledge integration. Despite encouraging results, reusability remains limited, impeding comparison and knowledge transfer. This review outlines opportunities for progress: shareable knowledge assets, swap‑friendly personalization engines, and clinically bounded hybrid approaches, alongside cautious use of generative AI to accelerate design while maintaining safety and explainability. This classification framework and synthesis aims to guide more modular, comparable, and clinically aligned personalized SGG.

\end{abstract}


\begin{highlights}
	\item A unified three-tiered classification framework that structures how personalization is implemented in \gls{sgg} for health. 
\item A comprehensive synthesis of 50 personalized \gls{sgg} systems published between 2014-2025, spanning 6 health domains
\item A detailed examination of player and expert knowledge models, highlighting how user data, medical data, sensor data, and domain expertise are represented and integrated withing such \gls{sgg} systems.
\item  A cross-study analysis of personalization paradigms, revealing domain-specific trends and constraints.
\item The first focused review of reusability in personalized \gls{sgg}, identifying how reusable components, such as ontologies or modular architectures, are currently implemented and where the gaps remain. 
\end{highlights}



\maketitle
\section{Introduction}
\label{sec:rev_introduction}
The use of \gls{sgg} for healthcare is increasingly popular as its use has shown positive effects on treatment adherence, user motivation and patient education~\cite{damasevicius_serious_2023,alahaivala_understanding_2016,metwally_does_2021, fitzgerald_serious_2020, seyderhelm_towards_2019,graafland_how_2018}.  Gamification is the use of game elements, such as rewards and leader boards, in a non-gaming context. The choice of included game elements can range from a few elements to a more game-like experience~\cite{tondello_empirical_2019}. \gls{sgs}, on the other hand, are games with a primary objective other than pure entertainment, such as education or training~\cite{susi_serious_2015, ritterfeld_serious_2009}. \gls{sgg} are used in a wide range of health domains, for example, physical and cognitive rehabilitation~\cite{afyouni_adaptive_2020,afyouni_motion-based_2017,aguilar_adaptive_2019,gonzalez-gonzalez_serious_2019,lau_framework_2021, silva_spatial_2020,goumopoulos_ontology-driven_2021,martinho_systematic_2020,vermeir_effects_2020}, the education of health professionals and patients~\cite{haoran_serious_2019,gorbanev_systematic_2018,abraham_investigating_2020,sharifzadeh_health_2020,ricciardi_comprehensive_2014}, health behavior change, such as the cessation of substance abuse or the improvement of physical activity~\cite{hervas_gamification_2017,david_how_2020} and the treatment of mental health disorders, such as anxiety and depression~\cite{fitzgerald_serious_2020}. Results indicate that \gls{sgg} show promise in reducing issues with treatment adherence in healthcare and that they can be effective tools for health, however, research remains in its infancy, limited by design and evaluation challenges~\cite{sardi_systematic_2017,ricciardi_comprehensive_2014,sharifzadeh_health_2020,thomas_mapping_2020, hamari_does_2014, fitzgerald_serious_2020,david_how_2020, king_review_2021, sipiyaruk_rapid_2018}. \par
One of those challenges is that \gls{sgg} might not be sustainable as patients and users might lose interest over time, leading again to a decrease in treatment adherence and user engagement~\cite{sardi_systematic_2017}. Users of mobile applications all have their specific profile and their contexts might change and evolve, calling for a dynamic and adaptable approach to keep motivation high. Research has indicated that the \textit{one-size-fits-all} approach needs to be abandoned to shift towards more personalized \gls{sgg}, that are able to re-engage the user~\cite{tondello_empirical_2019,martinho_systematic_2020,vermeir_effects_2020,hamari_does_2014,sajjadi_individualization_2022, van_dooren_reflections_2019, verschueren_developing_2019, de_troyer_towards_2017,blatsios_towards_2019,lazzaro_why_2004}. Moreover, designing and implementing a personalized \gls{sg} is a costly and challenging process as it requires the same effort from multiple stakeholders, such as (game) developers, software engineers and domain experts,  all over again for each \gls{sg}~\cite{goumopoulos_ontology-driven_2021,streicher_personalized_2016}. While the development of gamified interventions can be considered slightly less cost-intensive as it does not require the development of a full-fledged game, it should be avoided to use gamification as chocolate-dipped-broccoli, i.e. applied as an afterthought, but to integrate it from the start in the design process~\cite{sanchez_gamification_2019}. To create effective \gls{sgs} and gamified mHealth, or mobile health applications, domain expertise from health professionals is needed, involving them in each step of the design and development process~\cite{verschueren_developing_2019,korhonen_multidisciplinary_2017}.  \par
Personalized \gls{sgg}, with a user-centered approach, show promise in improving performance outcomes and boosting engagement~\cite{chow_can_2020,wouters_meta-analysis_2013,gentry_serious_2019,wiemeyer_serious_2012,alahaivala_understanding_2016}. Research exists on personalized \gls{sgg}, the obtained results so far are promising but challenged by the uncertainty on how personalization can be integrated to increase health outcomes~\cite{vermeir_effects_2020,mora_quest_2019}. Several reviews on personalized \gls{sgg} exist, focusing on which player aspects are used for the individualization of \gls{sgs}~\cite{sajjadi_individualization_2022}, difficulty adaptation and procedural content generation~\cite{paraschos_game_2023}, how game elements have been chosen and used in personalized gamification~\cite{rodrigues_personalized_2021,klock_tailored_2020}, how machine learning and \gls{ai} and gamification can interact~\cite{khakpour_convergence_2021} and how player models and adaptation methods are integrated~\cite{hare_player_2023}. 
Despite the growing interest in personalized serious games and gamified health interventions, existing reviews typically address either gamification or serious games in isolation, emphasize specific components such as player models or difficulty adaptation, or lack a health-domain focus. As a result, the domain still lacks a unified perspective on how personalization is technically achieved across different health applications and how reusable components can support salable design. \par 
This review aims to fill this gap by going beyond earlier surveys that treat gamification and serious games separately or focus on isolated approaches. Instead, it maps 50 digital health interventions using a three-tiered classification framework and provides the first focused analysis of reusability in personalized \gls{sgg}.\par
 To support researchers, engineers, game designers and healthcare professionals in designing personalized game-based health interventions, this review makes the following contributions: 
 \begin{itemize}
 	\item A unified three-tiered classification framework that structures how personalization is implemented in \gls{sgg} for health. 
 	\item A comprehensive synthesis of 50 personalized \gls{sgg} systems published between 2014-2025, spanning 6 health domains based on the applicability and existing prototype implementations
 	\item A detailed examination of player and expert knowledge models, highlighting how user data, medical data, sensor data, and domain expertise are represented and integrated withing such \gls{sgg} systems.
 	\item  A cross-study analysis of personalization paradigms, revealing domain-specific trends and constraints.
 	\item The first focused review of reusability in personalized \gls{sgg}, identifying how reusable components, such as ontologies or modular architectures, are currently implemented and where the gaps remain.

 \end{itemize}

The remainder of the paper is structured as follows: First, Section~\ref{sec:rev_Background} provides an introduction to the spectrum of game-based solutions and a discussion on personalization terminology. This is followed by Section~\ref{sec:rev:Methods}, explaining the search strategy, the inclusion criteria and selection procedure of the identified records. Next, Section~\ref{sec:rev:overview} provides an overview of included studies. The results of the three-tiered classification are discussed in Sections~\ref{sec:rev:model},\ref{sec:rev:paradigm} and \ref{sec:rev:algorithmic}. Next, the prevalence of reusability is discussed in Section~\ref{sec:rev:reuse}, followed by the findings (Section~\ref{sec:rev:findings}), future opportunities (Section~\ref{sec:rev:future}) and finally conclusion (Section~\ref{sec:rev:conclusion}).  

\section{Background}
\label{sec:rev_Background}
Although Section~\ref{sec:rev_introduction} provided a brief explanation of a \gls{sg}, no universally accepted definition of a serious game exists. Likewise, personalization is described in literature using a range of overlapping terms. The following sections elaborate on these concepts, clarify their distinctions, and establish how they are defined within the context of this review.

\subsection{In Search of a Definition}
\label{subsec:rev_definition}
 According to some researchers, the seriousness of a game is in the nature of the game itself~\cite{crookall_serious_2010}, while others claim it can only be evident through the resulting changes in the player~\cite{haring_understanding_2011}. The latter implies that commercial off-the-shelf games can be considered \gls{sg}s if the game has enriched the player, more specifically, trained, educated or changed behaviors. Nonetheless, some agreement is found in defining a serious game as a game that has a purpose beyond pure entertainment, such as to educate or to train~\cite{blumberg_serious_2013}. \par
Gamification, on the other hand, is used when game elements, such as badges or leader boards, are applied in a non-game environment~\cite{deterding_game_2011}. Gamification is not a game, it is a tool used to motivate people to continue or to change certain behaviors such as smoking in exchange for points or badges~\cite{loh_serious_2015}. The definition of a \gls{sg} is similar to that of gamification in that they both incorporate game elements in a non-game context with a purpose other than entertainment and they aim to engage users. They differ however, in that \gls{sg}s incorporate a mixture of all these game elements in varying degrees, while gamification identifies and applies individual game elements to form a meaningful combination to motivate the user. Gamification thereby augments an existing process by adding game elements without creating a new game~\cite{landers_developing_2014}. \par
\begin{figure*}[]
	\centering
	\includegraphics[scale=0.5]{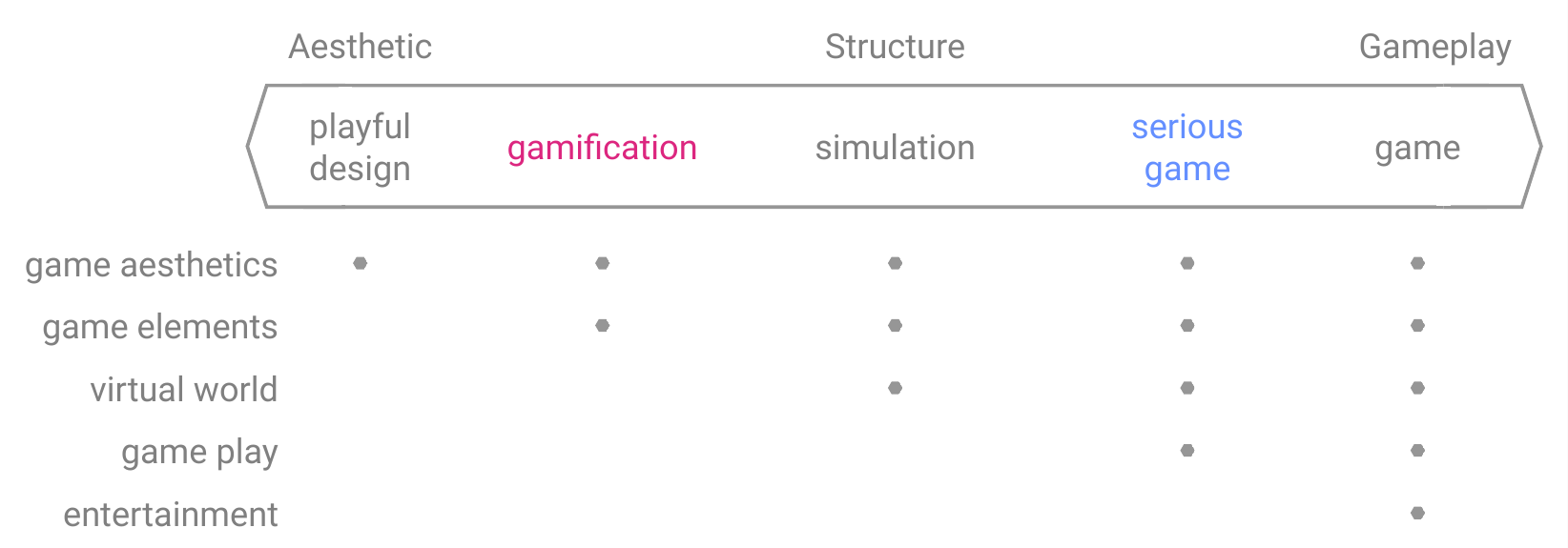}
	\caption{The game thinking spectrum, according to Marczewski~\cite{marczewski_even_2015}, visualizes the different relationships and characteristics between game-based solutions.}
	\label{fig:rev:game_thinking}
\end{figure*}

Marczewski~\cite{marczewski_even_2015} introduced the umbrella term \textit{game thinking} to also include the use of games and game-like solutions in non-game contexts, as depicted in Figure~\ref{fig:rev:game_thinking}. The left side of the spectrum considers the aesthetic of game-based solutions, inspired by the look and feel of games, i.e., game-like. The right side of the spectrum introduces solutions that are more game than game-like. In the middle, solutions will look like games and share some structural elements. \par 
For this review, both ends of the spectrum will be excluded as playful design solutions lack any actual game elements and full-fledged games prioritize entertainment above all else. 

\subsection{The Complexity of Personalization}
\label{subsec:rev_personalisation}
In digital healthcare, one of the challenges is often to support the targeted users long-term and continuously, irregardless of the user's evolving abilities or current context. For such a dynamic state, the possibility of a different configuration or altered support might be necessary. Many researchers have indicated the need for personalization in \gls{sgg}s to sustain long-term treatment adherence and retain user engagement. When games are adapted to individual users, performance outcomes improve, underscoring the importance of personalization in effective \gls{sgg} design. \par
In literature, different terms have been used to indicate the tailoring of serious games to users~\cite{carlier_software_2023}, such as adaptability, adaptivity, personalization, contextualization and customization. When no distinction between these terms is needed, the term \textit{individualization} can be used as an umbrella term~\cite{sajjadi_individualization_2022}. Personalization is, however, a complex subject and can take many forms and shapes depending on the moment of personalization and the type of personalization. When a game is adapted to the user's needs at design time, this is called static personalization. Dynamic personalization is when the adaptation happens while playing the game. Adaptability is considered as the possibility to change an environment based on the changing needs of the user, while adaptivity is the dynamic or automatic adjustment of game elements to the performance or actions of a player~\cite{sajjadi_individualization_2022,streicher_personalized_2016}. Next, personalization is defined by the, often automatic, adaptation of the game based on the profile or context of the user~\cite{sajjadi_individualization_2022}. Customization can then be seen as changing the system based on user preferences, manual or automatic, and is often related to changes in appearance and content~\cite{sajjadi_individualization_2022,turkay_effects_2014}. Streicher and Smeddinck~\cite{streicher_personalized_2016} consider personalization to be a specific form of customization, and adaptivity and adaptability as a way to achieve customization or personalization.  \par
This review includes concepts related to dynamic and automatic personalization, while studies on static personalization, such as player-centered design approaches that tailor games to user needs during the design phase, are excluded.

\section{Method}
\label{sec:rev:Methods}
The following paragraphs provide an overview of the search strategy that was used to identify the analyzed articles and second, a discussion of defined inclusion criteria and how the final studies were selected. Finally, Section~\ref{sec:rev:classification} explains the three different levels used for classifying the identified personalization methods. 
\subsection{Search Strategy}
The search was conducted in October 2025, using two databases, namely Web of Science and PubMed. To ensure high quality search results in the interdisciplinary domain of  \gls{sgg} for health, these two primary databases were selected. Web Of Science captures both technical innovations in algorithmic personalization, while PubMed covers the clinical validity of health-related interventions. This structured literature search was preceded by an exploratory search using Google Scholar to define the keywords to be used in the search. Table~\ref{table:rev:query} gives an overview of the used query and keywords with the respective number of articles that were retrieved from Web of Science and PubMed. Seven other articles were included in the results that were identified during the analysis of the found records. For the title it was required that some keyword referring to personalization and gamification and/or \gls{sgs} was included. Furthermore, for the topic of the paper, i.e. abstract and title, the domain of `healthcare' is delineated by all papers that refer to the health or well-being of patients, thereby excluding education, more specifically education of healthcare professionals and education of people with specific learning disorders. The publication year spans from 2014 to 2025, to ensure the review remained focused, manageable and reflective of current technologies and design practices in personalised \gls{sgg}. As the aim is to provide an overview of personalization algorithms and strategies for \gls{sgg} from the last decade, all review papers were excluded, as these were analyzed separately and reported upon above, in Section~\ref{sec:rev_introduction}.

\begin{table}[width=.9\linewidth,cols=2,h]
\caption{Search query and number of articles retrieved from the different databases.}\label{table:rev:query}
\begin{tabular*}{\tblwidth}{  LR }
\toprule
\multicolumn{2}{L}{Search Query}\\
\hline
\multicolumn{2}{L}{\makecell{Title=personali* OR adapt* OR context* OR individu*  \\ OR tailored OR intelligent OR "player model*" OR \\ "user model*" OR ontology OR customi*}}\\
\multicolumn{2}{L}{\makecell{AND Title="serious game*" OR gamification OR \\ gamified OR exergame*}}\\
\multicolumn{2}{L}{\makecell{AND Title or abstract=health* OR rehabilitation \\ OR treatment OR disorder OR "behavior change" \\OR disease OR "physical activity" OR "fitness" \\OR therapy OR "mental health" OR intervention*\\ OR chronic condition* OR self-management}}\\
\multicolumn{2}{L}{AND title= NOT review}\\
\multicolumn{2}{L}{AND Publication Year=2014-2025}\\

\hline
Database & Number of records\\
Web of Science & 288 \\
PubMed & 105 \\

Other sources & 7\\
\bottomrule
\end{tabular*}
\end{table}

\subsection{Inclusion Criteria and Study Selection}
\begin{figure}[]
	\centering
		\includegraphics[scale=1.4]{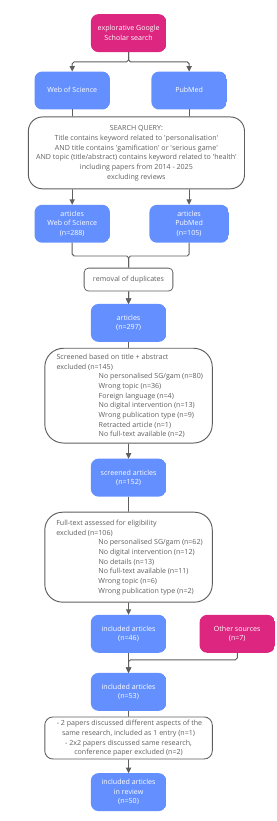}
	\caption{Flow chart of the literature search procedure.}
	\label{fig:rev:search}
\end{figure}

Figure~\ref{fig:rev:search} displays the number of publications identified, screened and excluded at each stage of the literature search and selection process. The structured search resulted in 288 articles from Web of Sciences and 105 articles from PubMed. After the removal of duplicate articles, 297 articles remained. These articles were screened based on title and abstract. After this first screening, the full text of the remaining 152 articles was analyzed. Seven more records were identified from other sources based on the expertise of the authors. In total, 50 articles are included in this literature review. Articles were excluded from the analysis if they described a gamified solution or serious game that was not personalized (screening n=80, full-text n=62) or if it did not include a digital intervention (screening n=13, full-text n=12). Furthermore, papers were excluded from the results if the topic was not health-related (screening n=36, full-text n=6) or if they were the wrong publication type, namely reviews or editorials (screening n=9, full-text n=2). Four articles were excluded due to not being available in English and of 13 articles no full text was found. Next, 13 papers were excluded after full-text assessment due to lack of details on the used personalization strategies. Finally, 2 papers that discussed different aspects of the same research have been included as 1 entry, and for 4 papers that reported the same research, the conference papers have been excluded (n=2). This brings the total of included articles to 50.

\subsection{Classification on Three Levels}
\label{sec:rev:classification}
To provide an in-depth and multi-dimensional analysis of how personalization is implemented in \gls{sgg} for health, this review employ a three-tiered classification. A distinction is made between the representation of the user (Model Level), the source of the adaptation logic (Paradigm Level), and the specific computational mechanism used (Algorithmic Level). A summary of this classification can be found in Table~\ref{table:rev:taxonomies}. \par
The model level defines if a model is used to structure and update specific user information within the system, i.e. it defines the "lens" through which the game perceives the user. The model level is divided into two categories: the player model and the expert knowledge model. The player model represents the individual's characteristics, behaviors or preferences, such as personality traits or in-game performance metrics. This captures who the player is. The expert knowledge model represents external, standardized benchmarks or clinical gold standards. It is based on domain expertise, such as kinematic limb movements for stroke rehabilitation or nutritional guidelines for diabetes management. This model often represents the therapeutic target. \par

The paradigm level refers to the source behind the adaption as it describes how the system decides to change based on the models. Three types of personalization can be identified: knowledge-driven, data-driven and hybrid personalization. Knowledge-driven personalization uses information, extracted from experts, such as clinicians, to make decisions. Data-driven personalization derives its logic from the analysis of patterns within the data. Finally, hybrid personalization, combines the two, often using data-driven methods to make predictions regarding user states, utilizing knowledge-driven guidelines or thresholds to ensure that the resulting changes remain within predefined limits. \par
 
The last and most detailed level, is the algorithmic level, which specifies the computational methods used to execute the personalization defined in the levels above. In this review, 5 categories were identified: Logic-based methods, learning-based, optimization \& search-based, probabilistic \& statistical and hybrid methods.  Logic-based methods employ formal logic and discrete rules, such as IF-THEN statements, decision trees or ontologies. There are often used for knowledge driven systems. Learning-based methods use \gls{ml}, such as supervised learning, unsupervised learning or reinforcement learning approaches. Optimization \& search-based methods treat personalization as a search problem within a complex space to find the most optimal game configuration. Examples are \gls{mcts} or genetic algorithms. Probabilistic \& statistical methods focus on modeling uncertainty and predicting future states. Examples of such methods include Bayesian Networks and Markov chains. Hybrid methods use a combination of the previously mentioned methods to execute the required personalization.

\begin{table*}[]
	
	\caption{An overview of the three-tiered classification used to review the personalization pipelines in the analyzed papers.}
	\label{table:rev:taxonomies}
	\resizebox{\textwidth}{!}{%
		\begin{tabular}{llll}
			\toprule
			\textbf{Classification}	& \textbf{Category} & \textbf{Characterized by...} & \textbf{Examples}  \\
			\midrule
			Model & Player Model & \makecell{a representation based on the user's individual characteristics, behaviors \\or preferences. } & \makecell{Hexad Player Model Types, \\ personality traits, fitness level} \\
			& Expert Knowledge Model & \makecell{a representation based on external, standardized expertise }&\makecell{ kinematic chain models, \\ nutritional guidelines}\\
			\midrule
			Personalization paradigm& Knowledge-driven &   when personalization is authored by expert logic, rules, human expertise. & rule-based expert systems \\
			& Data-driven & \makecell{personalization is derived from the analysis of data. The system learns to \\ adapt by finding patterns }& \gls{ann} \\
			& Hybrid & data-driven insights constrained by expert knowledge &  \\
			\midrule 
			Algorithmic Framework& Logic-based & formal logic and rule-following mechanisms& \makecell{ expert IF-THEN rules, \\ decision trees, ontologies} \\
			& Learning-based & \makecell{uses machine learning to map inputs to outputs.\\ Including supervised, unsupervised, and reinforcement learning} & \makecell{K-means clustering, \\ support vector machines, \\\gls{rl}} \\
			& Optimization \& Search-based & search for the best possible configuration in a complex space & \gls{mcts}, genetic algorithms \\
			& Probabilistic \& Statistical & \makecell{uses mathematical models of uncertainty to predict the likelihood of a \\ user's future state} & \makecell{Bayesian Networks or \\Markov Chains} \\
			& Hybrid & uses a combination of the above &  \\

			\bottomrule
		\end{tabular}
	}
\end{table*}


\section{Overview}
\label{sec:rev:overview}
This section provides an overview of the included papers. First, in Section~\ref{sec:rev:domains}, the identified domains are discussed, followed by an overview of the identified personalization goals, provided in Section~\ref{sec:rev:personalisation_goals}. Tables~\ref{table:rev:overview_gam} and \ref{table:rev:overview_SG} provide an overview of the included papers and their identified objectives. Furthermore, a summary of the study design, study output and a detailed domain description was provided.

\subsection{Domains}
\label{sec:rev:domains}
 Six domains have been identified:  \textit{education \& training}, \textit{technical/methodological}, \textit{rehabilitation \& therapy}, \textit{treatment \& disease management}, \textit{diagnosis \& assessment} and finally, \textit{prevention \& wellness}. A description of each domain can be found in Table~\ref{table:rev:domains}. \par
\begin{figure}[]
	\centering
		\includegraphics[scale=0.5]{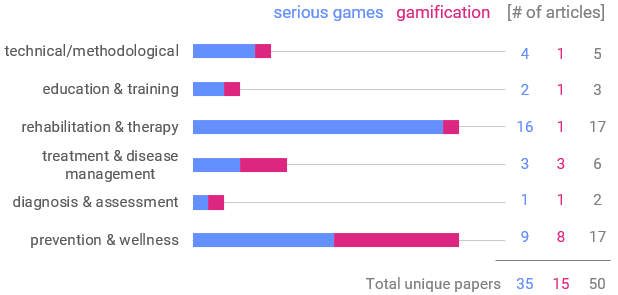}
	\caption[Identified healthcare domains]{Six domains within health care have been identified, of which 'rehabilitation \& therapy' and 'prevention \& wellness' contained the most articles.}
	\label{fig:rev:domain}
\end{figure}

 Of the 50 included papers, 35 discuss a serious game, while the other 15 articles research gamification. Figure~\ref{fig:rev:domain} indicates that \textit{rehabilitation \& therapy} and \textit{prevention \& wellness} are the leading domains, with both 17 entries. Important to note is that the domain of \textit{rehabilitation \& therapy} almost exclusively covers serious games, with only one gamification entry. \textit{Diagnosis \& assessment} is the smallest domain, including only 1 serious game and 1 gamified intervention, closely followed by \textit{education \& training}, with only 3 entries. The latter can be explained by the decision to only focus on education and training of the patient and/or family and exclude all research on the education of health professionals, as this can be classified on the larger domain of \gls{sgg} for education in general.  \textit{Prevention \& wellness} mostly focuses on increasing physical activity or changing nutritional habits, while systems for rehabilitation target a range of domains, namely neck and wrist or upper-limb rehabilitation, neuro-rehabilitation and post-stroke patients, both cognitive and physical rehabilitation.

\begin{table*}[]
	
	\caption{The six identified domains and their description. }
	\label{table:rev:domains}
	\resizebox{\textwidth}{!}{%
		\begin{tabular}{ll}
			\toprule
			\textbf{Domains}	 & \textbf{Description}   \\
			\midrule
			Technical/methodological &  \makecell{Focuses more on the technological and/or methodological aspects of designing \\ a gamified system, less on the health goals.} \\
			Education \& training & \makecell{Focuses on transferring medical knowledge and skills, e.g. to improve health literacy of the patient.}\\
			Rehabilitation \& therapy &\makecell{Focuses on restoring function, i.e. physical, cognitive, or psychological, that has been \\ lost due to injury, surgery or disease.}\\
			Treatment \& disease management & \makecell{Focuses on helping patients manage existing, often chronic, conditions. }\\
			Diagnosis \& assessment & \makecell{Focuses on applications designed to screen for, diagnose, or asses the state or progress \\ of a health condition.}\\
			Prevention \& wellness & \makecell{Focuses on proactive health management for the general population or at-risk individuals. \\ The goal is to prevent illness and promote a healthy lifestyle.} \\

			\bottomrule
		\end{tabular}
	}
\end{table*}

\subsection{Personalization Objectives}
\label{sec:rev:personalisation_goals}
The papers were reviewed based on three personalization goals, more specifically, increasing user engagement, increasing treatment adherence, and/or improving user performance. Some articles provided more specific objectives, such as increasing knowledge on a certain topic or implementing sustainable behavior change. For this review, the objectives were classified into the three aforementioned categories.  
Most studies, namely 34 out of 50, explicitly state that they want to increase user engagement by including personalized gamification or \gls{sgs}, while \gls{sgs} that focus on \textit{rehabilitation \& therapy} often not only aim to improve engagement but also increase the performance of the user (10 out of 16).  

\begin{table*}[h]
	\caption{In total, 50 articles have been identified of which 15 reported on gamification. An overview is provided of their personalization goal, study design, output type and application domain.}
	\label{table:rev:overview_gam}
	\resizebox{\textwidth}{!}{%
		\begin{threeparttable}
			\begin{tabular}{@{}cccllll@{}}
				\toprule
				\multicolumn{3}{c}{Personalization Goal}  & Study design  & Output & Domain (Detailed) & Ref \\
				 Engag. & Adh. & Perform.   &   &  &  &  \\
				\midrule
				\checkmark &- & -& mixed method & framework (FrameworkL) & technical/methodological (healthy habits)  & \cite{de_oliveira_gamification-based_2020} \\ 
				\checkmark &-&\checkmark & intervention (12 participants) & web application (CoaFeld) & \makecell{treatment \& disease management \\ (physical activity elderly)} & \cite{martinho_effects_2023} \\
				\checkmark &- &- & intervention (176 participants) & mHealth app (GameBus) & prevention \& wellness (physical activity) & \cite{nuijten_evaluating_2022}\\
				\checkmark & \checkmark & -& design \& development  & \makecell{m Health framework\\ (CarpeDiem app)} & prevention \& wellness (nutrition) & \cite{silvia_tailored_2023} \\
				- & \checkmark & -& intervention (61 children) & mHealth app & prevention \& wellness (physical activity) & \cite{schafer_study_2018} \\
				\checkmark & &\checkmark & intervention (40 participants) & mHealth app & prevention \& wellness (physical activity) & \cite{zhao_effects_2020} \\
				\checkmark & \checkmark &-& prototype evaluation (44 students) & chatbot app (CiboPoli) & education \& training (nutrition) & \cite{fadhil_adaptive_2017} \\
				\checkmark &- &- &design \& development & recommendation tool & \makecell{treatment \& disease management \\ (patient coaching)} & \cite{pardos_enriching_2023} \\
				\checkmark & \checkmark & -& intervention (28 participants) & mobile survey app & diagnosis \& assessment (surveys for health) & \cite{carlier_investigating_2021} \\
				\checkmark & - & \checkmark & intervention (10 people - not elderly) & Physical Activity Trainer (PAT) & prevention \& wellness (physical activity elderly) & \cite{mocanu_kinect_2016}\\
				\checkmark &- & \checkmark & intervention (21 participants) & exergame (SperyRacer on ExerCube) & prevention \& wellness (physical activity) & \cite{martin-niedecken_comparing_2021} \\
				\checkmark &- & \checkmark & RCT (3 arms, 136 participants) & gamified cognitive training program & \makecell{treatment \& disease management \\ (cognitive bias modification)} & \cite{shen_gamified_2025} \\
				- & -&\checkmark & \makecell{questionnaire responses (178) \\+ pilot study (5)} & exergame player modeling system & prevention \& wellness (physical activity) & \cite{zhao_physical_2020} \\
				\checkmark &- &- &2 experiments (378 smartwatch users) & \makecell{user segmentation for tailored \\ smartwatch-based gamification} & prevention \& wellness (physical activity) & \cite{yao_smartwatch-based_2025}\\
				\checkmark & \checkmark &-& pilot study (3 patients with MS) & \makecell{StepAR: a personalized exergame \\ using video mapping} & \makecell{ rehabilitation \& therapy (gait \\rehabilitation Multiple Sclerosis)} & \cite{amiri_stepar_2022} \\

				\bottomrule
			\end{tabular}
			\begin{tablenotes}
				\small
				\item gam = gamification
				\item Engag. = Engagement
				\item Adh. = Adherence
				\item Perform. = Performance
			\end{tablenotes}
	\end{threeparttable}}
\end{table*}

\begin{table*}[h]
\caption{In total, 50 articles have been identified of which 35 on serious games. An overview is provided of their personalization goal, study design, output type and application domain.}
\label{table:rev:overview_SG}
\resizebox{\textwidth}{!}{%
\begin{threeparttable}
\begin{tabular}{@{}cccllll@{}}
\toprule
 \multicolumn{3}{c}{Personalization Goal}  & Study design  & Output & Domain (Detailed) & Ref \\
   Engag. & Adh. & Perform.   &   &  &  &  \\
 \midrule
  
 \checkmark & \checkmark & -& prototype evaluation (6 experts) & \makecell{asynchronous multiplayer \\ exergame (GardenQuest)} & prevention \& wellness (physical activity) & \cite{meschtscherjakov_gardenquest_2023}\\
 \checkmark & \checkmark &-& intervention (29 participants) & \makecell{serious game (Express\\ Cooking Train)} & education \& training (nutrition) & \cite{mitsis_ontology-based_2019,mitsis_evaluation_2019} \\ 
 - & -&\checkmark & simulator-based validation & serious games (inLife platform) & \makecell{technical/methodological (sustainable \\ behavior + social skills children ASD)} & \cite{semet_artificial_2019} \\
 \checkmark &-&-&experiment (16 children) & KeepAttention game & \makecell{rehabilitation \& therapy (attention \\training for ADHD)} & \cite{hocine_personalized_2019} \\
 \checkmark & -& -&experiment (11 children) & \makecell{task-oriented design \\framework (KeepAttention game)} & \makecell{rehabilitation \& therapy (attention \\ training for ADHD)} & \cite{hocine_keep_nodate} \\
 - &- &\checkmark & prototype implementation & \makecell{conceptual architecture \\for smart serious games} & technical/methodological (physical activity) & \cite{ahmad_architecting_2022} \\
 \checkmark &- &- & intervention (21 participants) & first person shooter game (PC) & technical/methodological & \cite{alves_flow_2018} \\
 \checkmark & \checkmark &-& validation with sample dataset & serious game & \makecell{treatment \& disease management \\(cancer)} & \cite{brown_intelligent_2014} \\
 - &-&\checkmark & experiment (15 children with asthma) & e-learning platform (KidBreath) & education \& training (asthma) & \cite{delmas_fostering_2018} \\
 \checkmark &-&-& \makecell{intervention (37 participants\\ without cognitive impairment)} & \makecell{intelligent assistive system \\ with AR mobile serious game} & \makecell{treatment \& disease management \\(cognitive impairment elderly)} & \cite{ghorbani_towards_2022} \\
 - &-&\checkmark & intervention (10 participants) & RehaBot framework (VR) & rehabilitation \& therapy (neck) & \cite{afyouni_adaptive_2020} \\
 - &\checkmark & \checkmark & experiments (4 healthy participants) & \makecell{ wrist rehabilitation robot \\ \& serious game (Nuts Catcher)} & rehabilitation \& therapy (wrist) & \cite{andrade_dynamic_2014} \\
 - &\checkmark &\checkmark & design \& development & serious game (ReHabGame) & rehabilitation \& therapy (motor impairment) & \cite{esfahlani_adaptive_2017} \\
 \checkmark &- &\checkmark & \makecell{intervention (20 post-stroke \\ participants)} & serious game (ReHabGame) & rehabilitation \& therapy (neurological) & \cite{esfahlani_rehabgame_2018} \\
 \checkmark & \checkmark &- & three-fold validation & \makecell{exergame-based rehabilitation \\ system (TANGO:H)} & rehabilitation \& therapy (cognitive \& physical) & \cite{gonzalez-gonzalez_serious_2019} \\
 - &-&\checkmark & \makecell{intervention (7 post-stroke \\ participants, 3 therapists)} & serious game (Prehab) & rehabilitation \& therapy (post-stroke) & \cite{hocine_adaptation_2015} \\
 \checkmark & \checkmark &- &design \& development & serious game (InMotion) & rehabilitation \& therapy (upper limb) & \cite{pinto_adaptive_2018} \\
 \checkmark &-&\checkmark & design discussion & serious game (InMotion) & technical/methodological (upper limb) & \cite{alves_towards_2019}\\
- & -& -& \makecell{ intervention (20 post-stroke \\ participants)} &\makecell{ tele-rehabilitation system based \\ on serious games and in-cloud \\ data analytics services} & rehabilitation \& therapy (post-stroke) & \cite{caggianese_serious_2019} \\
\checkmark &-&\checkmark & intervention (25 elderly) & AR serious game & \makecell{prevention \& wellness (cognitive \& \\ physical abilities elderly)} & \cite{eun_artificial_2023}\\
\checkmark &-&\checkmark & blind experiment (42 participants) &\makecell{serious game (Wake Up For \\ The Future!)} & \makecell{treatment \& disease management \\ (obstructive sleep apnea)} & \cite{mitsis_procedural_2020} \\
- &-&\checkmark & intervention (52 participants) & \makecell{serious game (Fruit-Collection \\ and avatar maneuvering)} & rehabilitation \& therapy (neurorehabilitation) & \cite{sadeghi_esfahlani_fusion_2019}\\
- &-& \checkmark & RCT (35 older adults (age 50+) & \makecell{exergame-integrated IoT-based \\ergometer system (EIoT-ergo)} & prevention \& wellness (physical activity elderly) & \cite{lin_exergame-integrated_2023} \\
 - & - &\checkmark & \makecell{pilot experiment 3 able-bodied \\ participants + 1 pediatric patient \\with genu recurvatum} & \makecell{biofeedback game with \\knee exoskeleton} & \makecell{ rehabilitation \& therapy (physical therapy \\ children with gait impairments)} & \cite{nathella_challenge-based_2025} \\
 \checkmark & \checkmark & \checkmark & subjective evaluation (11 stroke patients) & adaptive semi-immersive VR SG & rehabilitation \& therapy (hand after stroke) & \cite{bouatrous_new_2023}\\
 \checkmark & - & \checkmark & \makecell{experimental validation  \\40 children with ASD and ADHD) }& adaptive neuro-affective system & \makecell{rehabilitation \& therapy (neurocognitive \\ training for neurodivergent children)} & \cite{faria_adaptive_2025}\\  
\checkmark & - &- & experiment (15 participants) & VR exergame: Militant of the Maze & \makecell{prevention \& wellness (prevent \\ work related musculoskeletal disorders)} & \cite{stranick_adaptive_2022} \\  
 \checkmark &-&\checkmark &  \makecell{proof-of-concept (24 participants \\ subacute and chronic stroke)} & \makecell{robot SG using Robot \\Assisted Therapy (ROBiGAME)} & \makecell{ rehabilitation \& therapy (upper limb recovery \\ \& cognitive rehabilitation after stroke)} & \cite{doumas_clinical_2025} \\
 \checkmark & - &- & design and development & Snowballz VR exergame & prevention \& wellness (physical activity) & \cite{yoo_designing_2017} \\
 -& \checkmark& - & mixed methods user study (n=46) & social exergame (Step Heroes) & prevention \& wellness (physical activity) & \cite{gray_improving_2023} \\
 \checkmark & - & \checkmark & 16 physically active adults & \makecell{cardio-exergame (Letterbird) \\ with cycle ergometer (ErgoActive)} & prevention \& wellness (cardiotraining) & \cite{hoffmann_personalized_2015}\\
 -&-&\checkmark & design and development & inhibitory control task in VR SG & \makecell{diagnosis \& assessment (cognitive\\ impairments - inhibitory control)} & \cite{forgiarini_probabilistic_nodate}\\ 
 \checkmark &-&\checkmark& \makecell{data collection for training (n=18)\& \\ evaluation of heuristic vs ANN (n=2x4)}& exergame (Pathologys) & prevention \& wellness (physical activity) & \cite{aguilar_proposal_2022}\\
 \checkmark & \checkmark & - & \makecell{evaluation with general public (n=15)\\ \&  physiotherapists (n=8)} & SG for motor rehabilitation (MazeOut) & \makecell{rehabilitation \& therapy \\(upper limb motor rehabilitation)} & \cite{kira_approach_2024} \\
 -& -& \checkmark & study 20 older adults & REAsmash immersive VR & \makecell{ prevention \& wellness (motor learning \\                                                                                                                                                                                        in older adults)} & \cite{everard_self-adaptive_2025}\\
   \bottomrule
\end{tabular}
\begin{tablenotes}
      \small
      \item SG = serious game
      \item Engag. = Engagement
      \item Adh. = Adherence
      \item Perform. = Performance
    \end{tablenotes}
\end{threeparttable}}
\end{table*}

\section{The Model Level}
\label{sec:rev:model}
Some studies use models to structure and update specific user information. This information can be limited to player or user data, which can consist of personal information, such as age or game progression data, sensor data, i.e., data collected via sensors or external data, such as heart rate or contextual data such as weather reports. A last type of user information that is sometimes collected is medical data, which we define as data that has been handled, or inputted by health professionals, such as results from medical tests or a set of rehabilitation exercises. \par 
This section provides a more detailed discussion of the identified models, following the classification outlined in Section~\ref{sec:rev:classification} and the used user information, i.e. general user and game information, sensor data, medical data. 
An overview of the results is presented in Table~\ref{table:rev:gamification} for gamification and Table~\ref{table:rev:serious games} for serious games. The tables report the use of a model, expert knowledge and/or player model, only when the study explicitly stated that a model was constructed using the collected user information.

\subsection{Hexad Player Model}
\begin{figure}[]
	\centering
		\includegraphics[scale=0.5]{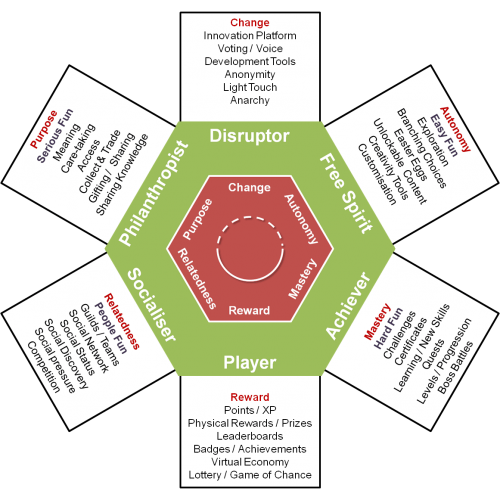}
	\caption[Hexad Player Framework]{The Hexad Player Framework maps game elements on 6 distinct player types, based on the intrinsic or extrinsic motivation of users~\cite{marczewski_even_2015}.}
	\label{fig:rev:Hexad}
\end{figure}

Six studies~\cite{de_oliveira_gamification-based_2020,fadhil_adaptive_2017,zhao_effects_2020,meschtscherjakov_gardenquest_2023,carlier_investigating_2021, zhao_physical_2020} use the Hexad Player Type Model to classify users according to their player type, which has been designed specifically for the design of gameful systems tailored to their users~\cite{tondello_gamification_2016}.  Six player types are defined based on their intrinsic or extrinsic motivation, namely, achiever, free spirit, philanthropist, disruptor, player and socializer. The Hexad framework proposes an empirically validated mapping of several game elements on the 6 player types, as shown in Figure~\ref{fig:rev:Hexad}~\cite{marczewski_even_2015}. \par
de Oliveira et al.~\cite{de_oliveira_gamification-based_2020} investigated how the user's player type can be incorporated to include the correct game elements for each player in a self-care application. The study uses the Hexad Player Model to classify the users according to their type, but they take into account that users and their preferences can change, meaning that their player type and game elements preferences can change too. To accommodate for this change, they include an artificial neural network (ANN) that classifies the user in the case of loss of interest in current game elements. Zhao et al.~\cite{zhao_effects_2020,zhao_physical_2020} compile a player model out of four submodels to create a personalized gamified fitness recommender system. The player model consists of an activity recognition model to track the activities of the player, a general model, which includes personal information of the user, an exerciser type model, an expert knowledge model for recommending specific activities and finally, a player-type model, based on the Hexad Player model. Similarly, Fadhil et al.~\cite{fadhil_adaptive_2017}, Carlier et al.~\cite{carlier_investigating_2021} and Chan et al.~\cite{meschtscherjakov_gardenquest_2023} use the Hexad Player Model to include user-specific game elements in the proposed system. Fadhil et al.~\cite{fadhil_adaptive_2017} included personalized gamification into a chatbot game to teach children about healthy lifestyles and habits. 
In our previous research~\cite{carlier_investigating_2021} we designed an application for increasing engagement and respondent behavior for health surveys, using personalized gamification. With the GardenQuest game, Chan et al.~\cite{meschtscherjakov_gardenquest_2023} aims to increase users' exercise adherence by creating a social multiplayer exergame that groups patients according to their respective player types. 

\subsection{Ontology}
\begin{figure*}[]
	\centering
		\includegraphics[scale=0.37]{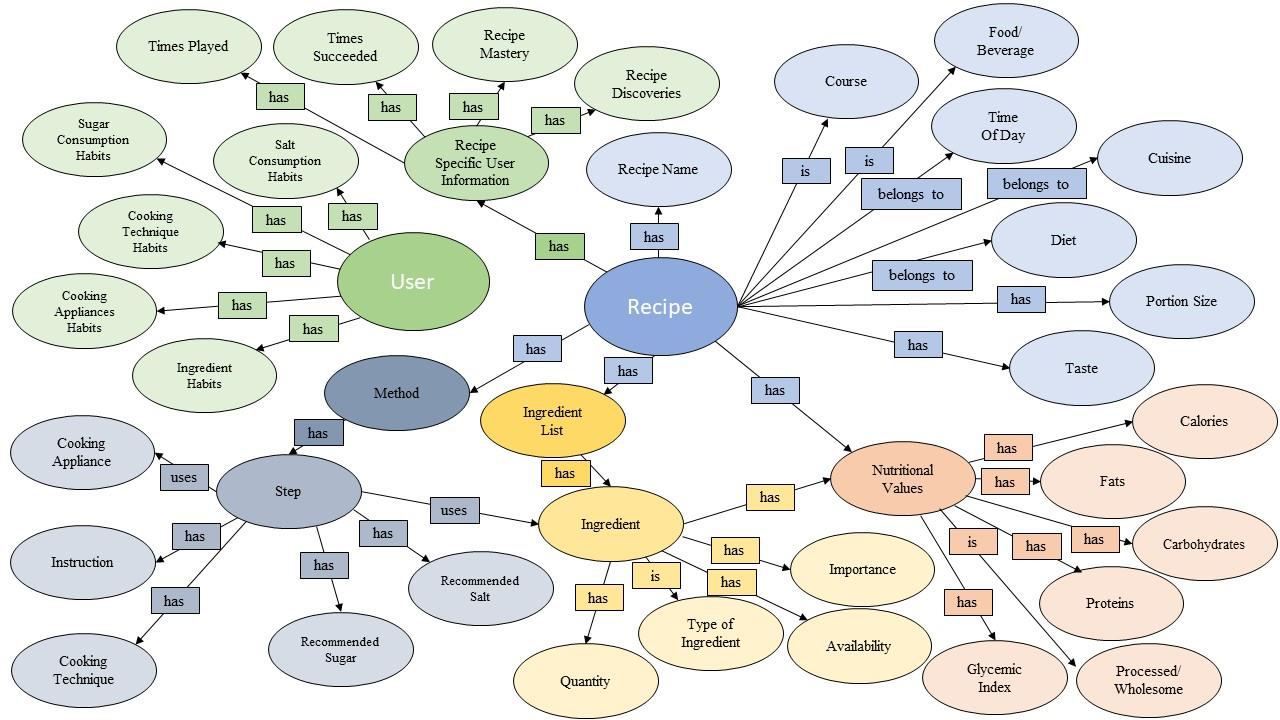}
	\caption[Recipe ontology]{Recipe ontology used for modeling user and expert knowledge in the serious game ``Express Cooking Train'' by Mitsis et al.~\cite{mitsis_ontology-based_2019}.}
	\label{fig:rev:ontology}
\end{figure*}

Two studies use ontologies, an expert knowledge model method, for player and expert knowledge modeling~\cite{mitsis_ontology-based_2019,caggianese_serious_2019}. Ontologies offer formal definitions of distinct concepts, their properties, and intricate relationships among these concepts, thereby establishing computer-readable classification systems~\cite{ashburner_gene_2000,dessimoz_gene_2017}. Figure~\ref{fig:rev:ontology} shows an example of such an ontology, more specifically, the recipe ontology included in the serious game on \gls{nl} and \gls{fl} skills by Mitsis et al.~\cite{mitsis_ontology-based_2019}. Using user game information and the knowledge contained in the ontology facilitates the personalization of the game as recipes can be suggested based on dietary needs and preferences and via a rule-based system, the user's cooking profile can be adapted according to their in-game performances. Caggianese et al.~\cite{caggianese_serious_2019} proposes a tele-rehabilitation system that utilizes different sources of information and data, namely game data, personal user information, Microsoft Kinect sensor data and input from health professionals to provide personalized decision support to the user. To model the required expert knowledge and user information, they used an ontological model, including both game description concepts and motor rehabilitation concepts. The system also provides an interface for health professionals to define each patient's rehabilitation goals, which include, amongst others, the anatomical problem for each motor district, e.g., left shoulder abduction. Due to the use of an ontology and hybrid production rules, i.e., the combination of ontological rules and fuzzy logic rules~\cite{wang_course_1997, zadeh_role_1994}, this diagnostical information can then be used in the decision support system for adapting the serious game and suggesting improvements in the offered therapy. 

\subsection{Kinematic Chain Model and Inverse Kinematic}
Another example of an expert knowledge model is a kinematic chain model, which describes the movement of a kinematic chain, i.e., the formulation of the translation, rotation, position and velocity of a body segment interconnected by joints, for a robot or animated character, e.g. human~\cite{borbely_real-time_2017,lura_creation_2012,papaleo_inverse_2012}. Three included studies from Esfahlani et al.~\cite{esfahlani_rehabgame_2018, esfahlani_adaptive_2017, sadeghi_esfahlani_fusion_2019} make use of a kinematic chain model to represent the mechanical structure of the user. These studies then use inverse kinematics to control and plan the motion of a desired position to achieve a specific task~\cite{lura_creation_2012}. The Microsoft Kinect sensor is used to track the user's skeleton joints in all three studies,in addition to a foot pedal~\cite{sadeghi_esfahlani_fusion_2019}, a Thalmic Myo armband~\cite{sadeghi_esfahlani_fusion_2019,esfahlani_rehabgame_2018,esfahlani_adaptive_2017}. This sensor data is then fed to the personalization methods to personalize the game and adapt to the difficulty of the conference rehabilitation exercises. The three studies investigate different approaches, namely fuzzy logic~\cite{esfahlani_adaptive_2017}, Monte Carlo Tree Search~\cite{esfahlani_rehabgame_2018} and a combination of fuzzy logic and an artificial neural network~\cite{sadeghi_esfahlani_fusion_2019}. 

\subsection{Expert IF-THEN Rules}
One gamified intervention~\cite{silvia_tailored_2023} and four serious games~\cite{ghorbani_towards_2022, alves_towards_2019, eun_artificial_2023, everard_self-adaptive_2025} mention the use of expert IF-THEN rules, which model expert domain knowledge. These rules act as the transparent translation of medical protocols or boundaries into game mechanics. In the domain of \textit{prevention \& wellness} for elderly, Eun et al.~\cite{eun_artificial_2023} and Everard et al.~\cite{everard_self-adaptive_2025}, use these rules to operationalize clinical protocols, ensuring that the intensity of the game only scales when specific safety thresholds are met. Similarly, Alves et al.~\cite{alves_towards_2019} uses expert rules to translate insights from the Five Factor Model into specific game difficulty adjustments, combining expert knowledge and player information into one model. The system aims to support patients with emotional instability as poor rehabilitation results or criticism by the therapist might easily demotivate them. The personality traits of the patient and in-game actions trigger specific responses to the game. These adaptation rules are used for the fuzzy logic model to provide personalized in-game support. An example of such a rule is: If a patient has High Neuroticism as a personality trait, and performs badly in the game, the game should respond by friendly encouraging the patient to try again~\cite{alves_towards_2019}.   
 Orte et al.~\cite{silvia_tailored_2023} designed a gamified mHealth application for nutritional behavior change that offers personalized dietary missions. To do so, information about the users' nutritional habits is gathered via questionnaires to build a nutritional behavior profile. Ghorbani et al.~\cite{ghorbani_towards_2022} evaluate an intelligent assistive system to support the elderly in their daily life activities using \gls{ar} and \gls{sgs}. To personalize the system, fuzzy rule bases are built, including the expert knowledge of therapists for each patient. \par

\subsection{Other model approaches}
\begin{figure*}[]
	\centering
		\includegraphics[scale=0.6]{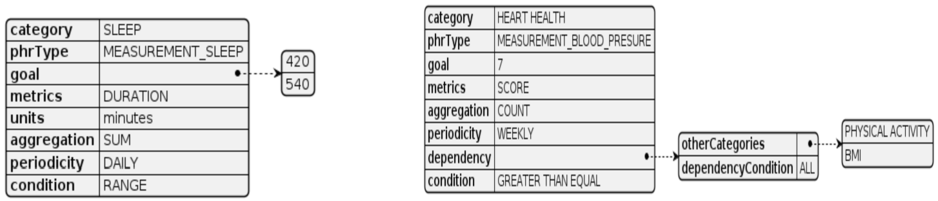}
	\caption[Example of a primitive element and a mixed element]{An example of a primitive element (a) and a mixed element (b) in the knowledge model of Pardos et al.~\cite{pardos_enriching_2023}}
	\label{fig:rev:multivariateobject}
\end{figure*}

The remaining gamified solutions employed different approaches, each dependent on the specific data sources required for the construction of the model: questionnaire responses~\cite{yao_smartwatch-based_2025}, physical activity data~\cite{schafer_study_2018}, and specific domain knowledge~\cite{pardos_enriching_2023}. Yao et al.~\cite{yao_smartwatch-based_2025} use Maximum Difference Scaling, i.e. a survey method to measure what people value most and least, as a segmentation method to incorporate individual user preferences of game elements in smartphone-based gamification for physical activity. Schäfer et al~\cite{schafer_study_2018} use smartphone sensor data to derive a physical activity model for children. This model is then used to personalize the application by using an avatar model that mirrors the children's physical activity level, i.e., sitting, standing, walking and intense. A Random Forest classifier has been used to classify the sensor data. Pardos et al.~\cite{pardos_enriching_2023} designed a remote patient monitoring and care platform that offers personalized gamified recommendations. The knowledge needed to recommend healthier habits to users includes official guidelines given by for example the WHO and the American Heart Association and is encoded by a set of multivariate objects and rules for each domain, as shown in Figure~\ref{fig:rev:multivariateobject}. Health professionals can then access the platform to create personalized rules for specific patients. \par

The remaining serious games use a variety of approaches for modeling the user. 
Alves et al.~\cite{alves_flow_2018} developed a first-person shooter video game that adapts the difficulty level to the mental state of the player. Consequently, a classification framework is developed that reads physiological signals, namely heart rate and beta bands of the brainwaves, and outputs the current mental state of the player, using \gls{mlp} Classification~\cite{kruse_multi-layer_2022}. Next, using a state machine, the difficulty level of the game is updated according to the current mental state. \par

Afyouni et al.~\cite{afyouni_adaptive_2020} introduce ``Rehabot'' for the adaptive generation of personalized \gls{sgs} for telerehabilitation. In order to provide personalized feedback and adapt the difficulty of the exercises to the user, expert knowledge regarding postures needs to be modeled. To that end, a therapist inputs a set of correct postures for the corresponding patient. The system translates this expert knowledge to a set of joints that are compared to the movements of the user, using the Microsoft Kinect and a posture-matching algorithm. \par
\begin{figure*}
	\centering
		\includegraphics[scale=0.6]{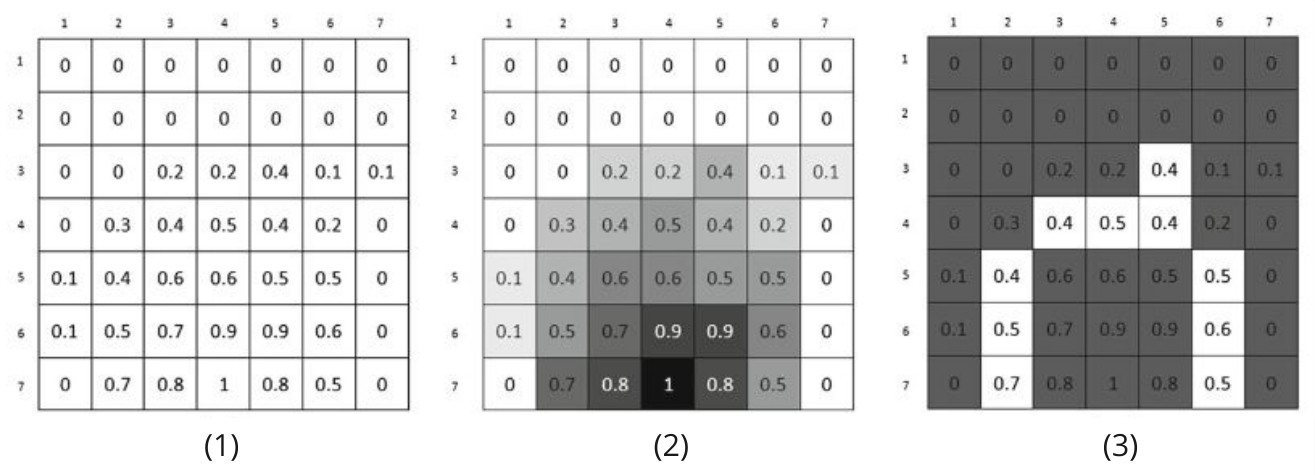}
	\caption[Player ability zone matrix]{An example of a player's ``ability zone'' matrix (1), the obtained image using gradients (2) and the detected edge of the ability zone (3).~\cite{hocine_adaptation_2015}}
	\label{fig:rev:matrix}
\end{figure*}

Another example of a personalized serious game for rehabilitation is the TANGO:H platform of González-González et al.~\cite{gonzalez-gonzalez_serious_2019}. The platform creates a user model that represents a set of data that characterizes the user at a specific moment in time. This user data includes explicit data, i.e. provided by the user, and implicit data, i.e., provided by their interaction with the system. Included in this user model is the system's estimation of the user's skill level. To suggest exercises to the user, the user's skill level is matched with the expected skill level of the rehabilitation exercises, using a recommender system. To update the skill level of the user, a heuristic approach is used, using a formula that considers certain expert intuitions on how the user's skill level should evolve over time. Another approach to modeling the player's skill level is seen in the work of Hocine et al.~\cite{hocine_adaptation_2015}. To model the player, and their motor abilities, they define the ``ability zone'', which represents the area where the patient can efficiently move on a 2D workspace, such as a graphical tablet. The ability zone is modeled using a $nxm$ matrix which maps the physical workspace and the virtual workspace (computer screen). Each matrix cell then includes information on the performed movements of the patient. Post-stroke patients move the computer mouse within the workspace and the system uses these mouse coordinates to calculate the resulting ability zone. During an assessment exercise, the ability zone matrix of each player is constructed and continuously updated during the playing sessions. This matrix is then used for the adaptation of the game to identify challenging areas for the patient as shown on Figure~\ref{fig:rev:matrix}. The ability zone matrix (Figure~\ref{fig:rev:matrix}-1) is transformed to an image by assigning gradients to each cell value (Figure~\ref{fig:rev:matrix}-2), this is then used to compute the edge of the matrix (Figure~\ref{fig:rev:matrix}-3). Targets that are situated inside this edge will be easy, while targets outside the ability zone's edge will be linked to a higher difficulty level. Similarly, Bouatrous et al~\cite{bouatrous_new_2023} utilize the initial motor capacity of the patient as a calibration step to calculate the patient-specific thresholds to initialize the game difficulty. The serious game of Hoffman et al.~\cite{hoffmann_personalized_2015} also uses a calibration phase, in which participants play the game to obtain heart rate information to configure the rest of the game. \par 
Some serious games use a multidimensional player model that includes and updates multiple aspects of player performance, using expert knowledge and player knowledge. Examples are a multi-modal engagement scoring model, based on game score, \gls{eeg} concentration and facial emotion~\cite{faria_adaptive_2025}, an adaptive-individual-task model, combining kinematic data and game performance data~\cite{stranick_adaptive_2022} or a user model that combines a game performance model and a fitness model based heart rate sensor data~\cite{yoo_designing_2017}.  \par 
 Finally, Forgiarini et al.~\cite{forgiarini_probabilistic_nodate} model their \gls{sg} as a Discrete-Time Markov Chain, where varying game play scenario's are linked to key actions with specific expert-defined probabilities, based on the patient's cognitive impairment stage. 

\begin{table*}[h]
\caption{An overview of the research on personalized gamification. For each entry, an overview of the applied models, integrated user information, computational method according to the algorithmic classification level, type of personalization paradigm, presence of reuse, dynamic difficulty balancing, and domain is provided.}
\label{table:rev:gamification}
\resizebox{\textwidth}{!}{%
\begin{threeparttable}

\begin{tabular}{@{}lcclccclcccl@{}}
\toprule
Ref &  \multicolumn{3}{c}{Model}& \multicolumn{3}{c}{User information}  & Computational method & Type & Reuse & \makecell{Dynamic \\ difficulty}& Domain\\
     & \makecell{Expert \\knowledge} & Player & information   & Gen. & Sens. & Medic.  &  & && \\
 \midrule
 \cite{de_oliveira_gamification-based_2020}  &-&\checkmark & Hexad Player Model & \checkmark &-&-& artificial neural network (ANN) &DD & \checkmark&-& technical/methodological \\
 \cite{martinho_effects_2023} &- &- &- &\checkmark & \checkmark & - & reinforcement learning (RL) & DD& -&- &\makecell{treatment \& disease \\management}  \\
\cite{nuijten_evaluating_2022} & - &- & -& \checkmark & \checkmark&- & rule-based system ( formula) & KD& \checkmark &-& prevention \& wellness  \\
 \cite{silvia_tailored_2023} & \checkmark & \checkmark & \makecell{nutritional behavior profile \\ \& expert IF-THEN rules} & \checkmark &- &-&  rule-based expert system& H &-&-& prevention \& wellness\\
 
 \cite{schafer_study_2018} &-&\checkmark & physical activity user model&-&\checkmark &- & random forest classification & DD&-&-& prevention \& wellness  \\
 
\cite{zhao_effects_2020}&\checkmark & \checkmark & \makecell{ 1. Hexad Player Model, \\  2. activity recognition model, \\ 3. general info model,\\ 4. exerciser type model} &\checkmark &\checkmark &-& decision trees & KD&-&-& prevention \& wellness \\

\cite{fadhil_adaptive_2017}&- & \checkmark & Hexad Player Model &-&-&-& - &DD&-&-& education \& training  \\
 \cite{pardos_enriching_2023}&\checkmark & - & \makecell{multivariate objects \\for expert knowledge } &\checkmark &\checkmark &\checkmark & rule-based expert system & KD& -&-&\makecell{ treatment \& disease \\management}\\
\cite{carlier_investigating_2021}&-& \checkmark &  Hexad Player Model & -&-&-&-&DD & \checkmark &-& diagnosis \& assessment \\
\cite{mocanu_kinect_2016} & \checkmark & - & set of exercises & \checkmark & -& \checkmark & cross-correlation for signal alignment &KD & - & \checkmark & prevention \& wellness \\
\cite{martin-niedecken_comparing_2021}& - & - & - & -& \checkmark &-& rule-based ( formula) &KD& - & \checkmark  & prevention \& wellness\\
\cite{shen_gamified_2025} & - & - & - & \checkmark & - &- &\makecell{ linear regression model \\using predefined difficulty curve}& H  &-&\checkmark &\makecell{treatment \& disease \\management}\\
\cite{zhao_physical_2020} & - & \checkmark & Hexad Player Model & \checkmark & - &- & \makecell{recommender system: collaborative \\filtering with binary classification model\\ (SVM) or K-means clustering model}& DD & \checkmark & -& prevention \& wellness\\
\cite{yao_smartwatch-based_2025} & - & \checkmark & Maximum Difference Scaling  & - & - & - & - &DD& \checkmark & - & prevention \& wellness \\
\cite{amiri_stepar_2022} & - & \checkmark & user performance model & \checkmark & - & -& \makecell{rule-based system \\(pre-defined difficulty logic)}&KD & - & \checkmark & rehabilitation \& therapy \\

   \bottomrule
\end{tabular}
\begin{tablenotes}
       \item Gen. = general user and game information
      \item Sens. = data collected via sensors, such as heart rate or contextual data and external data, such as weather reports 
      \item Medic. = medical data (input provided by health professionals or results from medical tests)
      \item KD =knowledge-driven, DD = data-driven, H= hybrid
    \end{tablenotes}
\end{threeparttable}
}
\end{table*}

\begin{table*}[h]
\caption{An overview of the research on personalized serious games. For each entry, an overview of the applied models, integrated user information, computational method according to the algorithmic classification level, type of personalization paradigm, presence of reuse and dynamic difficulty balancing and domain is provided.}
\label{table:rev:serious games}
\resizebox{\textwidth}{!}{%
\begin{threeparttable}
\begin{tabular}{@{}lcclccclcccl@{}}
\toprule
 Ref &  \multicolumn{3}{c}{Model}& \multicolumn{3}{c}{User information}  & Computational method &Type & Reuse & \makecell{Dynamic \\ difficulty}&  Domain\\
   & \makecell{Expert \\knowledge} & Player & information   & Gen. & Sens. & Medic.  & && &   & \\
 \midrule
 \cite{meschtscherjakov_gardenquest_2023}&-&\checkmark&Hexad Player Model &-&\checkmark&- & -& DD &- &-&prevention \& wellness\\
\cite{mitsis_ontology-based_2019,mitsis_evaluation_2019} & \checkmark & \checkmark  & ontology &\checkmark &-&- & rule-based expert system & KD&\checkmark &- & education \& training \\
 \cite{semet_artificial_2019}  & - & - & - & \checkmark & - &- & ant colony optimization& DD& \checkmark & - & technical/methodological \\
 \cite{hocine_personalized_2019}  &-&-&-&\checkmark&-&- & \makecell{rule based system \\(open learner model)} & KD &  - &-&rehabilitation \& therapy \\
 \cite{hocine_keep_nodate}  & - &-&-&\checkmark&-&\checkmark & rule-based system& KD & \checkmark & \checkmark& rehabilitation \&therapy  \\
 \cite{ahmad_architecting_2022}  & - & - & - &\checkmark & \checkmark & - & \makecell{deep learning \& deep RL \\ \& optimization (particle \\swarm optimization,\\ genetic algorithms) }& DD& \checkmark & - & technical/methodological \\
 \cite{alves_flow_2018} & - & \checkmark & \makecell{mental state model} & - & \checkmark &- & \makecell{ANN: multilayer perception (mental\\state model) \& state machine}& H &- &\checkmark  & technical/methodological \\
 \cite{brown_intelligent_2014}  & - & -& - &\checkmark &- &\checkmark & artificial neural network &DD& - &-& \makecell{treatment \& disease \\management} \\
 \cite{delmas_fostering_2018}&- & - &- &\checkmark&-&-&multi-arm bandit & DD& -&- & education \& training \\
 \cite{ghorbani_towards_2022}  &\checkmark& \checkmark &expert IF THEN rules& \checkmark &\checkmark &- & adaptive fuzzy logic model& H &- &- & \makecell{treatment \& disease\\management} \\
 \cite{afyouni_adaptive_2020}  & \checkmark&-& set of postures&\checkmark& \checkmark & \checkmark &\makecell{rule-based system \\\& ML enriched engine}& H & -&\checkmark & rehabilitation \& therapy \\
 \cite{andrade_dynamic_2014} &-&-&-&\checkmark & \checkmark & - & \makecell{reinforcement learning \\(Q-learning)} & DD &- &\checkmark & rehabilitation \& therapy\\
 \cite{esfahlani_adaptive_2017}  & \checkmark &-&  \makecell{kinematic chain model \\\& inverse kinematics} &\checkmark& \checkmark &- & fuzzy logic model& KD &- & \checkmark & rehabilitation \& therapy \\
 \cite{esfahlani_rehabgame_2018}  & \checkmark &-&  \makecell{kinematic chain model \\\& inverse kinematics} &-& \checkmark &- & Monte Carlo Tree Search & H&- & \checkmark & rehabilitation \& therapy \\
 \cite{gonzalez-gonzalez_serious_2019}  &\checkmark&\checkmark& updated by heuristic &\checkmark & \checkmark &- & \makecell{rule based expert system} &H& \checkmark &\checkmark& rehabilitation \& therapy \\
 \cite{hocine_adaptation_2015}  &-&\checkmark& \makecell{player's motor abilities\\( ``ability zone")} & \checkmark &- &-& \makecell{Monte Carlo Tree Search \& \\ procedural content generation}& H & \checkmark &\checkmark  & rehabilitation \& therapy \\
 \cite{pinto_adaptive_2018} &-&-&- & \checkmark& \checkmark &- & state-machine &KD&- &\checkmark & rehabilitation \& therapy \\
 \cite{alves_towards_2019} &\checkmark &\checkmark &\makecell{Five Factor Model \\ \& expert IF THEN rules} &\checkmark &-&- & rule-based expert system &KD&- &\checkmark & technical/methodological \\
 \cite{caggianese_serious_2019}  &\checkmark &\checkmark &ontology &\checkmark& \checkmark& \checkmark & \makecell{hybrid production \& and \\ontological rules \& fuzzy logic }&H& \checkmark &- & rehabilitation \& therapy \\
 \cite{eun_artificial_2023} &\checkmark&-&\makecell{expert IF-THEN rules}&\checkmark& \checkmark &-& rule-based expert system & KD &-&\checkmark & prevention \& wellness \\
 \cite{mitsis_procedural_2020}  & -&-&-&\checkmark&-&- & \makecell{search-based procedural \\content generation using \\ genetic algorithm }& DD &-&\checkmark& \makecell{treatment \& disease \\ management} \\
 \cite{sadeghi_esfahlani_fusion_2019} & \checkmark &-& \makecell{kinematic chain model \\\& inverse kinematics}&\checkmark & \checkmark &-  & \makecell{artificial neural network (ANN) \\ \& fuzzy logic model }& H&- &\checkmark & rehabilitation \& therapy\\
 \cite{lin_exergame-integrated_2023} & - & -& - & \checkmark& \checkmark&-& decision trees& DD & - & \checkmark & prevention \& wellness \\
 \cite{nathella_challenge-based_2025} & - & - & - & \checkmark & \checkmark & - & \makecell{Covariance Matrix Adaptation-\\Evolutionary Strategy (CMA-ES) \\ optimizer} & DD& - & \checkmark & rehabilitation \& therapy\\
 \cite{bouatrous_new_2023} & - & \checkmark & \makecell{initial motor \\ capacity (calibration)} & \checkmark &- &- & K-means clustering& DD & - & \checkmark & rehabilitation \& therapy \\
 
 \cite{faria_adaptive_2025} & \checkmark & \checkmark & \makecell{multimodal engagement \\ scoring model} & \checkmark & \checkmark & \checkmark & \makecell{Bayesian Immediate Feedback \\ Learning (BIFL)}& DD &-&-& rehabilitation \& therapy \\
 \cite{stranick_adaptive_2022} & \checkmark & \checkmark & \makecell{adaptive-individual-task \\ model} & \checkmark & - & - & \makecell{comparison of ML \\ approaches, SVM best accuracy}& DD& - & \checkmark & prevention \& wellness\\
 \cite{doumas_clinical_2025} & - &- & - & \checkmark & \checkmark & - & \makecell{rule-based system (continuous \\ \& multi-parametric formula)}&KD & - &\checkmark &  rehabilitation \& therapy \\
 \cite{yoo_designing_2017} & \checkmark & \checkmark & \makecell{user model: \\1. game performance model \\ 2. fitness model} & \checkmark & \checkmark & - & - & KD& - & \checkmark & prevention \& wellness\\
 \cite{gray_improving_2023} & - & - & -& \checkmark & \checkmark & - & Shapley Bandit & DD& - & - & prevention \& wellness \\
 \cite{hoffmann_personalized_2015} & \checkmark & \checkmark & \makecell{ predictive user model\\ (calibration)}& \checkmark & \checkmark & - & \makecell{linear regression (user \\ model) \& rule-based system\\ (formula)} &H&   - & \checkmark & prevention \& wellness \\
 \cite{forgiarini_probabilistic_nodate} & \checkmark & \checkmark & \makecell{Discrete-Time Markov Chain} & \checkmark & - & - & \makecell{rule-based system \\(thresholds defined by experts)}& H& - & \checkmark & diagnosis \& assessment \\
 \cite{aguilar_proposal_2022} & - & - & -& \checkmark & \checkmark & - & decision tree vs ANN &DD& - & \checkmark & prevention \& wellness\\
 \cite{kira_approach_2024} & - & - & -& \checkmark & - & - & \makecell{rule-based system \\ (procedural content generation)}  &KD& - & \checkmark & rehabilitation \& therapy \\
 \cite{everard_self-adaptive_2025} & \checkmark & - & expert IF THEN rules & \checkmark & - &- & \makecell{rule-based expert system}&KD & - & \checkmark & prevention \& wellness \\
 
   \bottomrule
\end{tabular}
\begin{tablenotes}
      \item Gen. = general user and game information
      \item Sens. = data collected via sensors, such as heart rate or contextual data and external data, such as weather reports 
      \item Medic. = medical data (input provided by health professionals or results from medical tests)
      \item KD =knowledge-driven, DD = data-driven, H= hybrid
    \end{tablenotes}
\end{threeparttable}
}
\end{table*}

\begin{table*}[]
	
	\caption{An overview of the identified computational models for personalization, categorized by type of method.}
	\label{table:rev:pers_methods}
	\resizebox{\textwidth}{!}{%
	\begin{tabular}{lclcl}
		\toprule
		 & \multicolumn{2}{l}{\textbf{Gamification}} & \multicolumn{2}{l}{\textbf{Serious Games}}  \\
		\textbf{Algorithmic Framework Classification} & total & references & total & references\\ \midrule
		\makecell{\textbf{Logic-Based}\\ e.g., heuristics, expert rules \& ontologies} &6&  \cite{nuijten_evaluating_2022,silvia_tailored_2023,zhao_effects_2020,pardos_enriching_2023,martin-niedecken_comparing_2021,amiri_stepar_2022}  &14 & \cite{mitsis_ontology-based_2019,mitsis_evaluation_2019,hocine_personalized_2019, hocine_keep_nodate,ghorbani_towards_2022,esfahlani_adaptive_2017,gonzalez-gonzalez_serious_2019,pinto_adaptive_2018,alves_towards_2019,caggianese_serious_2019,eun_artificial_2023,doumas_clinical_2025,kira_approach_2024,everard_self-adaptive_2025}   \\
		
		 \makecell{\textbf{Learning-Based} \\e.g., neural networks, RL, classification \& clustering} & 4 &\cite{de_oliveira_gamification-based_2020,martinho_effects_2023,schafer_study_2018,zhao_physical_2020} & 6& \cite{brown_intelligent_2014,andrade_dynamic_2014,lin_exergame-integrated_2023,bouatrous_new_2023,stranick_adaptive_2022,aguilar_proposal_2022}  \\
		 \makecell{\textbf{Optimization \& Search-Based}\\e.g., genetic algorithms, MCTS, swarm intelligence} & -&- &6& \cite{semet_artificial_2019,esfahlani_rehabgame_2018,hocine_adaptation_2015,mitsis_procedural_2020,nathella_challenge-based_2025,delmas_fostering_2018}   \\ 
		
		 \makecell{\textbf{Probabilistic \& Statistical}\\e.g., regression, Markov Chains, Bayesian} &2& \cite{mocanu_kinect_2016,shen_gamified_2025}&-&-   \\
		 \makecell{\textbf{Hybrid Systems}\\combination of 2 or more types} &-& - &8& \cite{ahmad_architecting_2022,alves_flow_2018,afyouni_adaptive_2020,sadeghi_esfahlani_fusion_2019,gray_improving_2023,hoffmann_personalized_2015,forgiarini_probabilistic_nodate,faria_adaptive_2025}    \\

		\bottomrule
	\end{tabular}
}
\end{table*}

\section{Personalization Paradigm Level}
\label{sec:rev:paradigm}

The paradigm level captures the source of the adaptation logic, i.e., how the system decides to change in response to the models described in Section~\ref{sec:rev:model} and before specifying the concrete computational mechanisms, detailed in Section~\ref{sec:rev:algorithmic}. A distinction is made between knowledge-driven, data-driven and hybrid personalization. Knowledge-driven systems are often transparent and align with existing (clinical) protocols, showing predictable and explainable behavior. Constructing such roles or expert knowledge models is often labor-intensive and requires regular maintenance. Moreover, scaling to diverse users can be challenging. Data-driven approaches can capture nuanced behavioral or physiological patterns, and often have the flexibility to adapt as more data accumulates. They, however, require the presence of sufficient and qualitative data. Additionally, data-driven techniques can often operate as a ``black box'' offering less interpretability compared to knowledge-driven approaches. Hybrid systems can combine the learned sensitivity of data-driven approaches with the expert-defined boundaries defined by knowledge-driven systems. Such system can however be more complex as they require integration between expert knowledge models and learning components.  \par
\begin{figure}[]
	\centering
	\includegraphics[scale=0.5]{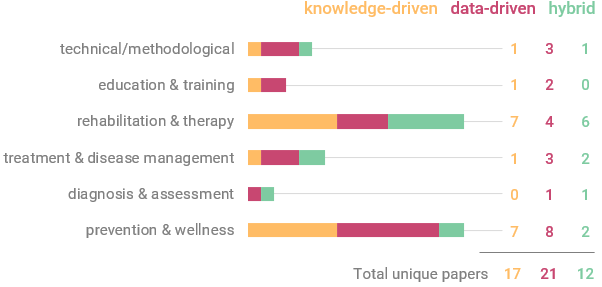}
	\caption{Distribution of knowledge-driven, data-driven and hybrid personalization across the six domains. Overall, hybrid personalization remains the minority.  }
	\label{fig:rev:domains_paradigm}
\end{figure}

 
 Tables~\ref{table:rev:gamification} and~\ref{table:rev:serious games} show the category for each of the included papers in the \textit{Type} column. Across the 50 included studies, data-driven approaches are most common, namely 21 studies out of 50, followed by knowledge-driven (17/50) and hybrid approaches (12/50). A strong presence of knowledge-driven (7/17) and hybrid (6/17) personalization approaches can be observed in the \textit{rehabilitation \& therapy} domain as shown in Figure~\ref{fig:rev:domains_paradigm}. Therapist knowledge and safety constraints are often operationalized as rules, ontologies or state machines and fuzzy logic, inverse kinematics or bounded search are used to adapt intensity or content. Hybrid approaches are more commonly used when performance- or sensor-based learning is restricted by expert-defined thresholds. In the \textit{prevention \& wellness} domain, it is a close comparison between knowledge-driven (8/17) and data-driven (7/17), with only 2 studies reporting on a hybrid approach (2/17). \par 
Knowledge-driven personalization often relies on expert knowledge models, such as ontologies and expert IF-THEN rules, sometimes complemented by lightweight player data. This trend can also be seen in the included studies, where 9 out of 17 knowledge-driven systems report of some kind of expert knowledge model. Data-driven approaches on the other hand might need player models populated by player performance and sensor data, which is also visible in our analysis, as 11 out of 21 data-driven approaches mention the use of a player model.

\section{Algorithmic Framework Level}
\label{sec:rev:algorithmic}
The algorithmic framework level specifies the computational methods used to execute the personalization. The following sections will discuss the 4 identified categories within this level, i.e. logic based, learning-based, optimization \& search-based, probabilistic \& statistical. Table~\ref{table:rev:pers_methods} provides an overview of the classification of the included papers, including those that apply a hybrid strategy. 
\subsection{Logic-based Methods}
Four distinct logic-based methods were identified, namely rule-based (expert) systems, fuzzy logic, finite state machines and decision trees. The methods will be briefly discussed in the following paragraphs. 
\subsubsection*{Rule-based systems} 

In the results, a distinction can be made between systems that use simple rule sets or pre-defined formula's~\cite{nuijten_evaluating_2022,martin-niedecken_comparing_2021,amiri_stepar_2022,doumas_clinical_2025,forgiarini_probabilistic_nodate, hocine_keep_nodate,kira_approach_2024, hocine_personalized_2019} and those employing rule-based expert systems~\cite{pardos_enriching_2023,mitsis_ontology-based_2019,mitsis_evaluation_2019,alves_towards_2019,eun_artificial_2023,everard_self-adaptive_2025,gonzalez-gonzalez_serious_2019}. Simple rule sets often rely on IF-THEN statements or pre-defined formula's that are used for linking specific player characteristics to game elements in a deterministic manner. These have limited flexibility and limited reasoning capacity. In contract, rule-based expert systems implement an inference engine that is decoupled from underlying data~\cite{liu_rule-based_2016,grosan_rule-based_2011}. Such a system uses a knowledge base to store domain knowledge as facts and rules~\cite{grosan_rule-based_2011}. An ontology is an example of such a knowledge base~\cite{mitsis_ontology-based_2019,mitsis_evaluation_2019}. The inference engine does the reasoning, using the facts and rules from the knowledge base. Two approaches can be discerned: forward or backward chaining. The first starts from known facts and applies rules to see what can be concluded. The latter starts with a goal or hypothesis, and works backwards to see if it is supported by the known facts. 
 
\subsubsection*{Fuzzy logic} 
Fuzzy Logic systems incorporate, similarly to rule-based systems, human logic and rules. However, unlike rule-based systems, the gradual transformation from one condition to another is possible, rather than strict true/false condition, which makes it possible to model uncertain information~\cite{wang_course_1997, zadeh_role_1994}. Fuzzy logic has been used in different \gls{sgs} for rehabilitation to analyze the player's achievements and suggest suitable adjustments to the physical rehabilitation exercises~\cite{ caggianese_serious_2019, esfahlani_adaptive_2017} or cognitive rehabilitation exercises~\cite{ghorbani_towards_2022}.

\subsubsection*{Finite State Machine} 
\begin{figure}[]
	\centering
	\includegraphics[scale=0.35]{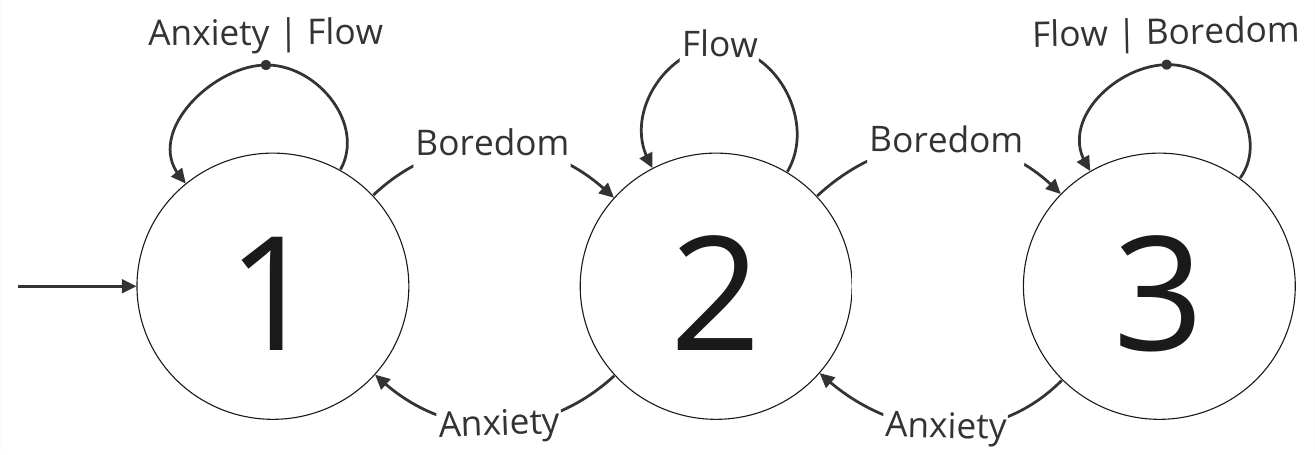}
	\caption[Example of a three-states-machine]{A three-states-machine that switches states if the user's current mental state changes~\cite{alves_flow_2018}.}
	\label{fig:rev:state-machine}
\end{figure}

\begin{figure}[]
	\centering
	\includegraphics[scale=0.5]{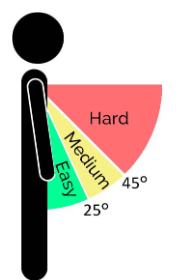}
	\caption[Example of three motion difficulty states]{An example of the three difficulty states for a specific mini-game in the InMotion game~\cite{pinto_adaptive_2018}.}
	\label{fig:rev:difficulty-states}
\end{figure}

A Finite State Machine consist of a set of states and transitions between these states~\cite{iovino_programming_2022}. The first-person shooter game of Alves et al.~\cite{alves_flow_2018} uses a finite state machine that transitions through the states based on the classification of the user's mental state, more specifically, boredom, anxiety or flow, as shown in Figure~\ref{fig:rev:state-machine}. The InMotion rehabilitation game~\cite{pinto_adaptive_2018}, on the other hand, uses the performance results of the user to transfer between different difficulty states, namely, easy, medium and hard, as shown in Figure~\ref{fig:rev:difficulty-states}. The thresholds for entering a different difficulty state differ for each minigame and are customized for each patient.

\subsubsection*{Decision trees}
\label{subsec:rev:decisiontrees}
\begin{figure}[]
	\centering
	\includegraphics[scale=0.7]{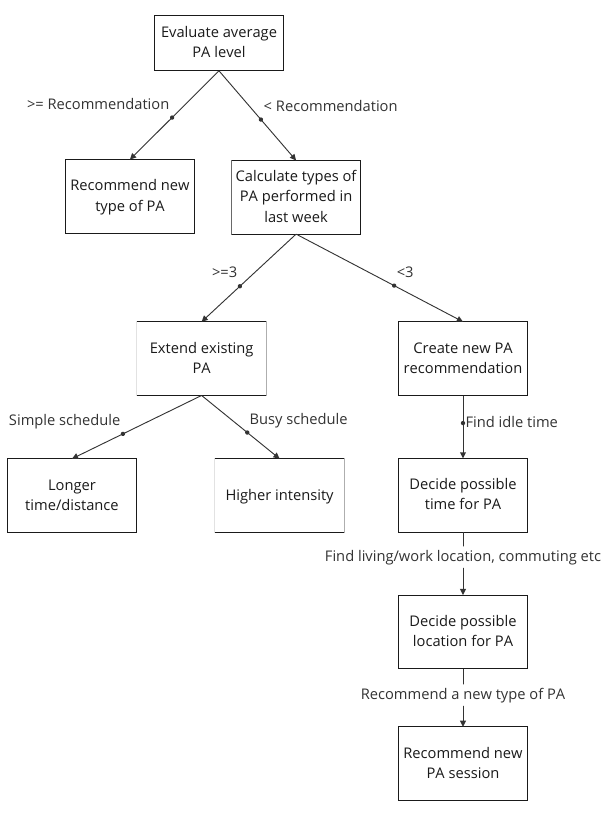}
	\caption[An example of a decision tree]{An example of a decision tree used by Zhao et al. in the recommendation engine~\cite{zhao_effects_2020}.}
	\label{fig:rev:decisiontree}
\end{figure}

A decision tree consists of decision nodes that specify conditions to with outgoing branches representing possible values resulting from that test. The leaves of the tree each specify a category or outcome~\cite{oliver_decision_1993}.  Zhao et al.~\cite{zhao_effects_2020} built a 4-layered model to represent the user in their personalized fitness recommender system, discussed in Section~\ref{sec:rev:model}. The recommendation engine is based on decision trees that incorporate all the user model information. The decision tree can suggest to extend an existing activity, recommend other types of activities, or recommend to fill some idle time with an activity. Figure~\ref{fig:rev:decisiontree} shows an example of such a decision tree.

\subsection{Learning-based Methods}

The included papers mentioned 5 different learning-based computational models for personalization, including \gls{ann}, Random Forest Classification, Deep Learning, \gls{rl} and decision trees. The following paragraphs briefly discuss these different methods. 
\subsubsection*{Artificial Neural Networks (ANN)} 
\gls{ann} are mathematical models that are able to detect complex non-linear correlations between data, using weights and biases~\cite{sadeghi_esfahlani_fusion_2019}. One gamified intervention~\cite{de_oliveira_gamification-based_2020} and two \gls{sgs}~\cite{brown_intelligent_2014, aguilar_proposal_2022} use a form of ANN to offer personalized support, while two other \gls{sgs} intervention consider a hybrid approach, combining an \gls{ann} with fuzzy logic~\cite{sadeghi_esfahlani_fusion_2019} or a finite state machine~\cite{alves_flow_2018}. de Oliveira et al.~\cite{de_oliveira_gamification-based_2020} use an \gls{ann} for the classification of the usage pattern of the user to assess if the player is still interested in the offered gamification elements or if updating them is required. In the case of a serious game for supporting Caribbean Men pre- and post- diagnosis of prostate cancer~\cite{brown_intelligent_2014}, an ANN is used for a computational intelligence predictor that predicts the risk of cancer for the user and then updates the offered information and support in the game for the user based on the outcome of the intelligence predictor. Esfahlani et al~\cite{sadeghi_esfahlani_fusion_2019} uses a hybrid combination of an ANN and fuzzy logic to adjust the difficulty of a serious game for neurorehabilitation. The ANN were used to detect complex non-linear correlations among player movement data and predict the player's improvement, while fuzzy logic was used to then personalize the offered rehabilitation exercises.

\subsubsection*{Random Forest Classification}
Random Forest Classicification is used for the classification of the activity model for children by Schäfer et al~\cite{schafer_study_2018}, which was explained in Section~\ref{sec:rev:model}. The authors compared two classification models, Support Vector Machines (SVM) and Random Forests (RF), with the latter reaching the highest accuracy.  

\subsubsection*{Deep Learning}
Deep learning is a form of machine learning that consists of multiple processing layers to learn complex patterns from high-dimensional data with multiple layers of abstraction~\cite{lecun_deep_2015,li_deep_2018}. Ahmad et al.~\cite{ahmad_architecting_2022} envision a platform for smart serious games that manage large volumes of real-time sensor data to make personalized decisions. The platform can use a variety of algorithms such as deep learning and deep reinforcement learning to analyze the contextual data and player history. The use of optimization algorithms, introducing a hybrid approach, can then optimize these results under specific constraints, such as age or medical history, using particle swarm optimization or genetic algorithms.

\subsubsection*{Reinforcement Learning}
\gls{rl} is a machine learning technique where an intelligent agent learns from its environment to maximize rewards and minimize punishment mechanisms~\cite{andrade_dynamic_2014}. Andrade et al.~\cite{andrade_dynamic_2014} use a form of reinforcement learning, namely Q-learning, which does not need a detailed environmental model and can be interpreted as a Markov decision process with unknown probabilities and rewards. In the Nut Catcher game, the Q-learning game is used to balance the game difficulty by maximizing the performance function and keeping the game challenging and entertaining while the user performs repetitive rehabilitation exercises. Martinho et al.~\cite{martinho_effects_2023} use reinforcement learning for gamified coaching to increase the physical activity of the elderly. Based on the user's performance, reinforcement learning is applied to decide what health challenges should be next sent to the user by the virtual coach and when. The platform for smart \gls{sgs} of Ahmad et al.~\cite{ahmad_architecting_2022}, explained in the previous paragraph, incorporates deep reinforcement learning, i.e., the combination of deep learning and reinforcement learning, as one of the intelligent algorithms~\cite{li_deep_2018}.

\subsubsection*{Decision Trees}

A distinction must be made based on the method of construction of a decision tree, although the resulting tree still provides human-readable logic. While some are manually authored by a clinical expert, as discussed in Section~\ref{subsec:rev:decisiontrees}, some are generated based by training a dataset~\cite{lin_exergame-integrated_2023, aguilar_proposal_2022}. 
Lin et al.~\cite{lin_exergame-integrated_2023} uses a decision tree for sequential discriminants, where the input variables were both objective  user data, such as previous workload level or physical data, and subjective data, such as perceived exertion rating. Based on this information, the decision tree is used to determine the workload level for the next exercises~\cite{lin_exergame-integrated_2023}. Aguilar et al.~\cite{aguilar_proposal_2022} uses a dataset to generate a decision tree that is able to decide difficulty perception level based on heart rate data.

\subsection{Optimization \& Search-based Methods}
Six different approaches were found that employ and optimization and search-based approach, namely, genetic algorithms, Particle Swarm Optimization, Ant Colony Optimization, Mone Carlo Tree Search, Multi-Armed Bandit and finally, Covariance Matrix Adaptation-Evolutionary Strategy Optimization. Each of these methods is briefly discussed in the following paragraphs. 

\subsubsection*{Genetic Algorithms} 
Genetic algorithms are heuristic search methods that use principles of natural selection and genetics to solve complex optimization problems~\cite{reeves_genetic_2018,sastry_genetic_2005}. Genetic algorithms are used in games for procedural content generation because they are able to generate highly customized content for a game, which keeps evolving according to the progress of the user~\cite{mitsis_procedural_2020}. The game ``Wake Up For the Future!''~\cite{mitsis_procedural_2020} uses procedural content generation based on a genetic algorithm to create educational content for obstructive sleep apnea. By automatically generating new \gls{npcs}, based on the user's in-game data and choices, the game difficulty can be dynamically adapted and educational content is personalized for each user. The platform for smart \gls{sgs} of Ahmad et al.~\cite{ahmad_architecting_2022}, which was already discussed earlier, suggests genetic algorithms can be used in the optimization module.

\subsubsection*{Particle Swarm Optimization} 
Particle Swarm Optimization is an optimization algorithm inspired by swarm behavior found in nature. It differs from genetic algorithms in the lack of a selection step as each member of the population survives~\cite{kumar_particle_2021}. Similarly to the previous paragraph, Particle Swarm Optimization can be used in the optimization module of the platform for smart \gls{sgs}~\cite{ahmad_architecting_2022}.

\subsubsection*{Ant Colony Optimization}
Ant Colony Optimization is also an optimization algorithm inspired by nature, more specifically, by the behavior of ants~\cite{stutzle_ant_2009}.  Semet et al.~\cite{semet_artificial_2019} apply the Ant Colony Optimization algorithm to achieve an intelligent and adaptive reward allocation system according to the performance of the user. 

\subsubsection*{Monte Carlo Tree Search (MCTS)}
\gls{mcts} is an optimization algorithm that takes random samples in the decision space and builds a search tree while doing so. It combines random simulation, i.e. Monte Carlo, with tree-based exploration~\cite{swiechowski_monte_2023,winands_monte-carlo_2024}. In the Rehabgame~\cite{esfahlani_rehabgame_2018} and the Prehab game~\cite{hocine_adaptation_2015}, MCTS is used to gradually control the intensity of the rehabilitation exercises based on the patient's previous performances, by generating the next set of tasks for the user's current skill level.  

\subsubsection*{Multi-Armed Bandit}
\gls{mab} is a decision-making and optimization algorithm that provides a simple model of the trade-off between exploration and exploitation to maximize gain~\cite{auer_nonstochastic_2002}. Multi-armed bandits are computationally efficient and rely on weak knowledge models, however, there is no long-term planning to find the optimal path~\cite{delmas_fostering_2018,auer_nonstochastic_2002}. The KidBreath~\cite{delmas_fostering_2018} serious game for children with Asthma uses an adaptation of the Multi-Armed Bandit algorithm to personalize the content of the health education game, based on the child's progression. Gray et al.~\cite{gray_improving_2023} propose a hybrid approach, the Shapley Bandit, which combines \gls{mab} with the Shapley Value to estimate the contributions of each of the players and prioritize players' preferences based on that contribution. The Shapley value is a method to determine the costs and rewards attributed to an individual working in a group~\cite{gray_improving_2023}. 

\subsubsection*{Covariance Matrix Adaptation-Evolutionary Strategy (CMA-ES) Optimizer}
Nathella et al.~\cite{nathella_challenge-based_2025} employ a human-in-the-loop optimization strategy to adapt parameters for a robotic knee exoskeleton and biofeedback serious game. Their approach relies on the CMA-ES optimizer, a black-box optimizer that iteratively samples and evaluates generations of parameter sets, ranks them according to a cost function, and updates its estimate of the underlying search space. CMA-ES is well suited to be used in time-varying spaces, such as rehabilitation, where the patient's response evolves over time, due to learning or motor adaptation. 

\subsection{Probabilistic \& Statistical Methods}
Only 4 probabilistic methods were identified, of which two are only used in a hybrid system. The different methods are discussed in the paragraphs below. 

\subsubsection*{Linear Regression}
Linear regression can be seen as a statistical method that predicts the value of a variable based on the value of another variable. Shen et al.~\cite{shen_gamified_2025} use a linear regression model to adjust the difficulty level according to a predefined difficulty curve. A hybrid approach is also possible, using linear regression to build a user model, and then controlling training load by a rule-based system~\cite{hoffmann_personalized_2015}. 

\subsubsection*{Cross-correlation}
Mocanu et al.~\cite{mocanu_kinect_2016} use cross-correlation for signal-processing, a statistical signal-similarity approach to quantify the similarity between two time-series signals at varying lags. These signals are used to evaluate the performance of the user relative to predefined exercises. 
\subsubsection*{Discrete Time Markov Chains}
A discrete time Markov chain models a system that moves between different states at fixed intervals, in this case one second of game-play in a 20 minute game session~\cite{forgiarini_probabilistic_nodate}. Forgiarini et al.~\cite{forgiarini_probabilistic_nodate} combine it in a hybrid system with expert-defined weights, to calculate the likelihood of game-play paths.

\subsubsection*{Bayesian Immediate Feedback Learning (BIFL)}
Faria et al.~\cite{faria_adaptive_2025} propose the hybrid approach of Bayesian Immediate Feedback Learning (BIFL) that combines Bayesian inference and \gls{mab} to dynamically select the most effective modality, namely visual, auditory, or textual feedback during game-play for neurodivergent children. Unlike \gls{rl}, which requires extensive training, BIFL is ideal when extensive datasets are limited. It uses Bayesian posterior updates of observed responses to update its decision-making strategy~\cite{faria_adaptive_2025}.

\section{Reusability and Standardization}
\label{sec:rev:reuse}

Reusable components shorten development cycles, lower integration costs and enable controlled comparisons of personalization strategies under similar constraints~\cite{carlier_software_2023}. In health contexts, reusability can also support explainability, i.e. traceable models and rules, safety, i.e. adaptation based on validated models and boundaries, and transferability of knowledge across different populations. \par  
Nonetheless, reusability remains highly under-addressed in personalized \gls{sgg} for health as only 12 out of 50 included studies explicitly design for reuse, as indicated in Tables~\ref{table:rev:gamification} and~\ref{table:rev:serious games}. Each of these 12 studies implement reuse of their \gls{sg} or parts of their implementation differently. \par 
This section will first provide a brief overview of how these 12 studies designed for reuse, followed by a discussion of possible implications for reuse. 

\subsection{Overview}
Carlier et al.~\cite{carlier_investigating_2021} designed a gamified app for health surveys. The app has been designed such that it can easily be reused for the gamification of other surveys. de Oliveira~\cite{de_oliveira_gamification-based_2020} created a framework, Framework L, that guides mobile health application developers in the creation of new mHealth apps for Self-care by selecting which categories should be included in the application and which data must be collected. The mHealth GameBus tool~\cite{nuijten_evaluating_2022} allows reuse for testing purposes, as the platform supports hosting multiple experimental designs and easy configuration of the gamification mechanisms.
Mitsis et al.~\cite{mitsis_ontology-based_2019} facilitate the reuse of their recipe and game ontology by focusing on reusability, extensibility and sustainability when designing their ontology. Semet et al.~\cite{semet_artificial_2019} consider reuse when drafting the requirements for their reward algorithm, as the algorithm should be generic to be used by other \gls{sgs} on the InLife platform in the future.  The Keep Attention serious game~\cite{hocine_keep_nodate} has been designed such that the tasks that consider the training objects are independent of the game elements, thereby facilitating the creation of different games. Similarly, the PRehab game decouples the game mechanics from the game graphics, so once rules and game behaviors are implemented, they can easily be reused in other games with different graphics~\cite{hocine_adaptation_2015}. Ahmet et al.~\cite{ahmad_architecting_2022} propose a modular architecture for smart \gls{sgs} that ensures high cohesion and low coupling. Developers can decide which contextual data is needed to be used for analysis for the game and other personalization strategies can easily be applied or added. The TANGO:H platform~\cite{gonzalez-gonzalez_serious_2019} allows health professionals to design different types of rehabilitation exercises and games using the Kinect. Similarly, the rehabilitation system of Caggianese et al.~\cite{caggianese_serious_2019} ensures reuse by introducing an \textit{adaptive game handler} component that decouples the \gls{sgs} from the rest of the system. This means that new \gls{sgs} can easily be added if they conform to the common interfaces.  

\subsection{Implications for Reuse}

On the model level, reuse depends on how player models and expert knowledge models are specified and decoupled from the game logic. Expert knowledge becomes reusable when ontologies and rule sets are modular, versioned and typed with stable interfaces, so these assets can be plugged into multiple games. Player models can become reusable when their model schema become explicit, i.e. fields, units, calibration procedures and update rules are defined and the data origin and flow, i.e., data provenance, are documented. Consequently, algorithms and data pipelines can be retrained or interchanged without redesign. Additionally, reuse can be facilitated by mapping tables that translate game-specific events into shared player model elements. \par 
On the personalization paradigm level, the approach has direct implications for reuse. Knowledge-driven systems are most reusable when expert knowledge is encapsulated in modular and shareable assets, such as ontologies or rule sets, that are decoupled from game logic. Data-driven systems, on the other hand, become reusable when models are retrainable across contexts or when pipelines become standardized. 
Hybrid approaches hold the potential for high reuse by sharing knowledge layers and/or swapping and comparing data-driven modules, as long as the interaction between these modules is well-defined. \par 
On the algorithmic framework level, reuse improves when personalization decisions and data pipelines are abstracted from content. Algorithms should have an abstraction interface that maps game states to personalization actions, such that algorithms can be swapped without the need for any game-specific knowledge. In addition to abstraction, meaningful reuse also requires shared benchmarking conditions. When different personalization methods can be executed on common inputs, their performance can be compared under controlled settings. Establishing such benchmarks would support systematic evaluation, enable transparent algorithm swaps and contribute to an evidence base for selecting personalization strategies for \gls{sgg}. 

\section{Findings}
\label{sec:rev:findings}
This review highlights overarching patterns in how personalization is approached within \gls{sgg} for health. The following sections discuss the most important findings from the three-tiered analysis. 
\subsection{Patterns Across Domains and Personalization Goals}
Personalization in \gls{sgg} is not applied uniformly across health domains. Rehabilitation \& therapy and prevention \& wellness dominate the current research, each 17 entries, reflecting the practical and relevance of adaptive systems for maintaining engagement and supporting behavior change. In total, 35 studies report on serious games and 15 on gamification. In terms of goals, engagement is the most frequently stated objective (34/50), and in rehabilitation \& therapy personalization is often explicitly linked to performance improvement rather than engagement alone (10/17). This confirms that health contexts demand both sustained motivation and measurable functional gains, with rehabilitation placing strong emphasis on performance-oriented adaptation. 

\subsection{Trends in Modeling and Personalization Approaches}
A key-finding is the co-existence of player-centered models, capturing preferences, performance, physiological signals, or behavior, and expert knowledge models, including clinical guidelines, therapeutic goals, or kinematic constraints. In gamified systems, Hexad Type profiling remains the most used player-type approach to select or rotate game elements. Next, sensor-rich serious games frequently incorporate motor-ability representations, mental-state estimators or calibration-derived thresholds to align difficulty with the user's capacity or functioning. On the expert knowledge side, systems use ontologies that formalize nutrition or rehabilitation concepts, kinematic chain models with inverse kinematics for movement modeling, and IF-THEN rule bases that are able to translate therapist knowledge into transparent difficulty adjustments and safety limits. Collectively, these models enable systems to personalize towards individual needs while remaining within clinical boundaries, which is an essential balance in health applications.  

\subsection{Gaps, Limitations and Opportunities for Reuse}
While the review reveals a rich variety of computational techniques, this diversity also reflects a lack of standardization. Many systems implement bespoke personalization pipelines that tightly couple user modeling, adaptation rules and game logic, which limits their reusability and reducing opportunities for systematic comparison. Only a minority of studies explicitly design for reuse through modular architectures, shareable ontologies or configurable personalization engines. Even when reuse is mentioned, implementations are often domain or game-specific rather than generalizable assets. \par 
This fragmentation signals and important research gap. Without interoperable components, consistent modeling schemes or reusable expert knowledge bases, it remains difficult to prototype new systems rapidly, benchmark alternative personalization methods or transfer therapeutic insights across applications. At the same time, the increasing availability of sensor data, learning-based methods and generative AI techniques present opportunities to develop more scalable and adaptable personalization pipelines, given that these remain grounded in transparent and clinically aligned modeling practices. 

\section{Research Gaps and Future Directions}
\label{sec:rev:future}
Despite encouraging progress, several gaps limit the scalability, comparability and adoption in real-world health settings of personalized \gls{sgg} in health. First, reusability remains the exception rather than the rule. Only a minority of systems are explicitly designed for reuse, and when they are, reuse tends to be limited, e.g., platform-bound ontologies, game-specific rule sets, or tightly coupled controllers. This hinders transfer across games and contexts, slowing comparative evaluation. Prioritizing shareable assets is a clear opportunity: (i) player-model schemas with explicit fields, units, and update procedures, and (ii) expert-knowledge assets, such as ontologies or rule sets, with maintenance procedures and data provenance. Decoupling these models from the game logic through stable interfaces would enable plug-and-play reuse and controlled algorithm swaps. \par 
A second challenge is the prevalence of bespoke personalization pipelines. Such pipelines are difficult to reproduce and compare. Modular personalization engines could drastically advance the field as computational methods could be swapped out under identical conditions, evaluating their effectiveness. This would improve replicability and accelerate method selection for specific health objectives.  \par 
Third, user profiles are often static or under-specified over time, despite the fact that abilities, preferences and contexts evolve, particularly in longer-term rehabilitation and chronic care. Moving towards more longitudinal, context-aware profiles that are able to update based on sensor data, clinical input or user interaction, is a promising direction. Such profiles should incorporate and model changes such as progression trends or fatigue patterns, and confidence estimates so that decisions can be calibrated to uncertainty as well as capability of the user. \par 
A fourth gap concerns the balance between data-driven sensitivity and clinical safety. Data-driven systems can personalize effectively but may lack transparency. Knowledge-driven systems are interpretable but can be rigid. Hybrid architectures, in which learned strategies operate within expert-defined bounds, e.g. intensity boundaries, progression rules, or protocol thresholds, offer a way forward. These hybrid systems should emphasize explainability to meet healthcare accountability demands and to support clinical oversight. \par 
Finally, generative AI holds the potential to reduce time to prototype in content generation, dynamic narratives and code generation. However, adoption should proceed with guardrails such as expert-approved constraints, human-in-the-loop verification, and prospective validation against clinical targets before deployment. Alongside this, the \gls{sgg} for health community should promote interoperability and sharing to enable external validation and comparisons, fostering a more cumulative science of personalised \gls{sgg}.

\section{Conclusion}
\label{sec:rev:conclusion}
This review aimed to clarify how personalization is achieved in Serious Games and Gamification for health by analyzing the model level, the personalization paradigm level and the algorithmic framework level. Fifty interventions, published between 2014 and 2025 were synthesized, consisting of 35 serious games and 15 gamified systems, spanning 6 domains. The domains of rehabilitation \& therapy and prevention \& wellness are most represented. The three-tiered classification framework offers a structured lens that makes the design choices of such personalized \gls{sgg} comparable across diverse health contexts. \par 
At the model level, their is an interplay between player-centered models and expert knowledge models. At the personalization paradigm level, data-driven approaches are the most common overall, yet knowledge-driven and hybrid systems dominate in rehabilitation. Algorithmically, a myriad of models have been identified and standardized personalization pipelines remain rare. \par 
A central challenge that emerges from the results of the analysis is reusability. Only a minority of studies mention reusable assets or modular personalization engines. Future progress in the field of personalized \gls{sgg} for health will depend on modularity, standardization and clinical alignment: reusable models and knowledge assets, plug-and-play engines, longitudinal profiling, hybrid decision architectures that are both data-sensitive and explainable, shared benchmarking conditions for comparing algorithms under controlled settings and user evaluations to characterize user experiences. Collectively, these directions provide a practical path to faster prototyping, more rigorous comparison and understanding of personalized \gls{sgg} in health. \par

\section*{Authors' Contributions}
\section*{Acknowledgements}
This research is funded by Research Foundation - Flanders (1248126N).
\section*{Declaration of generative AI and AI-assisted technologies in the manuscript preparation process}
During the preparation of this work the author(s) used Copilot in order to improve language and structure. After using this tool/service, the author(s) reviewed and edited the content as needed and take(s) full responsibility for the content of the publication.

\printcredits

\bibliographystyle{model1-num-names}
\bibliography{references,references2}

\begin{thebibliography}{129}
\expandafter\ifx\csname natexlab\endcsname\relax\def\natexlab#1{#1}\fi
\providecommand{\url}[1]{\texttt{#1}}
\providecommand{\href}[2]{#2}
\providecommand{\path}[1]{#1}
\providecommand{\DOIprefix}{doi:}
\providecommand{\ArXivprefix}{arXiv:}
\providecommand{\URLprefix}{URL: }
\providecommand{\Pubmedprefix}{pmid:}
\providecommand{\doi}[1]{\href{http://dx.doi.org/#1}{\path{#1}}}
\providecommand{\Pubmed}[1]{\href{pmid:#1}{\path{#1}}}
\providecommand{\bibinfo}[2]{#2}
\ifx\xfnm\relax \def\xfnm[#1]{\unskip,\space#1}\fi
\bibitem[{Damaševičius et~al.(2023)Damaševičius, Maskeliūnas, and
  Blažauskas}]{damasevicius_serious_2023}
\bibinfo{author}{R.~Damaševičius}, \bibinfo{author}{R.~Maskeliūnas},
  \bibinfo{author}{T.~Blažauskas},
\newblock \bibinfo{title}{Serious {Games} and {Gamification} in {Healthcare}:
  {A} {Meta}-{Review}},
\newblock \bibinfo{journal}{Information} \bibinfo{volume}{14}
  (\bibinfo{year}{2023}).
\bibitem[{Alahäivälä and
  Oinas-Kukkonen(2016)}]{alahaivala_understanding_2016}
\bibinfo{author}{T.~Alahäivälä}, \bibinfo{author}{H.~Oinas-Kukkonen},
\newblock \bibinfo{title}{Understanding persuasion contexts in health
  gamification: {A} systematic analysis of gamified health behavior change
  support systems literature},
\newblock \bibinfo{journal}{International Journal of Medical Informatics}
  \bibinfo{volume}{96} (\bibinfo{year}{2016}) \bibinfo{pages}{62--70}.
\bibitem[{Metwally et~al.(2021)Metwally, Chang, Wang, and
  Yousef}]{metwally_does_2021}
\bibinfo{author}{A.~H.~S. Metwally}, \bibinfo{author}{M.~Chang},
  \bibinfo{author}{Y.~Wang}, \bibinfo{author}{A.~M.~F. Yousef},
\newblock \bibinfo{title}{Does gamifying homework influence performance and
  perceived gameful experience?},
\newblock \bibinfo{journal}{Sustainability (Switzerland)} \bibinfo{volume}{13}
  (\bibinfo{year}{2021}) \bibinfo{pages}{4829}.
\bibitem[{Fitzgerald and Ratcliffe(2020)}]{fitzgerald_serious_2020}
\bibinfo{author}{M.~Fitzgerald}, \bibinfo{author}{G.~Ratcliffe},
\newblock \bibinfo{title}{Serious games, gamification, and serious mental
  illness: {A} scoping review},
\newblock \bibinfo{journal}{Psychiatric Services} \bibinfo{volume}{71}
  (\bibinfo{year}{2020}) \bibinfo{pages}{170--183}.
\bibitem[{Seyderhelm et~al.(2019)Seyderhelm, Blackmore, and
  Nesbitt}]{seyderhelm_towards_2019}
\bibinfo{author}{A.~J. Seyderhelm}, \bibinfo{author}{K.~L. Blackmore},
  \bibinfo{author}{K.~Nesbitt},
\newblock \bibinfo{title}{Towards {Cognitive} {Adaptive} {Serious} {Games}: {A}
  {Conceptual} {Framework}},
\newblock in: \bibinfo{booktitle}{Lecture {Notes} in {Computer} {Science}
  (including subseries {Lecture} {Notes} in {Artificial} {Intelligence} and
  {Lecture} {Notes} in {Bioinformatics})}, volume \bibinfo{volume}{11863 LNCS},
  \bibinfo{publisher}{Springer}, \bibinfo{year}{2019}, pp.
  \bibinfo{pages}{331--338}. \DOIprefix\doi{10.1007/978-3-030-34644-7_27}.
\bibitem[{Graafland and Schijven(2018)}]{graafland_how_2018}
\bibinfo{author}{M.~Graafland}, \bibinfo{author}{M.~Schijven},
\newblock \bibinfo{title}{How {Serious} {Games} {Will} {Improve} {Healthcare}},
\newblock \bibinfo{year}{2018}, pp. \bibinfo{pages}{139--157}.
\bibitem[{Tondello and {Others}(2019)}]{tondello_empirical_2019}
\bibinfo{author}{G.~F. Tondello}, \bibinfo{author}{{Others}},
\newblock \bibinfo{title}{Empirical validation of the {Gamification} {User}
  {Types} {Hexad} scale in {English} and {Spanish}},
\newblock \bibinfo{journal}{International Journal of Human Computer Studies}
  \bibinfo{volume}{127} (\bibinfo{year}{2019}) \bibinfo{pages}{95--111}.
\bibitem[{Susi et~al.(2015)Susi, Johannesson, and Backlund}]{susi_serious_2015}
\bibinfo{author}{T.~Susi}, \bibinfo{author}{M.~Johannesson},
  \bibinfo{author}{P.~Backlund},
\newblock \bibinfo{title}{Serious {Games} - {An} {Overview}}
  (\bibinfo{year}{2015}).
\bibitem[{Ritterfeld et~al.(2009)Ritterfeld, Cody, and
  Vorderer}]{ritterfeld_serious_2009}
\bibinfo{author}{U.~Ritterfeld}, \bibinfo{author}{M.~Cody},
  \bibinfo{author}{P.~Vorderer}, \bibinfo{title}{Serious {Games}: {Mechanisms}
  and {Effects}}, \bibinfo{publisher}{Routledge}, \bibinfo{year}{2009}.
\bibitem[{Afyouni et~al.(2020)Afyouni, Murad, and
  Einea}]{afyouni_adaptive_2020}
\bibinfo{author}{I.~Afyouni}, \bibinfo{author}{A.~Murad},
  \bibinfo{author}{A.~Einea},
\newblock \bibinfo{title}{Adaptive rehabilitation bots in serious games},
\newblock \bibinfo{journal}{Sensors (Switzerland)} \bibinfo{volume}{20}
  (\bibinfo{year}{2020}) \bibinfo{pages}{1--30}.
\bibitem[{Afyouni et~al.(2017)Afyouni, Qamar, Hussain, Rehman, Sadiq, and
  Murad}]{afyouni_motion-based_2017}
\bibinfo{author}{I.~Afyouni}, \bibinfo{author}{A.~M. Qamar},
  \bibinfo{author}{S.~O. Hussain}, \bibinfo{author}{F.~U. Rehman},
  \bibinfo{author}{B.~Sadiq}, \bibinfo{author}{A.~Murad},
\newblock \bibinfo{title}{Motion-based serious games for hand assistive
  rehabilitation},
\newblock in: \bibinfo{booktitle}{International {Conference} on {Intelligent}
  {User} {Interfaces}, {Proceedings} {IUI}}, \bibinfo{year}{2017}, pp.
  \bibinfo{pages}{133--136}. \URLprefix
  \url{http://dx.doi.org/10.1145/3030024.3040977}.
  \DOIprefix\doi{10.1145/3030024.3040977}.
\bibitem[{Aguilar et~al.(2019)Aguilar, Altamiranda, Diaz, De~Mesa, and
  Pinto}]{aguilar_adaptive_2019}
\bibinfo{author}{J.~Aguilar}, \bibinfo{author}{J.~Altamiranda},
  \bibinfo{author}{F.~Diaz}, \bibinfo{author}{J.~G. De~Mesa},
  \bibinfo{author}{A.~Pinto},
\newblock \bibinfo{title}{Adaptive plot system for serious emerging games based
  on the ant colony optimization algorithm},
\newblock in: \bibinfo{booktitle}{Proceedings - 2019 45th {Latin} {American}
  {Computing} {Conference}, {CLEI} 2019}, \bibinfo{publisher}{Institute of
  Electrical and Electronics Engineers Inc.}, \bibinfo{year}{2019}.
  \DOIprefix\doi{10.1109/CLEI47609.2019.235104}.
\bibitem[{González-González et~al.(2019)González-González, Toledo-Delgado,
  Muñoz-Cruz, and Torres-Carrion}]{gonzalez-gonzalez_serious_2019}
\bibinfo{author}{C.~S. González-González}, \bibinfo{author}{P.~A.
  Toledo-Delgado}, \bibinfo{author}{V.~Muñoz-Cruz}, \bibinfo{author}{P.~V.
  Torres-Carrion},
\newblock \bibinfo{title}{Serious games for rehabilitation: {Gestural}
  interaction in personalized gamified exercises through a recommender system},
\newblock \bibinfo{journal}{Journal of Biomedical Informatics}
  \bibinfo{volume}{97} (\bibinfo{year}{2019}) \bibinfo{pages}{103266}.
\bibitem[{Lau and Agius(2021)}]{lau_framework_2021}
\bibinfo{author}{S.~Y.~J. Lau}, \bibinfo{author}{H.~Agius},
\newblock \bibinfo{title}{A framework and immersive serious game for mild
  cognitive impairment},
\newblock \bibinfo{journal}{Multimedia Tools and Applications}
  \bibinfo{volume}{80} (\bibinfo{year}{2021}) \bibinfo{pages}{31183--31237}.
\bibitem[{Silva(2020)}]{silva_spatial_2020}
\bibinfo{author}{R.~J.~N. Silva}, \bibinfo{title}{Spatial {Augmented} {Reality}
  in {Serious} {Games} for {Cognitive} {Rehabilitation} of the {Elderly}},
  \bibinfo{year}{2020}. \URLprefix
  \url{https://estudogeral.sib.uc.pt/handle/10316/92257}.
\bibitem[{Goumopoulos and Igoumenakis(2021)}]{goumopoulos_ontology-driven_2021}
\bibinfo{author}{C.~Goumopoulos}, \bibinfo{author}{I.~Igoumenakis},
\newblock \bibinfo{title}{Ontology-{Driven} {Mental} {Healthcare}
  {Applications}: {A} {Case} {Study} on {Cognitive} {Rehabilitation} with
  {Serious} {Games}},
\newblock in: \bibinfo{booktitle}{Communications in {Computer} and
  {Information} {Science}}, volume \bibinfo{volume}{1387},
  \bibinfo{publisher}{Springer, Cham}, \bibinfo{year}{2021}, pp.
  \bibinfo{pages}{114--140}. \URLprefix
  \url{https://link.springer.com/chapter/10.1007/978-3-030-70807-8{\_}7}.
  \DOIprefix\doi{10.1007/978-3-030-70807-8_7}.
\bibitem[{Martinho et~al.(2020)Martinho, Carneiro, Corchado, and
  Marreiros}]{martinho_systematic_2020}
\bibinfo{author}{D.~Martinho}, \bibinfo{author}{J.~Carneiro},
  \bibinfo{author}{J.~M. Corchado}, \bibinfo{author}{G.~Marreiros},
\newblock \bibinfo{title}{A systematic review of gamification techniques
  applied to elderly care},
\newblock \bibinfo{journal}{Artificial Intelligence Review}
  \bibinfo{volume}{53} (\bibinfo{year}{2020}) \bibinfo{pages}{4863--4901}.
\bibitem[{Vermeir et~al.(2020)Vermeir, White, Johnson, Crombez, and van
  Ryckeghem}]{vermeir_effects_2020}
\bibinfo{author}{J.~F. Vermeir}, \bibinfo{author}{M.~J. White},
  \bibinfo{author}{D.~Johnson}, \bibinfo{author}{G.~Crombez},
  \bibinfo{author}{D.~M. van Ryckeghem},
\newblock \bibinfo{title}{The effects of gamification on computerized cognitive
  training: {Systematic} review and meta-analysis},
\newblock \bibinfo{journal}{JMIR Serious Games} \bibinfo{volume}{8}
  (\bibinfo{year}{2020}) \bibinfo{pages}{e18644}.
\bibitem[{Haoran et~al.(2019)Haoran, Bazakidi, and Zary}]{haoran_serious_2019}
\bibinfo{author}{G.~Haoran}, \bibinfo{author}{E.~Bazakidi},
  \bibinfo{author}{N.~Zary}, \bibinfo{title}{Serious {Games} in {Health}
  {Professions} {Education}: {Review} of {Trends} and {Learning} {Efficacy}},
  \bibinfo{year}{2019}. \URLprefix
  \url{http://www.thieme-connect.com/products/ejournals/html/10.1055/s-0039-1677904
  http://www.thieme-connect.de/DOI/DOI?10.1055/s-0039-1677904}.
\bibitem[{Gorbanev et~al.(2018)Gorbanev, Agudelo-Londoño, González, Cortes,
  Pomares, Delgadillo, Yepes, and Muñoz}]{gorbanev_systematic_2018}
\bibinfo{author}{I.~Gorbanev}, \bibinfo{author}{S.~Agudelo-Londoño},
  \bibinfo{author}{R.~A. González}, \bibinfo{author}{A.~Cortes},
  \bibinfo{author}{A.~Pomares}, \bibinfo{author}{V.~Delgadillo},
  \bibinfo{author}{F.~J. Yepes}, \bibinfo{author}{O.~Muñoz}, \bibinfo{title}{A
  systematic review of serious games in medical education: quality of evidence
  and pedagogical strategy}, \bibinfo{year}{2018}. \URLprefix
  \url{https://www.tandfonline.com/doi/abs/10.1080/10872981.2018.1438718}.
\bibitem[{Abraham et~al.(2020)Abraham, LeMay, Bittner, Thakur, Stafford, and
  Brown}]{abraham_investigating_2020}
\bibinfo{author}{O.~Abraham}, \bibinfo{author}{S.~LeMay},
  \bibinfo{author}{S.~Bittner}, \bibinfo{author}{T.~Thakur},
  \bibinfo{author}{H.~Stafford}, \bibinfo{author}{R.~Brown},
  \bibinfo{title}{Investigating serious games that incorporate medication use
  for patients: {Systematic} literature review}, \bibinfo{year}{2020}.
  \URLprefix \url{https://games.jmir.org/2020/2/e16096}.
\bibitem[{Sharifzadeh et~al.(2020)Sharifzadeh, Kharrazi, Nazari, Tabesh,
  Khodabandeh, Heidari, and Tara}]{sharifzadeh_health_2020}
\bibinfo{author}{N.~Sharifzadeh}, \bibinfo{author}{H.~Kharrazi},
  \bibinfo{author}{E.~Nazari}, \bibinfo{author}{H.~Tabesh},
  \bibinfo{author}{M.~E. Khodabandeh}, \bibinfo{author}{S.~Heidari},
  \bibinfo{author}{M.~Tara}, \bibinfo{title}{Health education serious games
  targeting health care providers, patients, and public health users: {Scoping}
  review}, \bibinfo{year}{2020}. \URLprefix
  \url{https://games.jmir.org/2020/1/e13459}.
\bibitem[{Ricciardi and De~Paolis(2014)}]{ricciardi_comprehensive_2014}
\bibinfo{author}{F.~Ricciardi}, \bibinfo{author}{L.~T. De~Paolis},
  \bibinfo{title}{A {Comprehensive} {Review} of {Serious} {Games} in {Health}
  {Professions}}, \bibinfo{year}{2014}. \URLprefix
  \url{https://dl.acm.org/doi/abs/10.1155/2014/787968}.
\bibitem[{Hervas et~al.(2017)Hervas, Ruiz-Carrasco, Bravo, and
  Mondejar}]{hervas_gamification_2017}
\bibinfo{author}{R.~Hervas}, \bibinfo{author}{D.~Ruiz-Carrasco},
  \bibinfo{author}{J.~Bravo}, \bibinfo{author}{T.~Mondejar},
\newblock \bibinfo{title}{Gamification mechanics for behavioral change: {A}
  systematic review and proposed taxonomy},
\newblock in: \bibinfo{booktitle}{{ACM} {International} {Conference}
  {Proceeding} {Series}}, \bibinfo{publisher}{Association for Computing
  Machinery}, \bibinfo{year}{2017}, pp. \bibinfo{pages}{395--404}. \URLprefix
  \url{https://doi.org/10.1145/}. \DOIprefix\doi{10.1145/3154862.3154939}.
\bibitem[{David et~al.(2020)David, Pop, Roxana, and Mogoaşe}]{david_how_2020}
\bibinfo{author}{O.~David}, \bibinfo{author}{C.~Pop},
  \bibinfo{author}{C.~Roxana}, \bibinfo{author}{C.~Mogoaşe},
\newblock \bibinfo{title}{How {Effective} are {Serious} {Games} for {Promoting}
  {Mental} {Health} and {Health} {Behavioral} {Change} in {Children} and
  {Adolescents}? {A} {Systematic} {Review} and {Meta}-analysis},
\newblock \bibinfo{journal}{Child and Youth Care Forum}
  (\bibinfo{year}{2020}).
\bibitem[{Sardi et~al.(2017)Sardi, Idri, and
  Fernández-Alemán}]{sardi_systematic_2017}
\bibinfo{author}{L.~Sardi}, \bibinfo{author}{A.~Idri}, \bibinfo{author}{J.~L.
  Fernández-Alemán}, \bibinfo{title}{A systematic review of gamification in
  e-{Health}}, \bibinfo{year}{2017}.
\bibitem[{Thomas et~al.(2020)Thomas, Sivakumar, Babichenko, Grieve, and
  Klem}]{thomas_mapping_2020}
\bibinfo{author}{T.~H. Thomas}, \bibinfo{author}{V.~Sivakumar},
  \bibinfo{author}{D.~Babichenko}, \bibinfo{author}{V.~L. Grieve},
  \bibinfo{author}{M.~L. Klem},
\newblock \bibinfo{title}{Mapping behavioral health serious game interventions
  for adults with chronic illness: {Scoping} review},
\newblock \bibinfo{journal}{JMIR Serious Games} \bibinfo{volume}{8}
  (\bibinfo{year}{2020}).
\bibitem[{Hamari et~al.(2014)Hamari, Koivisto, and Sarsa}]{hamari_does_2014}
\bibinfo{author}{J.~Hamari}, \bibinfo{author}{J.~Koivisto},
  \bibinfo{author}{H.~Sarsa},
\newblock \bibinfo{title}{Does gamification work? - {A} literature review of
  empirical studies on gamification},
\newblock in: \bibinfo{booktitle}{Proceedings of the {Annual} {Hawaii}
  {International} {Conference} on {System} {Sciences}},
  \bibinfo{publisher}{IEEE Computer Society}, \bibinfo{year}{2014}, pp.
  \bibinfo{pages}{3025--3034}. \DOIprefix\doi{10.1109/HICSS.2014.377}.
\bibitem[{King et~al.(2021)King, Marsh, and Akcay}]{king_review_2021}
\bibinfo{author}{M.~King}, \bibinfo{author}{T.~Marsh},
  \bibinfo{author}{Z.~Akcay},
\newblock \bibinfo{title}{A {Review} of {Indie} {Games} for {Serious} {Mental}
  {Health} {Game} {Design}},
\newblock in: \bibinfo{booktitle}{Lecture {Notes} in {Computer} {Science}
  (including subseries {Lecture} {Notes} in {Artificial} {Intelligence} and
  {Lecture} {Notes} in {Bioinformatics})}, volume \bibinfo{volume}{12945 LNCS},
  \bibinfo{publisher}{Springer Science and Business Media Deutschland GmbH},
  \bibinfo{year}{2021}, pp. \bibinfo{pages}{138--152}. \URLprefix
  \url{https://link.springer.com/chapter/10.1007/978-3-030-88272-3{\_}11}.
  \DOIprefix\doi{10.1007/978-3-030-88272-3_11}.
\bibitem[{Sipiyaruk et~al.(2018)Sipiyaruk, Gallagher, Hatzipanagos, and
  Reynolds}]{sipiyaruk_rapid_2018}
\bibinfo{author}{K.~Sipiyaruk}, \bibinfo{author}{J.~E. Gallagher},
  \bibinfo{author}{S.~Hatzipanagos}, \bibinfo{author}{P.~A. Reynolds},
\newblock \bibinfo{title}{A rapid review of serious games: {From} healthcare
  education to dental education},
\newblock \bibinfo{journal}{European Journal of Dental Education}
  \bibinfo{volume}{22} (\bibinfo{year}{2018}) \bibinfo{pages}{243--257}.
\bibitem[{Sajjadi et~al.(2022)Sajjadi, Ewais, and
  De~Troyer}]{sajjadi_individualization_2022}
\bibinfo{author}{P.~Sajjadi}, \bibinfo{author}{A.~Ewais},
  \bibinfo{author}{O.~De~Troyer}, \bibinfo{title}{Individualization in serious
  games: {A} systematic review of the literature on the aspects of the players
  to adapt to}, \bibinfo{year}{2022}.
\bibitem[{van Dooren et~al.(2019)van Dooren, Siriaraya, Visch, Spijkerman, and
  Bijkerk}]{van_dooren_reflections_2019}
\bibinfo{author}{M.~M. van Dooren}, \bibinfo{author}{P.~Siriaraya},
  \bibinfo{author}{V.~Visch}, \bibinfo{author}{R.~Spijkerman},
  \bibinfo{author}{L.~Bijkerk},
\newblock \bibinfo{title}{Reflections on the design, implementation, and
  adoption of a gamified {eHealth} application in youth mental healthcare},
\newblock \bibinfo{journal}{Entertainment Computing} \bibinfo{volume}{31}
  (\bibinfo{year}{2019}) \bibinfo{pages}{100305}.
\bibitem[{Verschueren et~al.(2019)Verschueren, Buffel, and
  Stichele}]{verschueren_developing_2019}
\bibinfo{author}{S.~Verschueren}, \bibinfo{author}{C.~Buffel},
  \bibinfo{author}{G.~V. Stichele}, \bibinfo{title}{Developing theory-driven,
  evidence-based serious games for health: {Framework} based on research
  community insights}, \bibinfo{year}{2019}. \URLprefix
  \url{https://games.jmir.org/2019/2/e11565}.
\bibitem[{De~Troyer(2017)}]{de_troyer_towards_2017}
\bibinfo{author}{O.~De~Troyer},
\newblock \bibinfo{title}{Towards effective serious games},
\newblock in: \bibinfo{booktitle}{2017 9th {International} {Conference} on
  {Virtual} {Worlds} and {Games} for {Serious} {Applications}, {VS}-{Games}
  2017 - {Proceedings}}, \bibinfo{publisher}{Institute of Electrical and
  Electronics Engineers Inc.}, \bibinfo{year}{2017}, pp.
  \bibinfo{pages}{284--289}. \DOIprefix\doi{10.1109/VS-GAMES.2017.8056615}.
\bibitem[{Blatsios and Refanidis(2019)}]{blatsios_towards_2019}
\bibinfo{author}{S.~Blatsios}, \bibinfo{author}{I.~Refanidis},
\newblock \bibinfo{title}{Towards an {Adaption} and {Personalisation}
  {Solution} {Based} on {Multi} {Agent} {System} {Applied} on {Serious}
  {Games}},
\newblock \bibinfo{journal}{IFIP Advances in Information and Communication
  Technology} \bibinfo{volume}{559} (\bibinfo{year}{2019})
  \bibinfo{pages}{584--594}.
\bibitem[{Lazzaro(2004)}]{lazzaro_why_2004}
\bibinfo{author}{N.~Lazzaro},
\newblock \bibinfo{title}{Why {We} {Play} {Games}: {Four} {Keys} to {More}
  {Emotion} {Without} {Story}},
\newblock in: \bibinfo{booktitle}{Game {Developer} {Conference} ({GDC})},
  \bibinfo{year}{2004}, pp. \bibinfo{pages}{1--8}. \URLprefix
  \url{www.xeodesign.com
  http://www.citeulike.org/group/596/article/436449{\%}5Cnhttp://www.xeodesign.com/xeodesign{\_}whyweplaygames.pdf}.
  \DOIprefix\doi{10.1111/j.1464-410X.2004.04896.x}.
\bibitem[{Streicher and Smeddinck(2016)}]{streicher_personalized_2016}
\bibinfo{author}{A.~Streicher}, \bibinfo{author}{J.~D. Smeddinck},
\newblock \bibinfo{title}{Personalized and adaptive serious games},
\newblock \bibinfo{journal}{Lecture Notes in Computer Science (including
  subseries Lecture Notes in Artificial Intelligence and Lecture Notes in
  Bioinformatics)} \bibinfo{volume}{9970 LNCS} (\bibinfo{year}{2016})
  \bibinfo{pages}{332--377}.
\bibitem[{Sanchez et~al.(2019)Sanchez, van Oostendorp, Fijnheer, and
  Lavoué}]{sanchez_gamification_2019}
\bibinfo{author}{E.~Sanchez}, \bibinfo{author}{H.~van Oostendorp},
  \bibinfo{author}{J.~D. Fijnheer}, \bibinfo{author}{E.~Lavoué},
\newblock \bibinfo{title}{Gamification},
\newblock in: \bibinfo{editor}{A.~Tatnall} (Ed.),
  \bibinfo{booktitle}{Encyclopedia of {Education} and {Information}
  {Technologies}}, \bibinfo{publisher}{Springer International Publishing},
  \bibinfo{address}{Cham}, \bibinfo{year}{2019}, pp. \bibinfo{pages}{1--11}.
  \URLprefix \url{https://doi.org/10.1007/978-3-319-60013-0_38-1}.
\bibitem[{Korhonen et~al.(2017)Korhonen, Halonen, Ravelin, Kemppainen, and
  Koskela}]{korhonen_multidisciplinary_2017}
\bibinfo{author}{T.~Korhonen}, \bibinfo{author}{R.~Halonen},
  \bibinfo{author}{T.~Ravelin}, \bibinfo{author}{J.~Kemppainen},
  \bibinfo{author}{K.~Koskela},
\newblock \bibinfo{title}{A {Multidisciplinary} {Approach} {To} {Serious}
  {Game} {Development} in the {Health} {Sector}},
\newblock \bibinfo{journal}{The 11th Mediterranean Conference on Information
  Systems (MCIS), Genoa, Italy,}  (\bibinfo{year}{2017}) \bibinfo{pages}{15}.
\bibitem[{Chow et~al.(2020)Chow, Riantiningtyas, Kanstrup, Papavasileiou, Liem,
  and Olsen}]{chow_can_2020}
\bibinfo{author}{C.~Y. Chow}, \bibinfo{author}{R.~R. Riantiningtyas},
  \bibinfo{author}{M.~B. Kanstrup}, \bibinfo{author}{M.~Papavasileiou},
  \bibinfo{author}{G.~D. Liem}, \bibinfo{author}{A.~Olsen},
\newblock \bibinfo{title}{Can games change children's eating behaviour? {A}
  review of gamification and serious games},
\newblock \bibinfo{journal}{Food Quality and Preference} \bibinfo{volume}{80}
  (\bibinfo{year}{2020}) \bibinfo{pages}{103823}.
\bibitem[{Wouters et~al.(2013)Wouters, van Nimwegen, van Oostendorp, and van
  Der~Spek}]{wouters_meta-analysis_2013}
\bibinfo{author}{P.~Wouters}, \bibinfo{author}{C.~van Nimwegen},
  \bibinfo{author}{H.~van Oostendorp}, \bibinfo{author}{E.~D. van Der~Spek},
\newblock \bibinfo{title}{A meta-analysis of the cognitive and motivational
  effects of serious games},
\newblock \bibinfo{journal}{Journal of Educational Psychology}
  \bibinfo{volume}{105} (\bibinfo{year}{2013}) \bibinfo{pages}{249--265}.
\bibitem[{Gentry et~al.(2019)Gentry, Gauthier, Ehrstrom, Wortley, Lilienthal,
  Car, Dauwels-Okutsu, Nikolaou, Zary, Campbell, and Car}]{gentry_serious_2019}
\bibinfo{author}{S.~V. Gentry}, \bibinfo{author}{A.~Gauthier},
  \bibinfo{author}{B.~L. Ehrstrom}, \bibinfo{author}{D.~Wortley},
  \bibinfo{author}{A.~Lilienthal}, \bibinfo{author}{L.~T. Car},
  \bibinfo{author}{S.~Dauwels-Okutsu}, \bibinfo{author}{C.~K. Nikolaou},
  \bibinfo{author}{N.~Zary}, \bibinfo{author}{J.~Campbell},
  \bibinfo{author}{J.~Car},
\newblock \bibinfo{title}{Serious gaming and gamification education in health
  professions: systematic review},
\newblock \bibinfo{journal}{Journal of Medical Internet Research}
  \bibinfo{volume}{21} (\bibinfo{year}{2019}) \bibinfo{pages}{e12994}.
\bibitem[{Wiemeyer and Kliem(2012)}]{wiemeyer_serious_2012}
\bibinfo{author}{J.~Wiemeyer}, \bibinfo{author}{A.~Kliem},
\newblock \bibinfo{title}{Serious games in prevention and rehabilitation-a new
  panacea for elderly people?},
\newblock \bibinfo{journal}{European Review of Aging and Physical Activity}
  \bibinfo{volume}{9} (\bibinfo{year}{2012}) \bibinfo{pages}{41--50}.
\bibitem[{Mora et~al.(2019)Mora, Tondello, Calvet, González, Arnedo-Moreno,
  and Nacke}]{mora_quest_2019}
\bibinfo{author}{A.~Mora}, \bibinfo{author}{G.~F. Tondello},
  \bibinfo{author}{L.~Calvet}, \bibinfo{author}{C.~González},
  \bibinfo{author}{J.~Arnedo-Moreno}, \bibinfo{author}{L.~E. Nacke},
\newblock \bibinfo{title}{The quest for a better tailoring of gameful design:
  {An} analysis of player type preferences},
\newblock \bibinfo{journal}{ACM International Conference Proceeding Series}
  (\bibinfo{year}{2019}).
\bibitem[{Paraschos and Koulouriotis(2023)}]{paraschos_game_2023}
\bibinfo{author}{P.~D. Paraschos}, \bibinfo{author}{D.~E. Koulouriotis},
\newblock \bibinfo{title}{Game {Difficulty} {Adaptation} and {Experience}
  {Personalization}: {A} {Literature} {Review}},
\newblock \bibinfo{journal}{International Journal of Human–Computer
  Interaction} \bibinfo{volume}{39} (\bibinfo{year}{2023})
  \bibinfo{pages}{1--22}. \bibinfo{note}{\_eprint:
  https://doi.org/10.1080/10447318.2021.2020008}.
\bibitem[{Rodrigues et~al.(2021)Rodrigues, Toda, Palomino, Oliveira, and
  Isotani}]{rodrigues_personalized_2021}
\bibinfo{author}{L.~Rodrigues}, \bibinfo{author}{A.~M. Toda},
  \bibinfo{author}{P.~T. Palomino}, \bibinfo{author}{W.~Oliveira},
  \bibinfo{author}{S.~Isotani},
\newblock \bibinfo{title}{Personalized gamification: {A} literature review of
  outcomes, experiments, and approaches},
\newblock in: \bibinfo{booktitle}{Eighth {International} {Conference} on
  {Technological} {Ecosystems} for {Enhancing} {Multiculturality}}, {TEEM}'20,
  \bibinfo{publisher}{Association for Computing Machinery},
  \bibinfo{address}{New York, NY, USA}, \bibinfo{year}{2021}, pp.
  \bibinfo{pages}{699--706}. \URLprefix
  \url{https://dl.acm.org/doi/10.1145/3434780.3436665}.
  \DOIprefix\doi{10.1145/3434780.3436665}.
\bibitem[{Klock et~al.(2020)Klock, Gasparini, Pimenta, and
  Hamari}]{klock_tailored_2020}
\bibinfo{author}{A.~C.~T. Klock}, \bibinfo{author}{I.~Gasparini},
  \bibinfo{author}{M.~S. Pimenta}, \bibinfo{author}{J.~Hamari},
\newblock \bibinfo{title}{Tailored gamification: {A} review of literature},
\newblock \bibinfo{journal}{International Journal of Human-Computer Studies}
  \bibinfo{volume}{144} (\bibinfo{year}{2020}) \bibinfo{pages}{102495}.
\bibitem[{Khakpour and Colomo-Palacios(2021)}]{khakpour_convergence_2021}
\bibinfo{author}{A.~Khakpour}, \bibinfo{author}{R.~Colomo-Palacios},
\newblock \bibinfo{title}{Convergence of {Gamification} and {Machine}
  {Learning}: {A} {Systematic} {Literature} {Review}},
\newblock \bibinfo{journal}{Technology, Knowledge and Learning}
  \bibinfo{volume}{26} (\bibinfo{year}{2021}) \bibinfo{pages}{597--636}.
\bibitem[{Hare and Tang(2023)}]{hare_player_2023}
\bibinfo{author}{R.~Hare}, \bibinfo{author}{Y.~Tang},
\newblock \bibinfo{title}{Player {Modeling} and {Adaptation} {Methods} {Within}
  {Adaptive} {Serious} {Games}},
\newblock \bibinfo{journal}{IEEE Transactions on Computational Social Systems}
  \bibinfo{volume}{10} (\bibinfo{year}{2023}) \bibinfo{pages}{1939--1950}.
\bibitem[{Crookall(2010)}]{crookall_serious_2010}
\bibinfo{author}{D.~Crookall},
\newblock \bibinfo{title}{Serious {Games}, {Debriefing}, and
  {Simulation}/{Gaming} as a {Discipline}},
\newblock \bibinfo{journal}{Simulation \& Gaming} \bibinfo{volume}{41}
  (\bibinfo{year}{2010}) \bibinfo{pages}{898--920}.
\bibitem[{Haring et~al.(2011)Haring, Chakinska, and
  Ritterfeld}]{haring_understanding_2011}
\bibinfo{author}{P.~Haring}, \bibinfo{author}{D.~Chakinska},
  \bibinfo{author}{U.~Ritterfeld},
\newblock \bibinfo{title}{Understanding {Serious} {Gaming}: {A} {Psychological}
  {Perspective}},
\newblock \bibinfo{year}{2011}, pp. \bibinfo{pages}{413--43}.
\bibitem[{Blumberg et~al.(2013)Blumberg, Almonte, Anthony, and
  Hashimoto}]{blumberg_serious_2013}
\bibinfo{author}{F.~Blumberg}, \bibinfo{author}{D.~Almonte},
  \bibinfo{author}{J.~Anthony}, \bibinfo{author}{N.~Hashimoto},
\newblock \bibinfo{title}{Serious {Games}: {What} {Are} {They}? {What} {Do}
  {They} {Do}? {Why} {Should} {We} {Play} {Them}?},
\newblock \bibinfo{year}{2013}, pp. \bibinfo{pages}{334--351}.
\bibitem[{Deterding et~al.(2011)Deterding, Dixon, Khaled, and
  Nacke}]{deterding_game_2011}
\bibinfo{author}{S.~Deterding}, \bibinfo{author}{D.~Dixon},
  \bibinfo{author}{R.~Khaled}, \bibinfo{author}{L.~Nacke},
\newblock \bibinfo{title}{From {Game} {Design} {Elements} to {Gamefulness}:
  {Defining} {Gamification}},
\newblock volume~\bibinfo{volume}{11}, \bibinfo{year}{2011}, pp.
  \bibinfo{pages}{9--15}. \DOIprefix\doi{10.1145/2181037.2181040}.
\bibitem[{Loh et~al.(2015)Loh, Yanyan, and Ifenthaler}]{loh_serious_2015}
\bibinfo{author}{C.~Loh}, \bibinfo{author}{S.~Yanyan},
  \bibinfo{author}{D.~Ifenthaler},
\newblock \bibinfo{title}{Serious {Games} {Analytics}: {Theoretical}
  {Framework}},
\newblock \bibinfo{year}{2015}, pp. \bibinfo{pages}{3--30}.
\bibitem[{Landers(2014)}]{landers_developing_2014}
\bibinfo{author}{R.~N. Landers},
\newblock \bibinfo{title}{Developing a {Theory} of {Gamified} {Learning}:
  {Linking} {Serious} {Games} and {Gamification} of {Learning}},
\newblock \bibinfo{journal}{Simulation \& Gaming} \bibinfo{volume}{45}
  (\bibinfo{year}{2014}) \bibinfo{pages}{752--768}.
\bibitem[{Marczewski(2015)}]{marczewski_even_2015}
\bibinfo{author}{A.~Marczewski}, \bibinfo{title}{Even {Ninja} {Monkeys} {Like}
  to {Play}: {Gamification}, {Game} {Thinking} and {Motivational} {Design}},
  \bibinfo{publisher}{CreateSpace Independent Publishing Platform},
  \bibinfo{year}{2015}.
\bibitem[{Carlier et~al.(2023)Carlier, Naessens, Backere, and
  Turck}]{carlier_software_2023}
\bibinfo{author}{S.~Carlier}, \bibinfo{author}{V.~Naessens},
  \bibinfo{author}{F.~D. Backere}, \bibinfo{author}{F.~D. Turck},
\newblock \bibinfo{title}{A {Software} {Engineering} {Framework} for {Reusable}
  {Design} of {Personalized} {Serious} {Games} for {Health}: {Development}
  {Study}},
\newblock \bibinfo{journal}{JMIR Serious Games} \bibinfo{volume}{11}
  (\bibinfo{year}{2023}) \bibinfo{pages}{e40054}.
\bibitem[{Turkay and Kinzer(2014)}]{turkay_effects_2014}
\bibinfo{author}{S.~Turkay}, \bibinfo{author}{C.~K. Kinzer},
\newblock \bibinfo{title}{The effects of avatar: {Based} customization on
  player identification},
\newblock \bibinfo{journal}{International Journal of Gaming and
  Computer-Mediated Simulations} \bibinfo{volume}{6} (\bibinfo{year}{2014})
  \bibinfo{pages}{1--25}.
\bibitem[{de~Oliveira and
  de~Carvalho(2020)}]{de_oliveira_gamification-based_2020}
\bibinfo{author}{L.~W. de~Oliveira}, \bibinfo{author}{S.~T. de~Carvalho},
\newblock \bibinfo{title}{A gamification-based framework for {mHealth}
  developers in the context of self-care},
\newblock in: \bibinfo{editor}{A.~DeHerrera}, \bibinfo{editor}{A.~Gonzalez},
  \bibinfo{editor}{K.~Santosh}, \bibinfo{editor}{Z.~Temesgen},
  \bibinfo{editor}{B.~Kane}, \bibinfo{editor}{P.~Soda} (Eds.),
  \bibinfo{booktitle}{2020 {IEEE} {33RD} {INTERNATIONAL} {SYMPOSIUM} {ON}
  {COMPUTER}-{BASED} {MEDICAL} {SYSTEMS}({CBMS} 2020)}, {IEEE} {International}
  {Symposium} on {Computer}-{Based} {Medical} {Systems},
  \bibinfo{publisher}{IEEE}, \bibinfo{address}{345 E 47TH ST, NEW YORK, NY
  10017 USA}, \bibinfo{year}{2020}, pp. \bibinfo{pages}{138--141}.
  \DOIprefix\doi{10.1109/CBMS49503.2020.00033}.
\bibitem[{Martinho et~al.(2023)Martinho, Crista, Matsui, Marreiros, and
  Corchado}]{martinho_effects_2023}
\bibinfo{author}{D.~Martinho}, \bibinfo{author}{V.~Crista},
  \bibinfo{author}{K.~Matsui}, \bibinfo{author}{G.~Marreiros},
  \bibinfo{author}{J.~M. Corchado},
\newblock \bibinfo{title}{Effects of a gamified agent-based system for
  personalized elderly care: pilot usability study},
\newblock \bibinfo{journal}{JMIR SERIOUS GAMES} \bibinfo{volume}{11}
  (\bibinfo{year}{2023}).
\bibitem[{Nuijten et~al.(2022)Nuijten, Van~Gorp, Khanshan, Le~Blanc, van~den
  Berg, Kemperman, and Simons}]{nuijten_evaluating_2022}
\bibinfo{author}{R.~Nuijten}, \bibinfo{author}{P.~Van~Gorp},
  \bibinfo{author}{A.~Khanshan}, \bibinfo{author}{P.~Le~Blanc},
  \bibinfo{author}{P.~van~den Berg}, \bibinfo{author}{A.~Kemperman},
  \bibinfo{author}{M.~Simons},
\newblock \bibinfo{title}{Evaluating the impact of adaptive personalized goal
  setting on engagement levels of government staff with a gamified {mHealth}
  tool: results from a 2-month randomized controlled trial},
\newblock \bibinfo{journal}{JMIR MHEALTH AND UHEALTH} \bibinfo{volume}{10}
  (\bibinfo{year}{2022}).
\bibitem[{Silvia et~al.(2023)Silvia, Migliorelli, Sistach-Bosch,
  Gomez-Martinez, and Boque}]{silvia_tailored_2023}
\bibinfo{author}{O.~Silvia}, \bibinfo{author}{C.~Migliorelli},
  \bibinfo{author}{L.~Sistach-Bosch}, \bibinfo{author}{M.~Gomez-Martinez},
  \bibinfo{author}{N.~Boque},
\newblock \bibinfo{title}{A tailored and engaging mhealth gamified framework
  for nutritional behaviour change},
\newblock \bibinfo{journal}{NUTRIENTS} \bibinfo{volume}{15}
  (\bibinfo{year}{2023}).
\bibitem[{Schafer et~al.(2018)Schafer, Bachner, Pretscher, Groh, and
  Demetriou}]{schafer_study_2018}
\bibinfo{author}{H.~Schafer}, \bibinfo{author}{J.~Bachner},
  \bibinfo{author}{S.~Pretscher}, \bibinfo{author}{G.~Groh},
  \bibinfo{author}{Y.~Demetriou},
\newblock \bibinfo{title}{Study on motivating physical activity in children
  with personalized gamified feedback},
\newblock in: \bibinfo{booktitle}{{UMAP}'18: {ADJUNCT} {PUBLICATION} {OF} {THE}
  {26TH} {CONFERENCE} {ON} {USER} {MODELING}, {ADAPTATION} {AND}
  {PERSONALIZATION}}, \bibinfo{publisher}{ASSOC COMPUTING MACHINERY},
  \bibinfo{address}{1601 Broadway, 10th Floor, NEW YORK, NY, UNITED STATES},
  \bibinfo{year}{2018}, pp. \bibinfo{pages}{221--226}.
  \DOIprefix\doi{10.1145/3213586.3225227}.
\bibitem[{Zhao et~al.(2020)Zhao, Arya, Orji, and Chan}]{zhao_effects_2020}
\bibinfo{author}{Z.~Zhao}, \bibinfo{author}{A.~Arya},
  \bibinfo{author}{R.~Orji}, \bibinfo{author}{G.~Chan},
\newblock \bibinfo{title}{Effects of a personalized fitness recommender system
  using gamification and continuous player modeling: {System} design and
  long-term validation study},
\newblock \bibinfo{journal}{JMIR Serious Games} \bibinfo{volume}{8}
  (\bibinfo{year}{2020}).
\bibitem[{Fadhil and Villafiorita(2017)}]{fadhil_adaptive_2017}
\bibinfo{author}{A.~Fadhil}, \bibinfo{author}{A.~Villafiorita},
\newblock \bibinfo{title}{An adaptive learning with gamification \&
  conversational {UIs}: {The} rise of {CiboPoliBot}},
\newblock in: \bibinfo{booktitle}{{UMAP} 2017 - {Adjunct} {Publication} of the
  25th {Conference} on {User} {Modeling}, {Adaptation} and {Personalization}},
  \bibinfo{publisher}{ACM}, \bibinfo{address}{New York, NY, USA},
  \bibinfo{year}{2017}, pp. \bibinfo{pages}{408--412}. \URLprefix
  \url{https://doi.org/http://dx.doi.org/10.1145/3099023.3099112}.
  \DOIprefix\doi{10.1145/3099023.3099112}.
\bibitem[{Pardos et~al.(2023)Pardos, Gallos, Menychtas, Panagopoulos, and
  Maglogiannis}]{pardos_enriching_2023}
\bibinfo{author}{A.~Pardos}, \bibinfo{author}{P.~Gallos},
  \bibinfo{author}{A.~Menychtas}, \bibinfo{author}{C.~Panagopoulos},
  \bibinfo{author}{I.~Maglogiannis},
\newblock \bibinfo{title}{Enriching {Remote} {Monitoring} and {Care}
  {Platforms} with {Personalized} {Recommendations} to {Enhance} {Gamification}
  and {Coaching}},
\newblock in: \bibinfo{editor}{M.~Hagglund}, \bibinfo{editor}{S.~Pelayo},
  \bibinfo{editor}{A.~Moen}, \bibinfo{editor}{M.~Blusi},
  \bibinfo{editor}{S.~Bonacina}, \bibinfo{editor}{L.~Nilsson},
  \bibinfo{editor}{I.~Madsen}, \bibinfo{editor}{A.~Benis},
  \bibinfo{editor}{L.~Lindskold}, \bibinfo{editor}{P.~Gallos} (Eds.),
  \bibinfo{booktitle}{caring is sharing-exploiting the value in data for health
  and innovation-proceedings of mie 2023}, volume \bibinfo{volume}{302} of
  \textit{\bibinfo{series}{Studies in {Health} {Technology} and
  {Informatics}}}, \bibinfo{publisher}{IOS PRESS}, \bibinfo{address}{Amsterdam,
  The Netherlands}, \bibinfo{year}{2023}, pp. \bibinfo{pages}{332--336}.
  \DOIprefix\doi{10.3233/SHTI230129}.
\bibitem[{Carlier et~al.(2021)Carlier, Coppens, De~Backere, and
  De~Turck}]{carlier_investigating_2021}
\bibinfo{author}{S.~Carlier}, \bibinfo{author}{D.~Coppens},
  \bibinfo{author}{F.~De~Backere}, \bibinfo{author}{F.~De~Turck},
\newblock \bibinfo{title}{Investigating the influence of personalised
  gamification on mobile survey user experience},
\newblock \bibinfo{journal}{SUSTAINABILITY} \bibinfo{volume}{13}
  (\bibinfo{year}{2021}).
\bibitem[{Mocanu et~al.(2016)Mocanu, Marian, Rusu, and
  Arba}]{mocanu_kinect_2016}
\bibinfo{author}{I.~Mocanu}, \bibinfo{author}{C.~Marian},
  \bibinfo{author}{L.~Rusu}, \bibinfo{author}{R.~Arba},
\newblock \bibinfo{title}{A {Kinect} based adaptive exergame},
\newblock in: \bibinfo{booktitle}{2016 {IEEE} 12th {International} {Conference}
  on {Intelligent} {Computer} {Communication} and {Processing} ({ICCP})},
  \bibinfo{publisher}{IEEE}, \bibinfo{address}{Cluj-Napoca, Romania},
  \bibinfo{year}{2016}, pp. \bibinfo{pages}{117--124}. \URLprefix
  \url{http://ieeexplore.ieee.org/document/7737132/}.
  \DOIprefix\doi{10.1109/ICCP.2016.7737132}.
\bibitem[{Martin-Niedecken et~al.(2021)Martin-Niedecken, Schwarz, and
  Schättin}]{martin-niedecken_comparing_2021}
\bibinfo{author}{A.~L. Martin-Niedecken}, \bibinfo{author}{T.~Schwarz},
  \bibinfo{author}{A.~Schättin},
\newblock \bibinfo{title}{Comparing the {Impact} of {Heart} {Rate}-{Based}
  {In}-{Game} {Adaptations} in an {Exergame}-{Based} {Functional}
  {High}-{Intensity} {Interval} {Training} on {Training} {Intensity} and
  {Experience} in {Healthy} {Young} {Adults}},
\newblock \bibinfo{journal}{Frontiers in Psychology} \bibinfo{volume}{12}
  (\bibinfo{year}{2021}) \bibinfo{pages}{572877}.
\bibitem[{Shen et~al.(2025)Shen, Jiao, Zhang, Liu, Liu, Li, Zhang, Li, and
  Hao}]{shen_gamified_2025}
\bibinfo{author}{D.~Shen}, \bibinfo{author}{J.~Jiao},
  \bibinfo{author}{L.~Zhang}, \bibinfo{author}{Y.~Liu},
  \bibinfo{author}{X.~Liu}, \bibinfo{author}{Y.~Li},
  \bibinfo{author}{T.~Zhang}, \bibinfo{author}{D.~Li},
  \bibinfo{author}{W.~Hao},
\newblock \bibinfo{title}{Gamified {Adaptive} {Approach} {Bias} {Modification}
  in {Individuals} {With} {Methamphetamine} {Use} {History} {From}
  {Communities} in {Sichuan}: {Pilot} {Randomized} {Controlled} {Trial}},
\newblock \bibinfo{journal}{JMIR Serious Games} \bibinfo{volume}{13}
  (\bibinfo{year}{2025}) \bibinfo{pages}{e56978--e56978}.
\bibitem[{Zhao et~al.(2020)Zhao, Arya, Orji, and Chan}]{zhao_physical_2020}
\bibinfo{author}{Z.~Zhao}, \bibinfo{author}{A.~Arya},
  \bibinfo{author}{R.~Orji}, \bibinfo{author}{G.~Chan},
\newblock \bibinfo{title}{Physical {Activity} {Recommendation} for {Exergame}
  {Player} {Modeling} using {Machine} {Learning} {Approach}},
\newblock in: \bibinfo{booktitle}{2020 {IEEE} 8th {International} {Conference}
  on {Serious} {Games} and {Applications} for {Health} ({SeGAH})},
  \bibinfo{publisher}{IEEE}, \bibinfo{address}{Vancouver, BC, Canada},
  \bibinfo{year}{2020}, pp. \bibinfo{pages}{1--9}. \URLprefix
  \url{https://ieeexplore.ieee.org/document/9201820/}.
  \DOIprefix\doi{10.1109/SeGAH49190.2020.9201820}.
\bibitem[{Yao et~al.(2025)Yao, Song, Xiao, and
  Zhao}]{yao_smartwatch-based_2025}
\bibinfo{author}{J.~Yao}, \bibinfo{author}{D.~Song}, \bibinfo{author}{T.~Xiao},
  \bibinfo{author}{J.~Zhao},
\newblock \bibinfo{title}{Smartwatch-{Based} {Tailored} {Gamification} and
  {User} {Modeling} for {Motivating} {Physical} {Exercise}: {Experimental}
  {Study} {With} the {Maximum} {Difference} {Scaling} {Segmentation} {Method}},
\newblock \bibinfo{journal}{JMIR Serious Games} \bibinfo{volume}{13}
  (\bibinfo{year}{2025}) \bibinfo{pages}{e66793}.
\bibitem[{Amiri et~al.(2022)Amiri, Sekhavat, and Goljaryan}]{amiri_stepar_2022}
\bibinfo{author}{Z.~Amiri}, \bibinfo{author}{Y.~A. Sekhavat},
  \bibinfo{author}{S.~Goljaryan},
\newblock \bibinfo{title}{{StepAR}: {A} personalized exergame for people with
  multiple sclerosis based on video-mapping},
\newblock \bibinfo{journal}{Entertainment Computing} \bibinfo{volume}{42}
  (\bibinfo{year}{2022}) \bibinfo{pages}{100487}.
\bibitem[{Chan et~al.(2023)Chan, Alslaity, Reen, Anukem, and
  Orji}]{meschtscherjakov_gardenquest_2023}
\bibinfo{author}{G.~Chan}, \bibinfo{author}{A.~Alslaity},
  \bibinfo{author}{J.~K. Reen}, \bibinfo{author}{S.~Anukem},
  \bibinfo{author}{R.~Orji},
\newblock \bibinfo{title}{{GardenQuest}: {Using} {Hexad} {Player} {Types} to
  {Design} a {Step}-{Based} {Multiplayer} {Persuasive} {Game} for {Motivating}
  {Physical} {Activity}},
\newblock in: \bibinfo{editor}{A.~Meschtscherjakov},
  \bibinfo{editor}{C.~Midden}, \bibinfo{editor}{J.~Ham} (Eds.),
  \bibinfo{booktitle}{Persuasive {Technology}}, volume \bibinfo{volume}{13832},
  \bibinfo{publisher}{Springer Nature Switzerland}, \bibinfo{address}{Cham},
  \bibinfo{year}{2023}, pp. \bibinfo{pages}{337--356}. \URLprefix
  \url{https://link.springer.com/10.1007/978-3-031-30933-5_22}.
\bibitem[{Mitsis et~al.(2019{\natexlab{a}})Mitsis, Zarkogianni, Bountouni,
  Athanasiou, and Nikita}]{mitsis_ontology-based_2019}
\bibinfo{author}{K.~Mitsis}, \bibinfo{author}{K.~Zarkogianni},
  \bibinfo{author}{N.~Bountouni}, \bibinfo{author}{M.~Athanasiou},
  \bibinfo{author}{K.~S. Nikita},
\newblock \bibinfo{title}{An ontology-based serious game design for the
  development of nutrition and food literacy skills},
\newblock in: \bibinfo{booktitle}{41st annual international conference of the
  {IEEE} engineering in medicine and biology society}, {IEEE} {Engineering} in
  {Medicine} and {Biology} {Society} {Conference} {Proceedings},
  \bibinfo{publisher}{IEEE}, \bibinfo{address}{New York, NY, USA},
  \bibinfo{year}{2019}{\natexlab{a}}, pp. \bibinfo{pages}{1405--1408}.
  \DOIprefix\doi{10.1109/embc.2019.8856604}.
\bibitem[{Mitsis et~al.(2019{\natexlab{b}})Mitsis, Zarkogianni, Dalakleidi,
  Mourkousis, and Nikita}]{mitsis_evaluation_2019}
\bibinfo{author}{K.~Mitsis}, \bibinfo{author}{K.~Zarkogianni},
  \bibinfo{author}{K.~Dalakleidi}, \bibinfo{author}{G.~Mourkousis},
  \bibinfo{author}{K.~S. Nikita},
\newblock \bibinfo{title}{Evaluation of a {Serious} {Game} {Promoting}
  {Nutrition} and {Food} {Literacy}: {Experiment} {Design} and {Preliminary}
  {Results}},
\newblock in: \bibinfo{booktitle}{2019 {IEEE} 19th {International} {Conference}
  on {Bioinformatics} and {Bioengineering} ({BIBE})},
  \bibinfo{publisher}{IEEE}, \bibinfo{address}{Athens, Greece},
  \bibinfo{year}{2019}{\natexlab{b}}, pp. \bibinfo{pages}{497--502}. \URLprefix
  \url{https://ieeexplore.ieee.org/document/8941930/}.
  \DOIprefix\doi{10.1109/BIBE.2019.00096}.
\bibitem[{Semet et~al.(2019)Semet, Marcon, Demestichas, Koutsouris, and
  Ascolese}]{semet_artificial_2019}
\bibinfo{author}{Y.~Semet}, \bibinfo{author}{B.~Marcon},
  \bibinfo{author}{K.~Demestichas}, \bibinfo{author}{N.~Koutsouris},
  \bibinfo{author}{A.~Ascolese},
\newblock \bibinfo{title}{Artificial ant colonies for adaptive rewards in
  serious games},
\newblock in: \bibinfo{editor}{H.~Fellermann}, \bibinfo{editor}{J.~Bacardit},
  \bibinfo{editor}{A.~GoniMoreno}, \bibinfo{editor}{R.~Fuchslin} (Eds.),
  \bibinfo{booktitle}{{ALIFE} 2019: {THE} 2019 {CONFERENCE} {ON} {ARTIFICIAL}
  {LIFE}}, \bibinfo{publisher}{MIT PRESS}, \bibinfo{address}{ONE ROGERS ST,
  CAMBRIDGE, MA 02142 USA}, \bibinfo{year}{2019}, pp.
  \bibinfo{pages}{533--540}.
\bibitem[{Hocine(2019)}]{hocine_personalized_2019}
\bibinfo{author}{N.~Hocine},
\newblock \bibinfo{title}{Personalized {Serious} {Games} for {Self}-regulated
  {Attention} {Training}},
\newblock \bibinfo{journal}{ACM UMAP 2019 Adjunct - Adjunct Publication of the
  27th Conference on User Modeling, Adaptation and Personalization}
  (\bibinfo{year}{2019}) \bibinfo{pages}{251--255}.
\bibitem[{Hocine et~al.(????)Hocine, Ameur, and Ziani}]{hocine_keep_nodate}
\bibinfo{author}{N.~Hocine}, \bibinfo{author}{M.~Ameur},
  \bibinfo{author}{W.~Ziani},
\newblock \bibinfo{title}{Keep {Attention}: {A} {Personalized} {Serious} {Game}
  for {Attention} {Training}}  (????).
\bibitem[{Ahmad et~al.(2022)Ahmad, Mehmood, Khan, and
  Whangbo}]{ahmad_architecting_2022}
\bibinfo{author}{S.~Ahmad}, \bibinfo{author}{F.~Mehmood},
  \bibinfo{author}{F.~Khan}, \bibinfo{author}{T.~K. Whangbo},
\newblock \bibinfo{title}{Architecting intelligent smart serious games for
  healthcare applications: {A} technical perspective},
\newblock \bibinfo{journal}{SENSORS} \bibinfo{volume}{22}
  (\bibinfo{year}{2022}).
\bibitem[{Alves et~al.(2018)Alves, Gama, and Melo}]{alves_flow_2018}
\bibinfo{author}{T.~Alves}, \bibinfo{author}{S.~Gama}, \bibinfo{author}{F.~S.
  Melo},
\newblock \bibinfo{title}{Flow adaptation in serious games for health},
\newblock in: \bibinfo{editor}{J.~Vilaca}, \bibinfo{editor}{T.~Grechenig},
  \bibinfo{editor}{D.~Duque}, \bibinfo{editor}{N.~Rodrigues},
  \bibinfo{editor}{N.~Dias} (Eds.), \bibinfo{booktitle}{2018 {IEEE} 6th
  international conference on serious games and applications for health
  ({SEGAH} `18)}, {IEEE} {International} {Conference} on {Serious} {Games} and
  {Applications} for {Health}, \bibinfo{publisher}{IEEE}, \bibinfo{address}{New
  York, NY, USA}, \bibinfo{year}{2018}.
\bibitem[{Brown et~al.(2014)Brown, Cosma, Acampora, Seymour-Smith, and
  Close}]{brown_intelligent_2014}
\bibinfo{author}{D.~Brown}, \bibinfo{author}{G.~Cosma},
  \bibinfo{author}{G.~Acampora}, \bibinfo{author}{S.~Seymour-Smith},
  \bibinfo{author}{A.~Close},
\newblock \bibinfo{title}{An intelligent serious game for supporting african
  and african caribbean men during pre- and post-diagnosis of prostate cancer},
\newblock in: \bibinfo{booktitle}{2014 international conference on interactive
  technologies and games ({ITAG} 2014)}, \bibinfo{publisher}{IEEE},
  \bibinfo{address}{New York, NY, USA}, \bibinfo{year}{2014}, pp.
  \bibinfo{pages}{20--27}. \DOIprefix\doi{10.1109/iTAG.2014.9}.
\bibitem[{Delmas et~al.(2018)Delmas, Clement, Oudeyer, and
  Sauzéon}]{delmas_fostering_2018}
\bibinfo{author}{A.~Delmas}, \bibinfo{author}{B.~Clement},
  \bibinfo{author}{P.-Y. Oudeyer}, \bibinfo{author}{H.~Sauzéon},
\newblock \bibinfo{title}{Fostering {Health} {Education} {With} a {Serious}
  {Game} in {Children} {With} {Asthma}: {Pilot} {Studies} for {Assessing}
  {Learning} {Efficacy} and {Automatized} {Learning} {Personalization}},
\newblock \bibinfo{journal}{Frontiers in Education} \bibinfo{volume}{3}
  (\bibinfo{year}{2018}).
\bibitem[{Ghorbani et~al.(2022)Ghorbani, Taghavi, and
  Delrobaei}]{ghorbani_towards_2022}
\bibinfo{author}{F.~Ghorbani}, \bibinfo{author}{M.~F. Taghavi},
  \bibinfo{author}{M.~Delrobaei},
\newblock \bibinfo{title}{Towards an intelligent assistive system based on
  augmented reality and serious games},
\newblock \bibinfo{journal}{ENTERTAINMENT COMPUTING} \bibinfo{volume}{40}
  (\bibinfo{year}{2022}).
\bibitem[{Andrade et~al.(2014)Andrade, Fernandes, Caurin, Siqueira, Romero, and
  Pereira}]{andrade_dynamic_2014}
\bibinfo{author}{K.~d.~O. Andrade}, \bibinfo{author}{G.~Fernandes},
  \bibinfo{author}{G.~A.~P. Caurin}, \bibinfo{author}{A.~A.~G. Siqueira},
  \bibinfo{author}{R.~A.~F. Romero}, \bibinfo{author}{R.~d.~L. Pereira},
\newblock \bibinfo{title}{Dynamic player modelling in serious games applied to
  rehabilitation robotics},
\newblock in: \bibinfo{editor}{F.~Osorio}, \bibinfo{editor}{R.~Romero},
  \bibinfo{editor}{V.~Grassi}, \bibinfo{editor}{D.~Wolf},
  \bibinfo{editor}{K.~Branco}, \bibinfo{editor}{M.~Becker} (Eds.),
  \bibinfo{booktitle}{2014 2nd brazilian robotics symposium ({SBR}) / 11th
  latin american robotics symposium ({LARS}) / 6th robocontrol workshop on
  applied robotics and automation}, \bibinfo{publisher}{IEEE},
  \bibinfo{year}{2014}, pp. \bibinfo{pages}{211--216}.
  \DOIprefix\doi{10.1109/SBR.LARS.Robocontrol.2014.41}.
\bibitem[{Esfahlani et~al.(2017)Esfahlani, Cirstea, Sanaei, and
  Wilson}]{esfahlani_adaptive_2017}
\bibinfo{author}{S.~S. Esfahlani}, \bibinfo{author}{S.~Cirstea},
  \bibinfo{author}{A.~Sanaei}, \bibinfo{author}{G.~Wilson},
\newblock \bibinfo{title}{An adaptive self-organizing fuzzy logic controller in
  a serious game for motor impairment rehabilitation},
\newblock in: \bibinfo{booktitle}{2017 {IEEE} 26th international symposium on
  industrial electronics ({ISIE})}, Proceedings of the {IEEE} {International}
  {Symposium} on {Industrial} {Electronics}, \bibinfo{publisher}{IEEE},
  \bibinfo{address}{New York, NY, USA}, \bibinfo{year}{2017}, pp.
  \bibinfo{pages}{1311--1318}.
\bibitem[{Esfahlani et~al.(2018)Esfahlani, Thompson, Parsa, Brown, and
  Cirstea}]{esfahlani_rehabgame_2018}
\bibinfo{author}{S.~S. Esfahlani}, \bibinfo{author}{T.~Thompson},
  \bibinfo{author}{A.~D. Parsa}, \bibinfo{author}{I.~Brown},
  \bibinfo{author}{S.~Cirstea},
\newblock \bibinfo{title}{{ReHabgame}: {A} non-immersive virtual reality
  rehabilitation system with applications in neuroscience},
\newblock \bibinfo{journal}{Heliyon} \bibinfo{volume}{4} (\bibinfo{year}{2018})
  \bibinfo{pages}{e00526}.
\bibitem[{Hocine et~al.(2015)Hocine, Gouaich, A.~Cerri, Mottet, Froger, and
  Laffont}]{hocine_adaptation_2015}
\bibinfo{author}{N.~Hocine}, \bibinfo{author}{A.~Gouaich},
  \bibinfo{author}{S.~A.~Cerri}, \bibinfo{author}{D.~Mottet},
  \bibinfo{author}{J.~Froger}, \bibinfo{author}{I.~Laffont},
\newblock \bibinfo{title}{Adaptation in serious games for upper-limb
  rehabilitation: an approach to improve training outcomes},
\newblock \bibinfo{journal}{USER MODELING AND USER-ADAPTED INTERACTION}
  \bibinfo{volume}{25} (\bibinfo{year}{2015}) \bibinfo{pages}{65--98}.
\bibitem[{Pinto et~al.(2018)Pinto, Carvalho, Chambel, Ramiro, and
  Goncalves}]{pinto_adaptive_2018}
\bibinfo{author}{J.~F. Pinto}, \bibinfo{author}{H.~R. Carvalho},
  \bibinfo{author}{G.~R.~R. Chambel}, \bibinfo{author}{J.~Ramiro},
  \bibinfo{author}{A.~Goncalves},
\newblock \bibinfo{title}{Adaptive gameplay and difficulty adjustment in a
  gamified upper-limb rehabilitation},
\newblock in: \bibinfo{editor}{J.~Vilaca}, \bibinfo{editor}{T.~Grechenig},
  \bibinfo{editor}{D.~Duque}, \bibinfo{editor}{N.~Rodrigues},
  \bibinfo{editor}{N.~Dias} (Eds.), \bibinfo{booktitle}{2018 {IEEE} 6th
  international conference on serious games and applications for health
  ({SEGAH}`18)}, {IEEE} {International} {Conference} on {Serious} {Games} and
  {Applications} for {Health}, \bibinfo{publisher}{IEEE}, \bibinfo{address}{New
  York, NY, USA}, \bibinfo{year}{2018}.
\bibitem[{Alves et~al.(2019)Alves, Martinho, and Prada}]{alves_towards_2019}
\bibinfo{author}{T.~Alves}, \bibinfo{author}{C.~Martinho},
  \bibinfo{author}{R.~Prada},
\newblock \bibinfo{title}{Towards incorporating personality in serious games
  for health},
\newblock in: \bibinfo{booktitle}{2019 11th international conference on virtual
  worlds and games for serious applications (vs-games)}, International
  {Conference} on {Games} and {Virtual} {Worlds} for {Serious} {Applications},
  \bibinfo{publisher}{IEEE}, \bibinfo{address}{New York, NY, USA},
  \bibinfo{year}{2019}, pp. \bibinfo{pages}{230--233}.
  \DOIprefix\doi{10.1109/vs-games.2019.8864521}.
\bibitem[{Caggianese et~al.(2019)Caggianese, Cuomo, Esposito, Franceschini,
  Gallo, Infarinato, Minutolo, Piccialli, and Romano}]{caggianese_serious_2019}
\bibinfo{author}{G.~Caggianese}, \bibinfo{author}{S.~Cuomo},
  \bibinfo{author}{M.~Esposito}, \bibinfo{author}{M.~Franceschini},
  \bibinfo{author}{L.~Gallo}, \bibinfo{author}{F.~Infarinato},
  \bibinfo{author}{A.~Minutolo}, \bibinfo{author}{F.~Piccialli},
  \bibinfo{author}{P.~Romano},
\newblock \bibinfo{title}{Serious games and in-cloud data analytics for the
  virtualization and personalization of rehabilitation treatments},
\newblock \bibinfo{journal}{IEEE transactions on industrial informatics}
  \bibinfo{volume}{15} (\bibinfo{year}{2019}) \bibinfo{pages}{517--526}.
\bibitem[{Eun et~al.(2023)Eun, Kim, and Kim}]{eun_artificial_2023}
\bibinfo{author}{S.-J. Eun}, \bibinfo{author}{E.~J. Kim},
  \bibinfo{author}{J.~Kim},
\newblock \bibinfo{title}{Artificial intelligence-based personalized serious
  game for enhancing the physical and cognitive abilities of the elderly},
\newblock \bibinfo{journal}{FUTURE GENERATION COMPUTER SYSTEMS-THE
  INTERNATIONAL JOURNAL OF ESCIENCE} \bibinfo{volume}{141}
  (\bibinfo{year}{2023}) \bibinfo{pages}{713--722}.
\bibitem[{Mitsis et~al.(2020)Mitsis, Kalafatis, Zarkogianni, Mourkousis, and
  Nikita}]{mitsis_procedural_2020}
\bibinfo{author}{K.~Mitsis}, \bibinfo{author}{E.~Kalafatis},
  \bibinfo{author}{K.~Zarkogianni}, \bibinfo{author}{G.~Mourkousis},
  \bibinfo{author}{K.~S. Nikita},
\newblock \bibinfo{title}{Procedural content generation based on a genetic
  algorithm in a serious game for obstructive sleep apnea},
\newblock in: \bibinfo{booktitle}{2020 {IEEE} {Conference} on {Games} ({CoG})},
  \bibinfo{publisher}{IEEE}, \bibinfo{address}{Osaka, Japan},
  \bibinfo{year}{2020}, pp. \bibinfo{pages}{694--697}. \URLprefix
  \url{https://ieeexplore.ieee.org/document/9231785/}.
  \DOIprefix\doi{10.1109/CoG47356.2020.9231785}.
\bibitem[{Sadeghi~Esfahlani et~al.(2019)Sadeghi~Esfahlani, Butt, and
  Shirvani}]{sadeghi_esfahlani_fusion_2019}
\bibinfo{author}{S.~Sadeghi~Esfahlani}, \bibinfo{author}{J.~Butt},
  \bibinfo{author}{H.~Shirvani},
\newblock \bibinfo{title}{Fusion of {Artificial} {Intelligence} in
  {Neuro}-{Rehabilitation} {Video} {Games}},
\newblock \bibinfo{journal}{IEEE Access} \bibinfo{volume}{7}
  (\bibinfo{year}{2019}) \bibinfo{pages}{102617--102627}.
\bibitem[{Lin et~al.(2023)Lin, Lin, Yeh, Huang, Kuo, and
  Su}]{lin_exergame-integrated_2023}
\bibinfo{author}{C.-C. Lin}, \bibinfo{author}{Y.-S. Lin},
  \bibinfo{author}{C.-H. Yeh}, \bibinfo{author}{C.-C. Huang},
  \bibinfo{author}{L.-C. Kuo}, \bibinfo{author}{F.-C. Su},
\newblock \bibinfo{title}{An {Exergame}-{Integrated} {IoT}-{Based} {Ergometer}
  {System} {Delivers} {Personalized} {Training} {Programs} for {Older} {Adults}
  and {Enhances} {Physical} {Fitness}: {A} {Pilot} {Randomized} {Controlled}
  {Trial}},
\newblock \bibinfo{journal}{Gerontology} \bibinfo{volume}{69}
  (\bibinfo{year}{2023}) \bibinfo{pages}{768--782}.
\bibitem[{Nathella et~al.(2025)Nathella, Ghonasgi, Harvey, Ting, Herrin, and
  Young}]{nathella_challenge-based_2025}
\bibinfo{author}{S.~R. Nathella}, \bibinfo{author}{K.~Ghonasgi},
  \bibinfo{author}{T.~A. Harvey}, \bibinfo{author}{L.~H. Ting},
  \bibinfo{author}{K.~R. Herrin}, \bibinfo{author}{A.~J. Young},
\newblock \bibinfo{title}{Challenge-{Based} {Adaptation} of {Exoskeleton}
  {Assistance} and {Gamified} {Biofeedback} {Enables} {Automated} {Gait}
  {Rehabilitation}},
\newblock in: \bibinfo{booktitle}{2025 {International} {Conference} {On}
  {Rehabilitation} {Robotics} ({ICORR})}, \bibinfo{publisher}{IEEE},
  \bibinfo{address}{Chicago, IL, USA}, \bibinfo{year}{2025}, pp.
  \bibinfo{pages}{567--572}. \URLprefix
  \url{https://ieeexplore.ieee.org/document/11062942/}.
  \DOIprefix\doi{10.1109/ICORR66766.2025.11062942}.
\bibitem[{Bouatrous et~al.(2023)Bouatrous, Meziane, Zenati, and
  Hamitouche}]{bouatrous_new_2023}
\bibinfo{author}{A.~Bouatrous}, \bibinfo{author}{A.~Meziane},
  \bibinfo{author}{N.~Zenati}, \bibinfo{author}{C.~Hamitouche},
\newblock \bibinfo{title}{A new adaptive {VR}-based exergame for hand
  rehabilitation after stroke},
\newblock \bibinfo{journal}{Multimedia Systems} \bibinfo{volume}{29}
  (\bibinfo{year}{2023}) \bibinfo{pages}{3385--3402}.
\bibitem[{Faria and Da~Silva~Ayrosa(2025)}]{faria_adaptive_2025}
\bibinfo{author}{D.~R. Faria}, \bibinfo{author}{P.~P. Da~Silva~Ayrosa},
\newblock \bibinfo{title}{Adaptive {Neuro}-{Affective} {Engagement} via
  {Bayesian} {Feedback} {Learning} in {Serious} {Games} for {Neurodivergent}
  {Children}},
\newblock \bibinfo{journal}{Applied Sciences} \bibinfo{volume}{15}
  (\bibinfo{year}{2025}) \bibinfo{pages}{7532}.
\bibitem[{Stranick and Lopez(2022)}]{stranick_adaptive_2022}
\bibinfo{author}{T.~Stranick}, \bibinfo{author}{C.~Lopez},
\newblock \bibinfo{title}{Adaptive {Virtual} {Reality} {Exergame}: {Promoting}
  {Physical} {Activity} {Among} {Workers}},
\newblock \bibinfo{journal}{Journal of Computing and Information Science in
  Engineering} \bibinfo{volume}{22} (\bibinfo{year}{2022})
  \bibinfo{pages}{031002}.
\bibitem[{Doumas et~al.(2025)Doumas, Lejeune, Edwards, Stoquart, Vandermeeren,
  Dehez, and Dehem}]{doumas_clinical_2025}
\bibinfo{author}{I.~Doumas}, \bibinfo{author}{T.~Lejeune},
  \bibinfo{author}{M.~Edwards}, \bibinfo{author}{G.~Stoquart},
  \bibinfo{author}{Y.~Vandermeeren}, \bibinfo{author}{B.~Dehez},
  \bibinfo{author}{S.~Dehem},
\newblock \bibinfo{title}{Clinical validation of an individualized
  auto-adaptative serious game for combined cognitive and upper limb motor
  robotic rehabilitation after stroke},
\newblock \bibinfo{journal}{Journal of NeuroEngineering and Rehabilitation}
  \bibinfo{volume}{22} (\bibinfo{year}{2025}) \bibinfo{pages}{10}.
\bibitem[{Yoo et~al.(2017)Yoo, Parker, and Kay}]{yoo_designing_2017}
\bibinfo{author}{S.~Yoo}, \bibinfo{author}{C.~Parker},
  \bibinfo{author}{J.~Kay},
\newblock \bibinfo{title}{Designing a {Personalized} {VR} {Exergame}},
\newblock in: \bibinfo{booktitle}{Adjunct {Publication} of the 25th
  {Conference} on {User} {Modeling}, {Adaptation} and {Personalization}},
  \bibinfo{publisher}{ACM}, \bibinfo{address}{Bratislava Slovakia},
  \bibinfo{year}{2017}, pp. \bibinfo{pages}{431--435}. \URLprefix
  \url{https://dl.acm.org/doi/10.1145/3099023.3099115}.
  \DOIprefix\doi{10.1145/3099023.3099115}.
\bibitem[{Gray et~al.(2023)Gray, Villareale, Fox, Dallal, Ontanon, Arigo,
  Jabbari, and Zhu}]{gray_improving_2023}
\bibinfo{author}{R.~C. Gray}, \bibinfo{author}{J.~Villareale},
  \bibinfo{author}{T.~B. Fox}, \bibinfo{author}{D.~H. Dallal},
  \bibinfo{author}{S.~Ontanon}, \bibinfo{author}{D.~Arigo},
  \bibinfo{author}{S.~Jabbari}, \bibinfo{author}{J.~Zhu},
\newblock \bibinfo{title}{Improving {Fairness} in {Adaptive} {Social}
  {Exergames} via {Shapley} {Bandits}},
\newblock in: \bibinfo{booktitle}{Proceedings of the 28th {International}
  {Conference} on {Intelligent} {User} {Interfaces}}, \bibinfo{publisher}{ACM},
  \bibinfo{address}{Sydney NSW Australia}, \bibinfo{year}{2023}, pp.
  \bibinfo{pages}{322--336}. \URLprefix
  \url{https://dl.acm.org/doi/10.1145/3581641.3584050}.
  \DOIprefix\doi{10.1145/3581641.3584050}.
\bibitem[{Hoffmann et~al.(2015)Hoffmann, Sportwiss, Hardy, Wiemeyer, and
  Göbel}]{hoffmann_personalized_2015}
\bibinfo{author}{K.~Hoffmann}, \bibinfo{author}{D.~Sportwiss},
  \bibinfo{author}{S.~Hardy}, \bibinfo{author}{J.~Wiemeyer},
  \bibinfo{author}{S.~Göbel},
\newblock \bibinfo{title}{Personalized {Adaptive} {Control} of {Training}
  {Load} in {Cardio}-{Exergames}—{A} {Feasibility} {Study}},
\newblock \bibinfo{journal}{Games for Health Journal} \bibinfo{volume}{4}
  (\bibinfo{year}{2015}) \bibinfo{pages}{470--479}.
\bibitem[{Forgiarini et~al.(????)Forgiarini, De, and
  Buttussi}]{forgiarini_probabilistic_nodate}
\bibinfo{author}{A.~Forgiarini}, \bibinfo{author}{E.~De},
  \bibinfo{author}{F.~Buttussi},
\newblock \bibinfo{title}{Probabilistic {Modeling} and {Verification} of an
  {Adaptive} {VR} {Serious} {Game} for {Patients} with {Cognitive}
  {Impairment}}  (????).
\bibitem[{Aguilar et~al.(2022)Aguilar, Guzman, Rengifo, Chalapud, and
  Guzman}]{aguilar_proposal_2022}
\bibinfo{author}{J.~D. Aguilar}, \bibinfo{author}{D.~E. Guzman},
  \bibinfo{author}{C.~F. Rengifo}, \bibinfo{author}{L.~M. Chalapud},
  \bibinfo{author}{J.~D. Guzman},
\newblock \bibinfo{title}{Proposal of a {Game} with {Dynamic} {Difficulty}
  {Adjustment} from {Physiological} {Signals} in the {Context} of an
  {Exergame}},
\newblock in: \bibinfo{booktitle}{2022 {IEEE} 40th {Central} {America} and
  {Panama} {Convention} ({CONCAPAN})}, \bibinfo{publisher}{IEEE},
  \bibinfo{address}{Panama, Panama}, \bibinfo{year}{2022}, pp.
  \bibinfo{pages}{1--6}. \URLprefix
  \url{https://ieeexplore.ieee.org/document/9997775/}.
  \DOIprefix\doi{10.1109/CONCAPAN48024.2022.9997775}.
\bibitem[{Kira et~al.(2024)Kira, Pontes, Araújo, Monteiro, Uribe-Quevedo, and
  Nunes}]{kira_approach_2024}
\bibinfo{author}{A.~Kira}, \bibinfo{author}{R.~G. Pontes},
  \bibinfo{author}{L.~V. Araújo}, \bibinfo{author}{C.~B.~M. Monteiro},
  \bibinfo{author}{A.~Uribe-Quevedo}, \bibinfo{author}{F.~L.~S. Nunes},
\newblock \bibinfo{title}{An {Approach} for {Automatic} {Adaptation} of
  {Serious} {Games} {Applied} to {Virtual} {Motor} {Rehabilitation}},
\newblock in: \bibinfo{booktitle}{2024 {IEEE} 12th {International} {Conference}
  on {Serious} {Games} and {Applications} for {Health} ({SeGAH})},
  \bibinfo{publisher}{IEEE}, \bibinfo{address}{Funchal, Portugal},
  \bibinfo{year}{2024}, pp. \bibinfo{pages}{1--8}. \URLprefix
  \url{https://ieeexplore.ieee.org/document/10639554/}.
  \DOIprefix\doi{10.1109/SeGAH61285.2024.10639554}.
\bibitem[{Everard et~al.(2025)Everard, Declerck, Lejeune, Edwards, Bogacki,
  Reiprich, Delvigne, Legrain, and Batcho}]{everard_self-adaptive_2025}
\bibinfo{author}{G.~Everard}, \bibinfo{author}{L.~Declerck},
  \bibinfo{author}{T.~Lejeune}, \bibinfo{author}{M.~G. Edwards},
  \bibinfo{author}{J.~Bogacki}, \bibinfo{author}{C.~Reiprich},
  \bibinfo{author}{K.~Delvigne}, \bibinfo{author}{N.~Legrain},
  \bibinfo{author}{C.~S. Batcho},
\newblock \bibinfo{title}{A {Self}-{Adaptive} {Serious} {Game} to {Improve}
  {Motor} {Learning} {Among} {Older} {Adults} in {Immersive} {Virtual}
  {Reality}: {Short}-{Term} {Longitudinal} {Pre}-{Post} {Study} on {Retention}
  and {Transfer}},
\newblock \bibinfo{journal}{JMIR Aging} \bibinfo{volume}{8}
  (\bibinfo{year}{2025}) \bibinfo{pages}{e64004}.
\bibitem[{Tondello et~al.(2016)Tondello, Wehbe, Diamond, Busch, Marczewski, and
  Nacke}]{tondello_gamification_2016}
\bibinfo{author}{G.~F. Tondello}, \bibinfo{author}{R.~R. Wehbe},
  \bibinfo{author}{L.~Diamond}, \bibinfo{author}{M.~Busch},
  \bibinfo{author}{A.~Marczewski}, \bibinfo{author}{L.~E. Nacke},
\newblock \bibinfo{title}{The {Gamification} {User} {Types} {Hexad} {Scale}},
\newblock in: \bibinfo{booktitle}{Proceedings of the 2016 {Annual} {Symposium}
  on {Computer}-{Human} {Interaction} in {Play}}, {CHI} {PLAY} '16,
  \bibinfo{publisher}{Association for Computing Machinery},
  \bibinfo{address}{New York, NY, USA}, \bibinfo{year}{2016}, pp.
  \bibinfo{pages}{229--243}. \URLprefix
  \url{https://dl.acm.org/doi/10.1145/2967934.2968082}.
  \DOIprefix\doi{10.1145/2967934.2968082}.
\bibitem[{Ashburner et~al.(2000)Ashburner, Ball, Blake, Botstein, Butler,
  Cherry, Davis, Dolinski, Dwight, Eppig, Harris, Hill, Issel-Tarver,
  Kasarskis, Lewis, Matese, Richardson, Ringwald, Rubin, and
  Sherlock}]{ashburner_gene_2000}
\bibinfo{author}{M.~Ashburner}, \bibinfo{author}{C.~A. Ball},
  \bibinfo{author}{J.~A. Blake}, \bibinfo{author}{D.~Botstein},
  \bibinfo{author}{H.~Butler}, \bibinfo{author}{J.~M. Cherry},
  \bibinfo{author}{A.~P. Davis}, \bibinfo{author}{K.~Dolinski},
  \bibinfo{author}{S.~S. Dwight}, \bibinfo{author}{J.~T. Eppig},
  \bibinfo{author}{M.~A. Harris}, \bibinfo{author}{D.~P. Hill},
  \bibinfo{author}{L.~Issel-Tarver}, \bibinfo{author}{A.~Kasarskis},
  \bibinfo{author}{S.~Lewis}, \bibinfo{author}{J.~C. Matese},
  \bibinfo{author}{J.~E. Richardson}, \bibinfo{author}{M.~Ringwald},
  \bibinfo{author}{G.~M. Rubin}, \bibinfo{author}{G.~Sherlock},
\newblock \bibinfo{title}{Gene {Ontology}: tool for the unification of
  biology},
\newblock \bibinfo{journal}{Nature Genetics} \bibinfo{volume}{25}
  (\bibinfo{year}{2000}) \bibinfo{pages}{25--29}.
\bibitem[{Dessimoz and Škunca(2017)}]{dessimoz_gene_2017}
\bibinfo{editor}{C.~Dessimoz}, \bibinfo{editor}{N.~Škunca} (Eds.),
  \bibinfo{title}{The {Gene} {Ontology} {Handbook}}, volume
  \bibinfo{volume}{1446} of \textit{\bibinfo{series}{Methods in {Molecular}
  {Biology}}}, \bibinfo{publisher}{Springer}, \bibinfo{address}{New York, NY},
  \bibinfo{year}{2017}. \URLprefix
  \url{http://link.springer.com/10.1007/978-1-4939-3743-1}.
\bibitem[{Wang(1997)}]{wang_course_1997}
\bibinfo{author}{L.-X. Wang}, \bibinfo{title}{A {Course} in {Fuzzy} {Systems}
  and {Control}}, \bibinfo{publisher}{Prentice Hall PTR}, \bibinfo{year}{1997}.
\bibitem[{Zadeh(1994)}]{zadeh_role_1994}
\bibinfo{author}{L.~Zadeh},
\newblock \bibinfo{title}{The role of fuzzy logic in modeling, identification
  and control},
\newblock \bibinfo{journal}{Modeling, Identification and Control}
  \bibinfo{volume}{15} (\bibinfo{year}{1994}).
\bibitem[{Borbély and Szolgay(2017)}]{borbely_real-time_2017}
\bibinfo{author}{B.~J. Borbély}, \bibinfo{author}{P.~Szolgay},
\newblock \bibinfo{title}{Real-time inverse kinematics for the upper limb: a
  model-based algorithm using segment orientations},
\newblock \bibinfo{journal}{Biomedical Engineering Online} \bibinfo{volume}{16}
  (\bibinfo{year}{2017}) \bibinfo{pages}{21}.
\bibitem[{Lura(2012)}]{lura_creation_2012}
\bibinfo{author}{D.~Lura},
\newblock \bibinfo{title}{The {Creation} of a {Robotics} {Based} {Human}
  {Upper} {Body} {Model} for {Predictive} {Simulation} of {Prostheses}
  {Performance}},
\newblock \bibinfo{journal}{USF Tampa Graduate Theses and Dissertations}
  (\bibinfo{year}{2012}).
\bibitem[{Papaleo et~al.(2012)Papaleo, Zollo, Sterzi, and
  Guglielmelli}]{papaleo_inverse_2012}
\bibinfo{author}{E.~Papaleo}, \bibinfo{author}{L.~Zollo},
  \bibinfo{author}{S.~Sterzi}, \bibinfo{author}{E.~Guglielmelli},
\newblock \bibinfo{title}{An inverse kinematics algorithm for upper-limb joint
  reconstruction during robot-aided motor therapy},
\newblock in: \bibinfo{booktitle}{2012 4th {IEEE} {RAS} \& {EMBS}
  {International} {Conference} on {Biomedical} {Robotics} and {Biomechatronics}
  ({BioRob})}, \bibinfo{year}{2012}, pp. \bibinfo{pages}{1983--1988}.
  \URLprefix \url{https://ieeexplore.ieee.org/document/6290861}.
  \DOIprefix\doi{10.1109/BioRob.2012.6290861}.
\bibitem[{Kruse et~al.(2022)Kruse, Mostaghim, Borgelt, Braune, and
  Steinbrecher}]{kruse_multi-layer_2022}
\bibinfo{author}{R.~Kruse}, \bibinfo{author}{S.~Mostaghim},
  \bibinfo{author}{C.~Borgelt}, \bibinfo{author}{C.~Braune},
  \bibinfo{author}{M.~Steinbrecher},
\newblock \bibinfo{title}{Multi-layer {Perceptrons}},
\newblock in: \bibinfo{editor}{R.~Kruse}, \bibinfo{editor}{S.~Mostaghim},
  \bibinfo{editor}{C.~Borgelt}, \bibinfo{editor}{C.~Braune},
  \bibinfo{editor}{M.~Steinbrecher} (Eds.), \bibinfo{booktitle}{Computational
  {Intelligence}: {A} {Methodological} {Introduction}},
  \bibinfo{publisher}{Springer International Publishing},
  \bibinfo{address}{Cham}, \bibinfo{year}{2022}, pp. \bibinfo{pages}{53--124}.
  \URLprefix \url{https://doi.org/10.1007/978-3-030-42227-1_5}.
\bibitem[{Liu et~al.(2016)Liu, Gegov, and Cocea}]{liu_rule-based_2016}
\bibinfo{author}{H.~Liu}, \bibinfo{author}{A.~Gegov},
  \bibinfo{author}{M.~Cocea},
\newblock \bibinfo{title}{Rule-based systems: a granular computing
  perspective},
\newblock \bibinfo{journal}{Granular Computing} \bibinfo{volume}{1}
  (\bibinfo{year}{2016}) \bibinfo{pages}{259--274}.
\bibitem[{Grosan and Abraham(2011)}]{grosan_rule-based_2011}
\bibinfo{author}{C.~Grosan}, \bibinfo{author}{A.~Abraham},
\newblock \bibinfo{title}{Rule-{Based} {Expert} {Systems}},
\newblock in: \bibinfo{editor}{C.~Grosan}, \bibinfo{editor}{A.~Abraham} (Eds.),
  \bibinfo{booktitle}{Intelligent {Systems}: {A} {Modern} {Approach}},
  \bibinfo{publisher}{Springer}, \bibinfo{address}{Berlin, Heidelberg},
  \bibinfo{year}{2011}, pp. \bibinfo{pages}{149--185}. \URLprefix
  \url{https://doi.org/10.1007/978-3-642-21004-4_7}.
  \DOIprefix\doi{10.1007/978-3-642-21004-4_7}.
\bibitem[{Iovino et~al.(2022)Iovino, Förster, Falco, Chung, Siegwart, and
  Smith}]{iovino_programming_2022}
\bibinfo{author}{M.~Iovino}, \bibinfo{author}{J.~Förster},
  \bibinfo{author}{P.~Falco}, \bibinfo{author}{J.~J. Chung},
  \bibinfo{author}{R.~Siegwart}, \bibinfo{author}{C.~Smith}, \bibinfo{title}{On
  the programming effort required to generate {Behavior} {Trees} and {Finite}
  {State} {Machines} for robotic applications}, \bibinfo{year}{2022}.
  \URLprefix \url{http://arxiv.org/abs/2209.07392}.
  \DOIprefix\doi{10.48550/arXiv.2209.07392}.
\bibitem[{Oliver(1993)}]{oliver_decision_1993}
\bibinfo{author}{J.~J. Oliver},
\newblock \bibinfo{title}{Decision {Graphs} - {An} {Extension} of {Decision}
  {Trees}},
\newblock \bibinfo{year}{1993}. \URLprefix
  \url{https://www.semanticscholar.org/paper/Decision-Graphs-An-Extension-of-Decision-Trees-Oliver/73f1d17df0e1232da9e2331878a802a941f351c6}.
\bibitem[{LeCun et~al.(2015)LeCun, Bengio, and Hinton}]{lecun_deep_2015}
\bibinfo{author}{Y.~LeCun}, \bibinfo{author}{Y.~Bengio},
  \bibinfo{author}{G.~Hinton},
\newblock \bibinfo{title}{Deep learning},
\newblock \bibinfo{journal}{Nature} \bibinfo{volume}{521}
  (\bibinfo{year}{2015}) \bibinfo{pages}{436--444}.
\bibitem[{Li(2018)}]{li_deep_2018}
\bibinfo{author}{Y.~Li}, \bibinfo{title}{Deep {Reinforcement} {Learning}},
  \bibinfo{year}{2018}. \URLprefix \url{http://arxiv.org/abs/1810.06339}.
  \DOIprefix\doi{10.48550/arXiv.1810.06339}.
\bibitem[{Reeves(2018)}]{reeves_genetic_2018}
\bibinfo{author}{C.~R. Reeves},
\newblock \bibinfo{title}{Genetic {Algorithms}},
\newblock in: \bibinfo{editor}{L.~Liu}, \bibinfo{editor}{M.~T. Özsu} (Eds.),
  \bibinfo{booktitle}{Encyclopedia of {Database} {Systems}},
  \bibinfo{publisher}{Springer}, \bibinfo{address}{New York, NY},
  \bibinfo{year}{2018}, pp. \bibinfo{pages}{1583--1587}. \URLprefix
  \url{https://doi.org/10.1007/978-1-4614-8265-9_562}.
\bibitem[{Sastry et~al.(2005)Sastry, Goldberg, and
  Kendall}]{sastry_genetic_2005}
\bibinfo{author}{K.~Sastry}, \bibinfo{author}{D.~Goldberg},
  \bibinfo{author}{G.~Kendall},
\newblock \bibinfo{title}{Genetic {Algorithms}},
\newblock in: \bibinfo{editor}{E.~K. Burke}, \bibinfo{editor}{G.~Kendall}
  (Eds.), \bibinfo{booktitle}{Search {Methodologies}: {Introductory}
  {Tutorials} in {Optimization} and {Decision} {Support} {Techniques}},
  \bibinfo{publisher}{Springer US}, \bibinfo{address}{Boston, MA},
  \bibinfo{year}{2005}, pp. \bibinfo{pages}{97--125}. \URLprefix
  \url{https://doi.org/10.1007/0-387-28356-0_4}.
\bibitem[{Kumar(2021)}]{kumar_particle_2021}
\bibinfo{author}{D.~P. Kumar},
\newblock \bibinfo{title}{Particle {Swarm} {Optimization}: {The} {Foundation}},
\newblock in: \bibinfo{editor}{B.~A. Mercangöz} (Ed.),
  \bibinfo{booktitle}{Applying {Particle} {Swarm} {Optimization}: {New}
  {Solutions} and {Cases} for {Optimized} {Portfolios}},
  \bibinfo{publisher}{Springer International Publishing},
  \bibinfo{address}{Cham}, \bibinfo{year}{2021}, pp. \bibinfo{pages}{97--110}.
  \URLprefix \url{https://doi.org/10.1007/978-3-030-70281-6_6}.
\bibitem[{Stützle(2009)}]{stutzle_ant_2009}
\bibinfo{author}{T.~Stützle},
\newblock \bibinfo{title}{Ant {Colony} {Optimization}},
\newblock in: \bibinfo{editor}{M.~Ehrgott}, \bibinfo{editor}{C.~M. Fonseca},
  \bibinfo{editor}{X.~Gandibleux}, \bibinfo{editor}{J.-K. Hao},
  \bibinfo{editor}{M.~Sevaux} (Eds.), \bibinfo{booktitle}{Evolutionary
  {Multi}-{Criterion} {Optimization}}, \bibinfo{publisher}{Springer},
  \bibinfo{address}{Berlin, Heidelberg}, \bibinfo{year}{2009}, pp.
  \bibinfo{pages}{2--2}. \DOIprefix\doi{10.1007/978-3-642-01020-0_2}.
\bibitem[{Świechowski et~al.(2023)Świechowski, Godlewski, Sawicki, and
  Mańdziuk}]{swiechowski_monte_2023}
\bibinfo{author}{M.~Świechowski}, \bibinfo{author}{K.~Godlewski},
  \bibinfo{author}{B.~Sawicki}, \bibinfo{author}{J.~Mańdziuk},
\newblock \bibinfo{title}{Monte {Carlo} {Tree} {Search}: {A} {Review} of
  {Recent} {Modifications} and {Applications}},
\newblock \bibinfo{journal}{Artificial Intelligence Review}
  \bibinfo{volume}{56} (\bibinfo{year}{2023}) \bibinfo{pages}{2497--2562}.
\bibitem[{Winands(2024)}]{winands_monte-carlo_2024}
\bibinfo{author}{M.~H.~M. Winands},
\newblock \bibinfo{title}{Monte-{Carlo} {Tree} {Search}},
\newblock in: \bibinfo{editor}{N.~Lee} (Ed.), \bibinfo{booktitle}{Encyclopedia
  of {Computer} {Graphics} and {Games}}, \bibinfo{publisher}{Springer
  International Publishing}, \bibinfo{address}{Cham}, \bibinfo{year}{2024}, pp.
  \bibinfo{pages}{1179--1184}. \URLprefix
  \url{https://doi.org/10.1007/978-3-031-23161-2_12}.
\bibitem[{Auer et~al.(2002)Auer, Cesa-Bianchi, Freund, and
  Schapire}]{auer_nonstochastic_2002}
\bibinfo{author}{P.~Auer}, \bibinfo{author}{N.~Cesa-Bianchi},
  \bibinfo{author}{Y.~Freund}, \bibinfo{author}{R.~E. Schapire},
\newblock \bibinfo{title}{The {Nonstochastic} {Multiarmed} {Bandit} {Problem}},
\newblock \bibinfo{journal}{SIAM Journal on Computing} \bibinfo{volume}{32}
  (\bibinfo{year}{2002}) \bibinfo{pages}{48--77}.

\end{thebibliography}





\end{document}